\documentclass[12pt,letterpaper]{article}
\pdfoutput=1
\usepackage[includeheadfoot,
            marginratio={1:1,2:3}, 
            width=412pt, 
            height=688pt,]{geometry}

\usepackage{amsmath}
\usepackage{amsfonts}
\usepackage{amssymb}
\usepackage{graphicx}
\usepackage{cite}

\newcommand{\eq}[1]{\begin{equation}
                     \begin{split} #1 \end{split}
                     \end{equation}}
\newcommand{\ov}{\overline}
\newcommand{\op}{\hspace{1pt}}
\newcommand{\parag}{\,\begin{matrix} \gtrsim \\[-0.3cm]{}_p \end{matrix}\,}
\newcommand{\paras}{\,\begin{matrix} \lesssim \\[-0.3cm]{}_p \end{matrix}\,}
\newcommand{\parasim}{\,\begin{matrix} \simeq \\[-0.4cm]{}_p \end{matrix}\,}
\newcommand{\numg}{\,\begin{matrix} \gtrsim \\[-0.3cm] {}_n \end{matrix}\,}

\allowdisplaybreaks[2]
\numberwithin{equation}{section}

\newcommand{\lab}{\mathsf }
\renewcommand*{\thefootnote}{\fnsymbol{footnote}}

\begin{document}

\vspace*{-1.5cm}
\begin{flushright}
  {\small
MPP-2015-44\\
LMU-ASC-17/15\\
DFPD-2015-TH-8\\
  }
\end{flushright}

\vspace{0.6cm}

\begin{center}
{\LARGE
A Flux-Scaling Scenario for High-Scale \\[0.1cm]
Moduli Stabilization in String Theory\\[0.3cm]
}
\end{center}

\vspace{0.45cm}

\begin{center}
Ralph Blumenhagen$^1$, Anamar\'{\i}a Font$^{1,2}$\footnote{On leave from Departamento de F\'{\i}sica, Facultad de Ciencias, Universidad Central de Venezuela}, Michael Fuchs$^1$, \\
Daniela Herschmann$^1$, Erik Plauschinn$^{3,4}$, Yuta Sekiguchi$^{1,2}$, Florian Wolf\op$^{1,2}$
\end{center}

\vspace{0.1cm}

\begin{center} 
\emph{$^{1}$ Max-Planck-Institut f\"ur Physik (Werner-Heisenberg-Institut), \\ 
F\"ohringer Ring 6,  80805 M\"unchen, Germany } \\[0.3cm] 
\emph{$^{2}$ Arnold Sommerfeld Center for Theoretical Physics,\\ 
LMU, Theresienstr.~37, 80333 M\"unchen, Germany} \\[0.3cm] 
\emph{$^{3}$ Dipartimento di Fisica e Astronomia ``Galileo Galilei'', \\
Universit\`a  di Padova, Via Marzolo 8, 35131 Padova, Italy}  \\[0.3cm] 
\emph{$^{4}$ INFN, Sezione di Padova, \\
Via Marzolo 8, 35131 Padova, Italy}  
\end{center} 

\vspace{0.6cm}

%%%%%%%%%%%%%%%%%%%%%%%%%%%%%%%%%%%%%%%%%%%%%%%
%%%%%%%%%%%%%%%%%%%%%%%%%%%%%%%%%%%%%%%%%%%%%%%
%%%%%%%%%%%%%%%%%%%%%%%%%%%%%%%%%%%%%%%%%%%%%%%
%%%%%%%%%%%%%%%%%%%%%%%%%%%%%%%%%%%%%%%%%%%%%%%
%%%%%%%%%%%%%%%%%%%%%%%%%%%%%%%%%%%%%%%%%%%%%%%
%%%%%%%%%%%%%%%%%%%%%%%%%%%%%%%%%%%%%%%%%%%%%%%
%%%%%%%%%%%%%%%%%%%%%%%%%%%%%%%%%%%%%%%%%%%%%%%
%%%%%%%%%%%%%%%%%%%%%%%%%%%%%%%%%%%%%%%%%%%%%%%

\renewcommand*{\thefootnote}{\arabic{footnote}}
\setcounter{footnote}{0}

\begin{abstract}
\noindent
Tree-level moduli stabilization via geometric and non-geometric fluxes
in type IIB orientifolds on Calabi-Yau manifolds is investigated. The
focus is on stable non-supersymmetric minima, where all moduli are
fixed except for some massless axions. The scenario includes the
purely axionic orientifold-odd moduli. A set of vacua allowing for
parametric control over the moduli vacuum expectation values and their
masses is presented, featuring a specific scaling with the fluxes.
Uplift mechanisms and supersymmetry breaking soft masses on MSSM-like
D7-branes are discussed as well. This scenario provides a complete
effective framework for realizing the idea of F-term axion monodromy
inflation in string theory. It is argued that, with
all masses close to the Planck and GUT scales, one is confronted with
working at the threshold of controlling all mass hierarchies.

\end{abstract}

\vspace*{0.4cm}

%%%%%%%%%%%%%%%%%%%%%%%%%%%%%%%%%%%%%%%%%%%%%%%
%%%%%%%%%%%%%%%%%%%%%%%%%%%%%%%%%%%%%%%%%%%%%%%

\clearpage

\tableofcontents

%%%%%%%%%%%%%%%%%%%%%%%%%%%%%%%%%%%%%%%%%%%%%%%
%%%%%%%%%%%%%%%%%%%%%%%%%%%%%%%%%%%%%%%%%%%%%%%
%%%%%%%%%%%%%%%%%%%%%%%%%%%%%%%%%%%%%%%%%%%%%%%
%%%%%%%%%%%%%%%%%%%%%%%%%%%%%%%%%%%%%%%%%%%%%%%
%%%%%%%%%%%%%%%%%%%%%%%%%%%%%%%%%%%%%%%%%%%%%%%
%%%%%%%%%%%%%%%%%%%%%%%%%%%%%%%%%%%%%%%%%%%%%%%
%%%%%%%%%%%%%%%%%%%%%%%%%%%%%%%%%%%%%%%%%%%%%%%
%%%%%%%%%%%%%%%%%%%%%%%%%%%%%%%%%%%%%%%%%%%%%%%

\section{Introduction}
\label{sec:intro}

The central element in relating string theory to the real world is
moduli stabilization, i.e. a dynamical mechanism that gives a mass to the 
ubiquitous massless scalar fields.
Most of the more detailed questions about string phenomenology and string cosmology 
can only be answered in  a framework of moduli stabilization.
Of course, it would be a big advance to isolate generic but specific 
predictions of string theory derived models, but so far there are only
very general predictions, such as the existence of supersymmetry at a
high scale, the existence  of axions,  gauge interactions with
chiral fermions and the existence of inflaton candidates.

The usual approach to 
moduli stabilization
\cite{Blumenhagen:2005mu, Grana:2005jc, Douglas:2006es, Blumenhagen:2006ci, Denef:2007pq, Ibanez:2012zz} 
is to start with
an ${\cal N}=1$  supersymmetric  compactification to four dimensions of one of the ten-dimensional
superstring theories, and then generate 
a scalar potential for  the many moduli by taking into account
additional ingredients. These are tree-level background fluxes as well as 
perturbative and non-perturbative corrections to the K\"ahler potential and the
superpotential. Once these data are specified, one can compute
the resulting scalar potential for the moduli and search for
minima, which can either preserve supersymmetry or break it spontaneously.
A {\it scenario} of moduli stabilization is a restricted  set-up, where
a certain type of minima is guaranteed to exist and where one has
parametric control over the emerging scales of the vacuum expectation
values and masses  of the moduli. 
It is fair to say that in view of the vast landscape, so far there
only exist 
few such scenarios. The most studied ones are
the racetrack, the KKLT \cite{Kachru:2003aw} 
the large volume scenario (LVS) \cite{Balasubramanian:2005zx,Conlon:2005ki}, and variations
thereof.

The aim of  this paper is to propose 
a scenario of moduli stabilization, which is 
entirely based on the tree-level flux induced scalar potential.
The motivation for this study is two-fold, and to appreciate our
approach and its historical embedding  let us elucidate this point further.

In the first run of LHC no direct indication of supersymmetry has been found, 
so that naturalness as a guiding principle is 
under pressure, and fine-tuning of the Higgs mass (in the string landscape)
might eventually be something we have to face.
In most approaches to string phenomenology, a supersymmetry breaking
scale of the order $M_{\rm susy}\sim 1\,$TeV was used as an input to fix the stringy
scales. Due to $M_{\rm Pl}/M_{\rm susy}\sim 10^{15}$, a moduli stabilization 
scenario, dynamically generating exponential hierarchies,
seemed very natural. This is precisely what the LVS achieves.
However, if $M_{\rm susy}$ is indeed much larger or even close to the
GUT scale, then scenarios generating only polynomial hierarchies 
might also be interesting to consider.

Furthermore, the BICEP2 claim \cite{Ade:2014xna} to have measured primordial 
B-modes with a large tensor-to-scalar
ratio of $r\sim 0.2$ has triggered quite some activity in realizing
large-field inflation models in string theory.  
Although by now there is agreement between the PLANCK and the BICEP2
collaboration that 
the main component of the B-modes is  due to dust in the foreground \cite{Adam:2014oea,Ade:2015tva,Ade:2015lrj},
BICEP2's initial results 
have led to a number of developments in string cosmology.
Invoking string theory is motivated because inflation is UV sensitive.
For instance, for chaotic inflation with a quadratic potential, the
mass scale of inflation is at
$M_{\rm inf}\sim 10^{16}\,$GeV, the Hubble scale of inflation at
$H_{\rm inf}\sim 10^{14}\,$GeV
and the mass of the inflaton is $m_\theta\sim 10^{13}\,$GeV.
Therefore, a mechanism such as the shift symmetry of an axion is
necessary to gain control over
higher-order Planck-suppressed operators.\footnote{
For other symmetry-based mechanisms to suppress higher-order corrections, see \cite{Burgess:2014tja}.}
Various scenarios for axion inflation have been proposed, such as
natural inflation~\cite{Freese:1990rb}, N-flation~\cite{Dimopoulos:2005ac}, 
or aligned inflation~\cite{Kim:2004rp}. During the last year,  it was
analyzed how these scenarios can be embedded into string theory 
\cite{Palti:2014kza,Grimm:2014vva,Kappl:2014lra,Ben-Dayan:2014zsa,Long:2014dta,Gao:2014uha,Ben-Dayan:2014lca,Bachlechner:2014gfa,Burgess:2014oma,Buchmuller:2015oma,Shiu:2015uva,Kappl:2015pxa,Shiu:2015xda,Montero:2015ofa,Brown:2015iha}.

A promising string theory approach, 
still allowing for some control over the higher-order corrections, is
axion monodromy inflation \cite{Silverstein:2008sg,McAllister:2008hb},
for which a field-theory version has been proposed in 
\cite{Kaloper:2008fb,Kaloper:2011jz} (for a review see for instance 
\cite{Westphal:2014ana}).
More recently,  axion monodromy inflation has been realized via the F-term scalar potential induced
by background fluxes \cite{Marchesano:2014mla,Blumenhagen:2014gta,Hebecker:2014eua}.
This has the advantage that supersymmetry is broken
spontaneously by the very same
effect by which usually moduli are stabilized. Various scenarios have
been studied, mostly in the type IIB context, which differ by what kind of axion is identified
as the inflaton.
The latter can be the universal axion \cite{Blumenhagen:2014gta}, a geometric axion like a
complex structure modulus or a ${\rm D}7$-brane deformation developing a shift symmetry in the
large complex structure regime \cite{Hebecker:2014eua,Arends:2014qca,Garcia-Etxebarria:2014wla},
a Higgs-like open string modulus 
\cite{Ibanez:2014kia},  or
the  Kalb-Ramond or Ramond-Ramond (R-R) two-form field
\cite{Marchesano:2014mla,McAllister:2014mpa}.
Moreover, in \cite{Hassler:2014mla} non-geometric fluxes
were employed and the inflaton was given by a K\"ahler modulus.

In view of single-field large-field inflation, a challenge for string theory is to find a scheme of moduli
stabilization such  that a single  axion $\theta$ is the lightest state,
beyond maybe some lighter axions providing candidates for the QCD
axion or dark radiation \cite{Cicoli:2012aq}. In fact,
the challenge is to fix the moduli such that during inflation the following sensitive
hierarchy of scales is  guaranteed 
\eq{ 
M_{\rm Pl}> M_{\rm s}>M_{\rm KK}> M_{\rm inf}\sim M_{\rm mod} >
H_{\inf} > |M_{\theta}|\, ,
}
where neighboring scales differ by (only) a factor of $O(10)$. 
An argument for the second-last  relation was presented  in \cite{Buchmuller:2014vda}.
Such a hierarchy can either appear just by numerical coincidence, 
or by having a parameter that controls the quotient of two scales. 
On the formal side such a string theory realization is constrained
by the no-go theorem of \cite{Conlon:2006tq}. It states that once an axion is
completely  unstabilized in a supersymmetry-preserving minimum, its saxionic partner is tachyonic. 
Therefore, non-supersymmetric minima are a better-suited
starting point.

In \cite{Blumenhagen:2014nba} it was analyzed whether the no-scale scalar potential for
the complex structure and axio-dilaton moduli in
type IIB orientifolds with NS-NS and \mbox{R-R} three-form fluxes admits
non-supersymmetric minima, where a single  axion can be  parametrically
lighter than the rest of the moduli (see also \cite{Hebecker:2014kva} for an
alternative approach invoking tunings in the string landscape). The procedure, that we will also
follow in this paper, is to first turn on large fluxes such that  all
moduli except  one axion are frozen. In a second step, we 
turn on order-one fluxes freezing also the last axion, which was
parametrically lighter than the rest. The main shortcoming of the original
approach  was to neglect the K\"ahler moduli. If they could be
stabilized by some subleading effect (as in KKLT or LVS) their
masses would be  parametrically lighter than the tree-level induced
inflaton mass. Since the F-term monodromy potential for the latter
is a tree-level effect, one should better stabilize all heavy moduli
by a flux induced tree-level potential.

Working in type IIB superstring theory, a superpotential for
the K\"ahler moduli is generated (at tree-level) by turning on non-geometric fluxes.
In this paper, following \cite{Grana:2005ny,Grana:2006hr}, we study  type IIB orientifolds on Calabi-Yau
three-folds and their flux induced scalar potential for the K\"ahler,
the complex structure and the axio-dilaton moduli.  Therefore, the induced
scalar potential is the one of (orientifolded) ${\cal N}=2$ gauged
supergravity \cite{D'Auria:2007ay}, where the
fluxes are considered as small perturbations around the Calabi-Yau
geometry. Vacua of this potential have been discussed in toroidal
backgrounds in 
\cite{Shelton:2005cf, Aldazabal:2006up, Villadoro:2006ia, Shelton:2006fd, Wecht:2007wu,
Gray:2008zs, Font:2008vd, Guarino:2008ik, deCarlos:2009fq, Caviezel:2009tu, deCarlos:2009qm,
Aldazabal:2011yz, Dibitetto:2011gm,Danielsson:2012by, Damian:2013dq, Damian:2013dwa, Blaback:2013ht}
and in more general CY three-folds \cite{Micu:2007rd,Palti:2007pm}.
Here, we examine this framework 
for the existence of a scheme (a subset of fluxes) such that the
following aspects are realized:
\begin{itemize}
\item{There exist non-supersymmetric minima stabilizing the saxions
in their perturbative regime.}
\item{All mass eigenvalues are positive semi-definite, where the
    massless states are only axions.}
\item{For both the values of the moduli in the minima and the mass 
of the heavy moduli one has parametric control in terms of ratios of 
fluxes.}
\item{One has either parametric or at least numerical control over the mass
of the lightest (massive) axion, i.e. the inflaton
candidate.\footnote{Axions staying massless at tree-level can only
  receive
tiny masses from non-perturbative effects and are expected not to interfere with
the moduli dynamics in the early universe.}}
\item{The moduli masses are smaller than the string and the
Kaluza-Klein scale.}
\end{itemize}
Since, as in the  LVS, we deal with a non-supersymmetric minimum,
once this is determined we can continue to study many string-phenomenological
and cosmological questions.
For instance, we can compute the effect of this supersymmetry breaking on the
MSSM-like theory on stacks of ${\rm D}7$-branes, wrapping a four-cycle
not forbidden by Freed-Witten anomalies.

Since we introduce non-geometric fluxes, let us mention some of 
the open questions and limitations of our approach: 
\begin{itemize}
\item{We work in the effective four-dimensional supergravity
    theory. The uplift of new minima of the scalar potential to
   genuine solutions of the ten-dimensional string equations of motion
   is a subtle problem.}
\item{It is often questioned whether there exists a clear separation
    of Kaluza-Klein and moduli mass scales
    that allow to only consider a finite number of modes in the
   effective theory.  }
\item{Here we are mostly interested in the scalar potential and its
    mathematical structure for a small treatable number of moduli. Therefore, we are not carefully specifying
    global Calabi-Yau geometries and  orientifold projections that
    concretely realize our supergravity models.}
\end{itemize}
Concerning the first item, we will  confirm that  for non-geometric fluxes a proper dilute-flux limit does not exist, 
so that backreaction is expected. Our point of view is
that not having an explicit uplift does not mean
that orientifolded  ${\cal N}=2$ gauged supergravity cannot be a consistent
truncation for the dynamics of the string modes kept in the model.

Having expressed our concerns, let us summarize how 
this paper is organized. In section~\ref{sec_theo} we describe the
string theory framework that will be used, i.e. type IIB
orientifolds on Calabi-Yau three-folds with various kinds of geometric
and non-geometric fluxes. These induce a scalar potential
which is the one of (orientifolded) ${\cal N}=2$ gauged supergravity.
This  section partially reviews known results from generalized geometry,
but  also adds some new aspects, like the couplings of orientifold-odd
moduli to geometric fluxes and the inclusion of non-geometric R-R
$P$-flux. We also derive the tadpole cancellation conditions and the  generalized Freed-Witten
anomaly conditions.  Generically, the fluxes fix all closed string moduli 
at tree-level.

In section~\ref{sec_simplemodel} we first present some simple
examples, with a small number of K\"ahler moduli and no
complex structure moduli, that show  a peculiar
pattern. Namely, for $n$ moduli  the superpotential  contains
$n+1$ terms so that the requirement that all terms scale in the same
way with the fluxes
uniquely fixes the scaling of the frozen moduli with the fluxes. By
minimizing the scalar potential we find that this intended scaling behavior
indeed shows up in the  minima. Generically, both supersymmetric AdS and stable
non-supersymmetric AdS minima appear, where the latter can be 
tachyon-free and are our main interest throughout this paper.
 The scaling behavior allows us to gain  parametric control over
the physics in these minima, in particular to fix the dilaton and the K\"ahler
moduli in their perturbative regime, or to adjust the relative sizes
of the string, Kaluza-Klein and moduli-mass scale. 
We also consider models which cannot be realized on tori,
having for instance a K\"ahler potential of swiss-cheese  or
K3-fibration type. Thus, the flux scaling behavior is the tool to design
certain properties in the minimum.

In the second part of section~\ref{sec_simplemodel} we generalize the construction to
models also having complex structure moduli. The more moduli we add, 
the more likely it becomes that one encounters tachyonic directions.
We also discuss more models with odd K\"ahler moduli and non-geometric
$P$-flux. At the end, we describe how we can systematically  search
for this kind of scaling minima.

 In section~\ref{sec_uplift} we investigate whether there exists
 a general mechanism to uplift tachyons, in particular those arising  when more
 than one K\"ahler modulus is involved. Indeed, we find that the D-term of an appropriate stack
of ${\rm D}7$-branes, subject to the generalized Freed-Witten anomalies, 
precisely adds a positive contribution to the mass-square of this
type of tachyons, while not affecting  the masses of the other moduli.
We believe that this is a fairly non-trivial result. In this section
we also discuss the  uplift of the cosmological constant by adding
a simple term of the form $\varepsilon/{\cal V}^\alpha$ with $\alpha>0$. We find 
that the uplift by anti ${\rm D}3$-branes does not work for these tree-level
models in the sense that  $\alpha$ has to be smaller than $4/3$. 

Section~\ref{sec_pheno} is devoted to discussing aspects related to string
phenomenology.
First, we comment on the generic issue of justifying
the existence of a string theory uplift of these flux vacua to
the full ten-dimensional string theory. Then, we concretely evaluate
the various resulting mass scales, which generically are only a few orders
below the Planck-scale. For that purpose we introduce a well-defined  notion of
parametric equality or inequality, respectively. 
We also estimate the tunneling amplitude between different branches
of the flux landscape.
Having a source of supersymmetry breaking in the bulk, we 
compute the gravity-mediated soft-masses on stacks of ${\rm D}7$-branes, both for  a bulk
and a sequestered set-up. For the  latter case anomaly-mediation is
the dominant contribution. Such a sequestered scenario is important for lowering
the supersymmetry breaking scale down to the intermediate or
even the TeV regime.

In section~\ref{sec_cosmo} we analyze the models with respect to the presence of
axions capable of realizing  F-term
axion monodromy inflation. We mostly consider the scenario where an axion
can gain a parametrically small mass via turning on additional fluxes.
In this case we follow the ideas put forward in \cite{Blumenhagen:2014nba} in the context of
no-scale models. Generically, a tension between this kind of
parametric control and the Kaluza-Klein scale shows up. 
In a separate article \cite{Blumenhagen:2015qda} we  discuss a toy
model for this kind of scenario, in which the backreaction
\cite{Dong:2010in}  of the
heavy moduli onto the flow of the inflaton can  be taken into account
analytically.

%%%%%%%%%%%%%%%%%%%%%%%%%%%%%%%%%%%%%%%%%%%%%%%
%%%%%%%%%%%%%%%%%%%%%%%%%%%%%%%%%%%%%%%%%%%%%%%
%%%%%%%%%%%%%%%%%%%%%%%%%%%%%%%%%%%%%%%%%%%%%%%
%%%%%%%%%%%%%%%%%%%%%%%%%%%%%%%%%%%%%%%%%%%%%%%
%%%%%%%%%%%%%%%%%%%%%%%%%%%%%%%%%%%%%%%%%%%%%%%
%%%%%%%%%%%%%%%%%%%%%%%%%%%%%%%%%%%%%%%%%%%%%%%
%%%%%%%%%%%%%%%%%%%%%%%%%%%%%%%%%%%%%%%%%%%%%%%
%%%%%%%%%%%%%%%%%%%%%%%%%%%%%%%%%%%%%%%%%%%%%%%

\section{Fluxes and branes in type IIB orientifolds}
\label{sec_theo}

In this section, we describe the set-up we will be employing in the
following: This is type IIB orientifolds with geometric
and non-geometric fluxes, where the latter are used to stabilize all
moduli at string tree-level. We also derive conditions arising
when fluxes and D-branes are present simultaneously, which
can be considered as generalized Freed-Witten anomaly
cancellation conditions.

%%%%%%%%%%%%%%%%%%%%%%%%%%%%%%%%%%%%%%%%%%%%%%%
%%%%%%%%%%%%%%%%%%%%%%%%%%%%%%%%%%%%%%%%%%%%%%%
%%%%%%%%%%%%%%%%%%%%%%%%%%%%%%%%%%%%%%%%%%%%%%%
%%%%%%%%%%%%%%%%%%%%%%%%%%%%%%%%%%%%%%%%%%%%%%%

\subsection{Orientifold compactifications}

The framework we are considering  is that of type IIB string theory
compactified on orientifolds of Calabi-Yau manifolds $\mathcal M$.
The orientifold projection $\Omega_{\rm P} (-1)^{F_{\rm L}} \sigma$ contains,
besides the world-sheet parity operator $\Omega_{\rm P}$ and the left-moving fermion
number $F_{\rm L}$,
a holomorphic involution $\sigma:{\cal M}\to {\cal M}$. We choose the latter to act on the 
K\"ahler form $J$ and the holomorphic $(3,0)$-form $\Omega_3$ of the Calabi-Yau three-fold $\mathcal M$ as
\eq{
  \label{op_01}
        \sigma^*: J\to +J\,,\hspace{50pt} \sigma^*:\Omega_3\to-\Omega_3\,.
}
The fixed loci of this involution correspond to O$7$- and O$3$-planes, which 
in general require the presence of D$7$- and D$3$-branes  to satisfy the tadpole cancellation
conditions.

%%%%%%%%%%%%%%%%%%%%%%%%%%%%%%%%%%%%%%%%%%%%%%%
%%%%%%%%%%%%%%%%%%%%%%%%%%%%%%%%%%%%%%%%%%%%%%%

\subsubsection{Cohomology}

In order to establish the conventions for our subsequent discussion, let us note the following.
We denote a symplectic basis for the third cohomology of the Calabi-Yau manifold $\mathcal M$ by
\eq{
  \{\alpha_{\Lambda},\beta^{\Lambda}\} \in H^3(\mathcal M) \,, \hspace{60pt}
  \Lambda =0,\ldots, h^{2,1} \,,
}
which can be chosen such that the only non-vanishing pairings satisfy
\eq{
  \label{symp_01}
  \int_{\mathcal M} \alpha_{\Lambda}\wedge \beta^{\Sigma} = \delta_{\Lambda}{}^{\Sigma} \,.
}
For the $(1,1)$- and $(2,2)$-cohomology of $\mathcal M$  we introduce bases of the form
\eq{
  \arraycolsep2pt
  \begin{array}{ccl}
  \{ \omega_{\lab A} \}  &\in&  H^{1,1}(\mathcal M) \,, \\[5pt]
  \{ \tilde\omega^{\lab A} \} & \in & H^{2,2}(\mathcal M) \,,
  \end{array}
  \hspace{50pt} \lab A = 1,\ldots, h^{1,1} \,,
}  
and for later convenience we also define $\{ \omega_{ A} \}  =  \{ 1, \omega_{\lab A}\}$ and
$\{ \tilde\omega^{ A} \} = \{ d{\rm vol}_6, \tilde\omega^{\lab A}\}$, with $ A = 0,\ldots, h^{1,1}$.
The latter two bases are chosen as
\eq{
  \int_{\mathcal M} \omega_{ A}\wedge \tilde\omega^{ B} = \delta_{ A}{}^{ B} \,.
}

Turning to the orientifold projection, we have to take into account that the holomorphic involution $\sigma$ shown in 
equation \eqref{op_01} 
splits the cohomology into even and odd parts. This means in particular that 
\eq{
 H^{p,q}(\mathcal M) = H^{p,q}_+(\mathcal M) \oplus H^{p,q}_-(\mathcal M) \,,
 \hspace{50pt}
 h^{p,q} = h^{p,q}_+ + h^{p,q}_-\,.
}
We also note that constants as well as the volume form $d{\rm vol}_6$ on $\mathcal M$ are always even
under the involution. For the other bases introduced above,
we employ the following notation
\eq{
  \label{basis_coho}
  \arraycolsep1.5pt
  \renewcommand{\arraystretch}{1.4}
  \begin{array}{rcl@{\hspace{12pt}}lcl@{\hspace{22.5pt}}rcl@{\hspace{12pt}}lcl}
  \{\omega_{\alpha}\} & \in & H^{1,1}_+(\mathcal M) & \alpha &=& 1,\ldots,h^{1,1}_+ ,&
  \{\omega_{a}\} & \in & H^{1,1}_-(\mathcal M) & a &=& 1,\ldots,h^{1,1}_-,
  \\
  \{\tilde\omega^{\alpha}\} & \in & H^{2,2}_+(\mathcal M) & \alpha &=& 1,\ldots,h^{1,1}_+ ,&
  \{\tilde\omega^{a}\} & \in & H^{2,2}_-(\mathcal M) & a &=& 1,\ldots,h^{1,1}_-,
  \\
  \{\alpha_{\hat\lambda},\beta^{\hat\lambda}\} & \in & H^{3}_+(\mathcal M)
  &\hat\lambda &=& 1,\ldots,h^{2,1}_+, &
  \{\alpha_{\lambda},\beta^{\lambda}\} & \in & H^{3}_-(\mathcal M)
  &\lambda &=& 0,\ldots,h^{2,1}_-.
  \end{array}
  \\[5pt]
}

%%%%%%%%%%%%%%%%%%%%%%%%%%%%%%%%%%%%%%%%%%%%%%%
%%%%%%%%%%%%%%%%%%%%%%%%%%%%%%%%%%%%%%%%%%%%%%%

\subsubsection{Moduli fields}

Compactifications of type IIB string theory on Calabi-Yau orientifolds with O$7$- and
O$3$-planes are well-studied. Here, we recall only some results
which are needed below; for more details we would like to refer the reader to the original papers
\cite{Brunner:2003zm,Grimm:2004uq,Jockers:2004yj}, and for a broader overview for instance to
\cite{Blumenhagen:2006ci,Plauschinn:2010zz}.

We first note that under the combined world-sheet parity and left-moving fermion number 
operator $\Omega_{\rm P}(-1)^{F_{\rm L}}$ the ten-dimensional bosonic fields in type IIB string theory transform
as 
\eq{
  \Omega_{\rm P}(-1)^{F_{\rm L}} = \left\{ \begin{array}{l@{\hspace{20pt}}l}
  g,\, \phi,\, C_0, \, C_4 & {\rm even}\,, \\[4pt]
  B_2, \, C_2 &{\rm odd}\,,
  \end{array}
  \right.
}
where $g$, $\phi$, $B_2$ are the metric, dilaton and Kalb-Ramond field,
and $C_p$ denote the Ramond-Ramond potentials.
The components of the ten-dimensional form fields which are purely 
in the six-dimensional space $\mathcal M$ can then be expanded as
\eq{
  \label{expansion_01}
  e^{-\phi/2}J = t^{\alpha} \omega_{\alpha} \,, \hspace{37pt}
  B_2 = b^a \omega_a\,, \hspace{37pt}
  C_2 = c^a \omega_a\,, \hspace{37pt}
  C_4 = \rho_{\alpha}\op \tilde\omega^{\alpha} \,,
}  
where the factor of $e^{-\phi/2}$ for $J$ has been included for later convenience.
It implies that $\{t^{\alpha}\}$ is expressed in Einstein frame.
We also note that the potential $C_4$ appears in the five-form field strength
as $\widetilde F_5 = dC_4 - C_2 \wedge d B_2$.
The moduli fields of the effective four-dimensional theory 
after compactification are summarized in table~\ref{table_moduli} (see  \cite{Grimm:2004uq} for more details).
%%%%%%%%%%%%%%
%%%%%%%%%%%%%%
\begin{table}[t]
\centering
\renewcommand{\arraystretch}{1.3}
\begin{tabular}{|c|l@{\hspace{1pt}}l|c|}
  \hline
   number & \multicolumn{2}{c|}{modulus} &  name \\
  \hline\hline
  $1$ & $S$&$=e^{-\phi}-i\op C_0$ & axio-dilaton \\
  $h^{2,1}_-$ & $U^i$&$=v^i+i\op u^i$ & complex structure\\
 $h^{1,1}_+$ & $T_\alpha$&$=\tau_\alpha+ i \op \rho_\alpha+ \ldots$ & K\"ahler \\
 $h^{1,1}_-$ & $G^a$&$=S\op b^a+i\op c^a$ & axionic odd\\
\hline
     \end{tabular} 
     \caption{\small Moduli in type IIB orientifold compactifications.}
      \label{table_moduli}
\end{table}
%%%%%%%%%%%%%%
%%%%%%%%%%%%%%
The full definition of the K\"ahler moduli $T_{\alpha}$ is 
given by
\eq{
\label{defkaehler}
    T_\alpha=\frac{1}{2}\op\kappa_{\alpha\beta\gamma} t^\beta t^\gamma
   +i\left(\rho_\alpha-\frac{1}{2}\op\kappa_{\alpha a b} c^a b^b\right)
  -\frac{1}{4} \op e^\phi  \kappa_{\alpha a b} {G}^a (G+\ov G)^b \,,
}
where the triple intersection numbers  are defined as
$\kappa_{\mathsf {ABC}} = \int_{\mathcal M} \omega_{\mathsf A}\wedge
\omega_{\mathsf B}\wedge\omega_{\mathsf C}$. Note that since 
the holomorphic involution $\sigma$ has to leave the constants $\kappa_{\mathsf{ABC}}$ 
invariant, it follows that 
the components $\kappa_{abc}$ and $\kappa_{a\beta\gamma}$ are vanishing.

The complex structure moduli $U^i$ are contained in the holomorphic three-form $\Omega_3$.
The latter can be expanded in the basis of odd three-forms shown in \eqref{basis_coho}
as follows
\eq{
  \label{exp_02}
  \Omega_3 = X^{\lambda} \alpha_{\lambda} - F_{\lambda} \op\beta^{\lambda} \,.
}
Usually, the periods $F_{\lambda}$ can be expressed as derivatives 
$F_{\lambda} = \partial F/\partial X^{\lambda}$ of a prepotential $F$. 
In the large complex structure limit ${\rm Re}\op U^i \gg 1$, the prepotential takes the form
\eq{
  \label{prepot}
   {F}=\frac{d_{ijk} \op{X}^i{X}^j{X}^k}{ {X}^0} \,, 
   \hspace{50pt}
   i = 1,\ldots,h^{2,1}_- \,,
} 
where the constants $d_{ijk}$ are symmetric in their indices.
In terms of the periods $X^{\lambda}$, the complex structure moduli are  given by
\eq{
  U^i = v^i + i\op u^i = -i\op\frac{X^i}{X^0} \,.
}
Note that  $X^0$ does not contain any physical information and can be chosen as $X^0=1$.
For later reference, we also note the following relation 
\eq{
  \label{choice_periods}
  0<- i \int_{\mathcal M} \Omega_3 \wedge \ov \Omega_3 =
 8\op \bigl| X^0 \bigr|^2\, d_{ijk}v^iv^jv^k 
  \hspace{40pt}{\rm for~}h^{2,1}_-\neq 0\,.
}
For $h^{2,1}_-=0$ the right hand side in \eqref{choice_periods}
is replaced by $ 2\op \bigl| X^0 \bigr|^2\: {\rm Im}\op \bigl( F_0/X^0 \bigr)$
and we usually take $F_0=i$.

%%%%%%%%%%%%%%%%%%%%%%%%%%%%%%%%%%%%%%%%%%%%%%%
%%%%%%%%%%%%%%%%%%%%%%%%%%%%%%%%%%%%%%%%%%%%%%%

\subsubsection{Scalar potential}

After compactifying the ten-dimensional theory on a Calabi-Yau orientifold, 
the F-term potential of the four-dimensional theory is given by the standard 
supergravity formula
\eq{
  \label{f_pot}
  V_F = \frac{M_\text{Pl}^4}{4 \pi} \,\, e^{K} \Bigl( K^{I\ov J} D_IW D_{\ov J}\ov W - 3 \op\bigl|W\bigr|^2 \Bigr) \,,
}  
expressed in terms of a K\"ahler potential $K$, the corresponding K\"ahler metric
$K_{I\ov J} = \partial_I \partial_{\ov J}\op K$
and a superpotential $W$. The K\"ahler-covariant derivative is given by
$D_I W = \partial_I W + (\partial_I K)\op W$, and the sum runs over all holomorphic and 
anti-holomorphic fields in the theory. At tree-level and in the large-volume regime, the K\"ahler potential reads
\eq{
  \label{k_pot}
      K=-\log\left(-i\int_{\mathcal M} \Omega\wedge \ov{\Omega}\right)-\log\bigl(S+\ov S\bigr)
     -2\log {\cal V} \,,
}
where ${\cal V}=\frac{1}{6} \op\kappa_{\alpha\beta\gamma} t^\alpha t^\beta
t^\gamma$ denotes the volume of the Calabi-Yau three-fold $\mathcal M$ in Einstein frame. It is expressed 
in terms of the two-cycle volumes $t^{\alpha}$ introduced in 
\eqref{expansion_01}.
Note that in order to write $\mathcal V$  in terms of the moduli fields $T_{\alpha}, G^a, S$, one has to
invert the relation \eqref{defkaehler}. 
We furthermore observe that the K\"ahler potential \eqref{k_pot}
satisfies a no-scale relation \cite{Grimm:2004uq}
\eq{
       K^{I\ov J} (\partial_I  K) (\partial_{\ov J}  K) =4\,,
}
where the sum runs over the axio-dilaton $S$, and the even and odd moduli
$T_\alpha$ and $G^a$. However, perturbative 
corrections to the K\"ahler potential will spoil this no-scale structure.

%%%%%%%%%%%%%%%%%%%%%%%%%%%%%%%%%%%%%%%%%%%%%%%
%%%%%%%%%%%%%%%%%%%%%%%%%%%%%%%%%%%%%%%%%%%%%%%
%%%%%%%%%%%%%%%%%%%%%%%%%%%%%%%%%%%%%%%%%%%%%%%
%%%%%%%%%%%%%%%%%%%%%%%%%%%%%%%%%%%%%%%%%%%%%%%

\subsection{Geometric and non-geometric fluxes}

The moduli fields shown in table~\ref{table_moduli} are a priori massless, and 
therefore are in conflict with
experimental observations. However, by turning on fluxes on the Calabi-Yau 
manifold $\mathcal M$, a mass term for the moduli can be generated.
Usually, one considers the three-form flux
\eq{
  G_3={\mathfrak F}-i \op S\op H \,,
}
where $\mathfrak F = \langle dC_2\rangle $ and $H = \langle dB_2\rangle $ are fluxes 
for the two-form potentials $C_2$ and $B_2$.
These fluxes can be expanded in the basis of three-forms as
\eq{
  \label{exp_01}
  \mathfrak F = -\tilde{\mathfrak f}^{\Lambda} \op \alpha_{\Lambda} +
  \mathfrak f_{\Lambda} \op\beta^{\Lambda}\,,
  \hspace{50pt}
  H = -\tilde{h}^{\Lambda} \op \alpha_{\Lambda} +
  h_{\Lambda} \op\beta^{\Lambda}\,.
}
In addition to the $\mathfrak F$- and $H$-flux, in this work we also take into account 
geometric and non-geometric fluxes
$F^I{}_{JK}$, $Q_I{}^{JK}$ and $R^{IJK}$.
In the context of type IIB orientifolds, these fluxes have been studied for instance in
\cite{Shelton:2005cf,Shelton:2006fd,Grana:2006hr,Micu:2007rd,Cassani:2007pq}; here we will not repeat this analysis but
recall only those expressions needed in our discussion.

%%%%%%%%%%%%%%%%%%%%%%%%%%%%%%%%%%%%%%%%%%%%%%%
%%%%%%%%%%%%%%%%%%%%%%%%%%%%%%%%%%%%%%%%%%%%%%%

\subsubsection{Twisted differential and Bianchi identities}

Following the approach of \cite{Shelton:2005cf,Shelton:2006fd,Grana:2006hr,Micu:2007rd}, 
let us introduce a twisted differential acting on $p$-forms. This differential contains
the constant fluxes $H$, $F$, $Q$ and $R$, and is given by
\eq{
  \label{d_operator_01}
  \mathcal D = d - H\wedge\: - F\circ\: - Q\bullet\: - R\,\llcorner \,,
}
where the operators appearing in \eqref{d_operator_01} implement the mapping
\eq{
  \renewcommand{\arraystretch}{1.2}
  \arraycolsep3pt
  \begin{array}{l@{\hspace{7pt}}c@{\hspace{12pt}}lcl}
  H\,\wedge & :& \mbox{$p$-form} &\to& \mbox{$(p+3)$-form} \,, \\
  F\,\circ & :& \mbox{$p$-form} &\to& \mbox{$(p+1)$-form} \,, \\
  Q\,\bullet & :& \mbox{$p$-form} &\to& \mbox{$(p-1)$-form} \,, \\  
  R\,\llcorner & :& \mbox{$p$-form} &\to& \mbox{$(p-3)$-form} \,.  
  \end{array}
}
For the present example of a Calabi-Yau three-fold, we can be more specific about the action of
$\mathcal D$.  Recalling our notation \eqref{basis_coho} and following  \cite{Grana:2006hr}, we introduce the geometric and non-geometric fluxes as
\eq{\label{deffluxes}
\arraycolsep2pt
\begin{array}{lcl@{\hspace{1.5pt}}l@{\hspace{40pt}}lcr@{\hspace{1.5pt}}l}
\mathcal D\alpha_\Lambda &=& q_{\Lambda}{}^{ A} \omega_{ A}
&+\,  f_{\Lambda \op A}\tilde\omega^{ A}\,,
&
\mathcal D\beta^\Lambda &=& \tilde q^{\Lambda \op A} \omega_{ A}
&+ \,\tilde f^{\Lambda}{}_ { A} \tilde\omega^{ A}\,, 
\\[8pt]
\mathcal D\omega_{ A}&=& \tilde f^{\Lambda}{}_{ A} \alpha_\Lambda &-  \,
f_{\Lambda  A}\beta^\Lambda\,,
&
\mathcal D\tilde\omega^{ A} &=& -\tilde q^{\Lambda\op  A} \alpha_\Lambda &+  
\,q_{\Lambda}{}^{ A} \beta^\Lambda\,.
\end{array}
}
Here, $f_{\Lambda \op A}$ and $\tilde f^{\Lambda}{}_ { A}$ denote the geometric
fluxes, while $q_{\Lambda}{}^{ A}$ and $\tilde q^{\Lambda\op  A}$ are the
non-geometric ones.
Moreover, we use the following convention  for the $H$- and $R$-flux
\eq{
\label{fluxzerocomp}
\arraycolsep2pt
\begin{array}{lcl@{\hspace{70pt}}lcl}
f_{\Lambda \op0}&=&h_\Lambda\,, & \tilde f^{\Lambda}{}_0&=&\tilde h^\Lambda\,,\\[5pt]
q_{\Lambda}{}^0&=&r_\Lambda\,,& \tilde q^{\Lambda\op 0}&=&\tilde r^\Lambda\, .
\end{array}
}
Imposing then a nilpotency condition of the form $\mathcal D^2=0$ leads to the 
well-known Bianchi identities for the fluxes \cite{Shelton:2005cf}
\eq{
\label{Bianchi}
\arraycolsep2pt
\begin{array}{lcl@{\hspace{50pt}}lcl}
0&=&\tilde q^{\Lambda \op A} \tilde f^\Sigma{}_{ A} -  \tilde f^\Lambda{}_{ A} \tilde q^{\Sigma \op  A}\,,
&
0&=& q_{\Lambda}{}^{ A}  f_{\Sigma\op  A} -  f_{\Lambda \op  A}  q_{\Sigma}{}^ { A}\,,
\\[7pt]
0&=&q_{\Lambda}{}^{ A}  \tilde f^\Sigma{}_{ A}  -   f_{\Lambda\op  A}  \tilde q^{\Sigma \op A}\,,
&
0&=& \tilde f^\Lambda{}_{ A} q_{\Lambda}{}^{ B} - f_{\Lambda\op A} \tilde q^{\Lambda\op   B}\,.
\\[7pt]
0&=&\tilde f^\Lambda{}_{ A} f_{\Lambda\op  B}- f_{\Lambda \op A} \tilde f^\Lambda{}_{ B}\,,
&
0&=& \tilde q^{\Lambda \op A} q_{\Lambda}{}^{ B}- q_{\Lambda}{}^{ A}  \tilde q^{\Lambda\op  B}\,.
\end{array}
}

We now want to take the orientifold projection into account. To do so, we first
note that under the combined world-sheet parity and left-moving fermion-number 
transformation, the five types of fluxes  behave as
\eq{
\Omega_{\rm P} (-1)^{F_{\rm L}}:
\renewcommand{\arraystretch}{1.2}
\arraycolsep2pt
\left\{
\begin{array}{lcc@{\hspace{1.75pt}}l@{}l}
  \:{\mathfrak F} &\to& -& {\mathfrak F} &\,,\\ 
  H &\to& - &H &\,,\\
  F&\to& &F &\,,\\
  Q&\to& -&Q &\,,\\
  R&\to& &R &\,.
\end{array}
\right.
}
Thus, under $\Omega_{\rm P} (-1)^{F_{\rm L}}$ only the fluxes $F$ and $R$ are even.
Including the holomorphic involution  $\sigma$
defined in \eqref{op_01} and recalling \eqref{basis_coho},
we can deduce the non-vanishing flux components as follows
\eq{
\label{op_02}
\renewcommand{\arraystretch}{1.2}
\begin{array}{l@{\hspace{6pt}}c@{\hspace{18pt}}llll}
{\mathfrak F}&:& && {\mathfrak f}_{\lambda}\,, & \tilde {\mathfrak f}^{\lambda} \,,\\
H&:& && h_{\lambda}\,, &  \tilde h^{\lambda} \,,\\
F&:& f_{\hat \lambda \,\alpha}\,, &  \tilde f^{\hat\lambda}{}_{\alpha}\,, & f_{\lambda \,a}\,, &  \tilde f^{\lambda}{}_{a}\,, \\
Q&:& q_{\hat\lambda}{}^a\,, & \tilde q^{\hat\lambda\,a}\,, & q_{\lambda}{}^{\alpha} \,, & \tilde q^{\lambda\,\alpha}\,, \\
R&:& r_{\hat\lambda} \,, & \tilde r^{\hat\lambda} \,.
\end{array}
}

%%%%%%%%%%%%%%%%%%%%%%%%%%%%%%%%%%%%%%%%%%%%%%%
%%%%%%%%%%%%%%%%%%%%%%%%%%%%%%%%%%%%%%%%%%%%%%%

\subsubsection{Superpotential}

Let us turn to the scalar F-term potential \eqref{f_pot}. The K\"ahler potential appearing
in $V_F$ is shown in equation \eqref{k_pot}, whereas the superpotential $W$ induced by
the background fluxes will be determined in the following.
For non-trivial fluxes $\mathfrak F$ and $H$, it has been shown 
in \cite{Benmachiche:2006df} that the superpotential takes the form
\eq{
  \label{s_pot_01}
  W^{(1)} = \int_{\mathcal M} \Bigl[ \mathfrak F + d_H \Phi^{\rm ev}_{\rm c} \Bigr]_{3} \wedge \Omega_3
  \,,
}
where, in the present conventions, the complex multi-form of even degree $\Phi^{\rm ev}_{\rm c}$ is 
defined as follows 
\eq{
  \Phi^{\rm ev}_{\rm c} = i\op S -i\op G^a \omega_a -i \op T_{\alpha} \op\tilde\omega^{\alpha} \,.
}
The subscript on the parentheses in \eqref{s_pot_01} means that the three-form part of a multi-form 
should be selected, and the operator $d_H$ is defined as $d_H = d - H\wedge$. 
Evaluating then \eqref{s_pot_01} leads to the familiar Gukov-Vafa-Witten superpotential~\cite{Gukov:1999ya}.

However, in order to account for other geometric as well as non-geometric fluxes, 
the authors in \cite{Shelton:2005cf} (see also \cite{Shelton:2006fd,Micu:2007rd,Cassani:2007pq}) proposed to
replace the operator $d_H$ in \eqref{s_pot_01} by $\mathcal D$ defined in \eqref{d_operator_01},
that is
\eq{
  \label{replace}
  d_H \to \mathcal D \,.
}
We mention that in \cite{Micu:2007rd}, the case $h^{1,1}_-=0$ was studied, which we generalize here
to $h^{1,1}_-\neq0$. The superpotential we are therefore considering is expressed as
\eq{
  \label{s_pot_02}
  W^{(2)} &= \int_{\mathcal M} \Bigl[ \,\mathfrak F + \mathcal D \op\Phi^{\rm ev}_{\rm c} \Bigr]_{3} \wedge \Omega_3
  \\[2pt]
  &= \int_{\mathcal M} \Bigl[\,\mathfrak F - i\op S \op H + i\op G^a \op (F \circ \omega_a)  
   + i\op T_{\alpha} \op \bigl( Q \bullet \tilde \omega^{\alpha}\bigr) \Bigr]_{3} \wedge \Omega_3
  \,.
}
Employing then the expansions \eqref{exp_02} and \eqref{exp_01} together with \eqref{symp_01},
and using the action of $\mathcal D$ on the cohomology defined in \eqref{deffluxes}, 
we find the following expression for the superpotential
\eq{
\label{thebigW}
  \arraycolsep2pt
  \begin{array}{rcl@{\hspace{1pt}}c@{\hspace{1.5pt}}c@{\hspace{1.5pt}}lcc@{\hspace{1.5pt}}c@{\hspace{1,5pt}}l}
  W^{(2)}=\:
  &-&&\bigl(&{\mathfrak f}_\lambda  &X^\lambda &-&\tilde {\mathfrak f}^\lambda & F_\lambda & \bigr) \\[4pt]
  &+&i\op S& \big(& h_\lambda&  X^\lambda &-& \tilde h^\lambda & F_\lambda &\bigr) \\[4pt]
  &-&i\op G^a & \bigl(& f_\lambda{}_a &  X^\lambda &-& \tilde f^\lambda{}_a & F_\lambda &\bigr) \\[4pt]
  &+&i\op T_{\alpha} & \bigl(& q_\lambda{}^\alpha &  X^\lambda &-& \tilde q^\lambda{}^\alpha & F_\lambda&\bigr)\,.
  \end{array}
}
Note that the fluxes are subject to the Bianchi identities
\eqref{Bianchi}. We observe  that the $R$-flux does not
appear in $W^{(2)}$ and that the geometric flux couples to the odd moduli $G^a$.
Furthermore, the peculiar feature of this superpotential is that it only depends
linearly on the three kinds of moduli $S,G^a,T_\alpha$.

%%%%%%%%%%%%%%%%%%%%%%%%%%%%%%%%%%%%%%%%%%%%%%%
%%%%%%%%%%%%%%%%%%%%%%%%%%%%%%%%%%%%%%%%%%%%%%%

\subsubsection{Contribution to the tadpoles}

The fluxes appearing in the superpotential \eqref{s_pot_02} contribute
to the tadpole cancellation conditions. In general, for a D$p$-brane charge the tadpole 
cancellation conditions take the form 
\eq{
  \label{tadpole_00}
  N^{\rm flux}_{{\rm D}p} + \sum_{
  \substack{\mbox{\scriptsize D-branes~} \\  \mbox{\scriptsize O-planes~}}i
  } Q_{{\rm D}p}^{(i)}
  =0 \,,
}
where the sum runs over all  D-branes and 
orientifold planes present in the setting.
The flux part can be derived by varying the type IIB action with respect to the 
R-R potentials $C_p$ (in the democratic formulation \cite{Bergshoeff:2001pv}).
We find that \cite{Plauschinn:2008yd}
\eq{
  \delta_{C_p} \,
  \mathcal{S}_{\rm IIB} 
  = \frac{1}{2\kappa_{10}^2} 
  \int_{\mathbb R^{3,1}\times \mathcal M}
   \frac{(-1)^{\frac{p}{2}}}{2}\op \, \delta C_p  \wedge \Bigl[ \, ( d - H_3\wedge ) \,\widetilde{\mathfrak F} \,\Bigr]_{10-p} \,,
}
where $\widetilde{\mathfrak F}_p=d\,C_{p-1}-H_3\wedge C_{p-3}$ denotes the generalized R-R field strength.
To obtain the contribution of the non-geometric fluxes, 
we then perform the replacement shown in \eqref{replace} (see also \cite{Micu:2007rd}).

The ${\rm D}3$-brane tadpole originates from the variation with respect to $C_4$, and leads to the familiar 
$H\wedge\mathfrak F$ expression. For the ${\rm D}5$- and ${\rm D}7$-brane tadpole 
we consider the variation with respect to $C_6$ and $C_8$, to which
non-geometric fluxes contribute. After a short computation, we obtain 
\eq{
  \label{tadpole_01}
  \arraycolsep2pt
  \renewcommand{\arraystretch}{1.4}
  \begin{array}{lcclcl}
   \hspace*{5pt}N^{\rm flux}_{{\rm D}3}&=&-&{\mathfrak f}_{\lambda} \op \tilde h^{\lambda} 
     &+& \tilde{\mathfrak f}^{\lambda} \op h_{\lambda} \,, \\
   \bigl[ N^{\rm flux}_{{\rm D}5}\bigr]_a &=& +&{\mathfrak f}_{\lambda} \op \tilde f^{\lambda}{}_a 
     &-& \tilde{\mathfrak f}^{\lambda}  \op    f_{\lambda \op a}   \,, \\
   \bigl[ N^{\rm flux}_{{\rm D}7}\bigr]^\alpha &=& -& {\mathfrak f}_{\lambda} \op \tilde q^{\lambda\op \alpha} 
     &+& \tilde{\mathfrak f}^{\lambda}\op  q_{\lambda}{}^\alpha \,.
  \end{array}                 
} 
Below we discuss the contribution to $Q_{{\rm D}p}$ due to magnetized ${\rm D}7$-branes.

%%%%%%%%%%%%%%%%%%%%%%%%%%%%%%%%%%%%%%%%%%%%%%%
%%%%%%%%%%%%%%%%%%%%%%%%%%%%%%%%%%%%%%%%%%%%%%%
%%%%%%%%%%%%%%%%%%%%%%%%%%%%%%%%%%%%%%%%%%%%%%%
%%%%%%%%%%%%%%%%%%%%%%%%%%%%%%%%%%%%%%%%%%%%%%%

\subsection{${\rm D}7$-branes, tadpoles and Freed-Witten anomalies}

When constructing models of particle physics in type IIB string theory, 
we are required to introduce D-branes which, in the present 
setting, are ${\rm D}3$- and magnetized 
${\rm D}7$-branes. Note that these D-branes contribute to the 
tadpole cancellation conditions \eqref{tadpole_00}, and that 
the mutual presence of fluxes and D-branes
gives rise to a number of consistency conditions.
The most famous one is the Freed-Witten anomaly condition \cite{Freed:1999vc}, 
following from the relation $d{\mathcal F}=H$ for the two-form gauge field 
\eq{
{\cal F}=F_2+B_2
}
on the D-brane. This implies that 
$\int_{\Gamma_3} H=0$ for every three-cycle $\Gamma_3$ in the D-brane
world-volume.
In this section, we discuss the contribution of D-branes to the tadpole cancellation 
conditions, and we derive generalized Freed-Witten anomalies for D-branes in 
non-geometric flux backgrounds.

%%%%%%%%%%%%%%%%%%%%%%%%%%%%%%%%%%%%%%%%%%%%%%%
%%%%%%%%%%%%%%%%%%%%%%%%%%%%%%%%%%%%%%%%%%%%%%%

\subsubsection{D-branes and tadpole contributions}

We begin by recalling some fact about D$7$-branes in type IIB orientifolds.
In order to keep our discussion general,
it is useful to work in the upstairs picture and carry
out the orientifold projection later. For that purpose,
upstairs, we introduce all objects such that
the orientifold projection leaves the whole configuration invariant.

The generic single-brane configuration we are considering consists of a D$7$-brane wrapping
a homological four-cycle $\Sigma$ in $\mathcal M$, and carries an abelian
gauge flux ${\cal E}=\langle \mathcal F\rangle$ along the four-cycle.
From the upstairs point of view, the orientifold projection  $\Omega_{\rm P}
(-1)^{F_{\rm L}}\sigma$  maps such a brane to an image 
brane, wrapping an image four-cycle $\Sigma'$ with 
an image gauge flux ${\cal E}'$.  Neither item of the data has to be orientifold invariant 
by itself.
Instead, for a generic $U(1)$ brane configuration we have 
\eq{
  \arraycolsep2pt
  \left(\begin{array}{lcl@{\hspace{2pt}}cl}  
  \Sigma &=&\Sigma_+& +& \Sigma_- \\[2pt]
  {\cal E} &=&{\cal E}_+ &+& {\cal E}_- 
  \end{array}\right)
  \qquad \xrightarrow{\hspace{15pt}\Omega_{\rm P}(-1)^{F_{\rm L}} \sigma\hspace{15pt}}\qquad
  \left(\begin{array}{lccl@{\hspace{2pt}}cl}  
  \Sigma' &=&&\Sigma_+& -& \Sigma_- \\[2pt]
  {\cal E}' &=&-&{\cal E}_+ &+& {\cal E}_- 
  \end{array}\right),
}
where the overall minus sign for ${\cal E}'$ comes from the fact that
the gauge field is odd under $\Omega_{\rm P} (-1)^{F_{\rm L}}$.
The fluxes as well as the Poincar\'e duals of the four-cycles are two-forms, that we
can expand in the basis \eqref{basis_coho} as
\eq{
  \arraycolsep2pt
  \begin{array}{ccc@{\hspace{1pt}}c}
       [\Sigma]_+&=&m^\alpha & \omega_\alpha\,, \\[2pt]
    {\cal  E}_+&=&e^\alpha & \omega_\alpha\,,
  \end{array}       
  \hspace{50pt}
  \begin{array}{ccc@{\hspace{1pt}}c}
       [\Sigma]_-&=&m^a & \omega_a\,, \\[2pt]
    {\cal  E}_-&=&e^a & \omega_a\,.
  \end{array}       
}
A stack of $N$ such brane--image-brane pairs contributes to the D$7$-, D$5$- and 
D$3$-brane tadpole equations \eqref{tadpole_00}. 
For the present setting, we can refer for instance to \cite{Plauschinn:2008yd,Blumenhagen:2008zz}
(see also \cite{Plauschinn:2010zz} for a more extended discussion) for the contribution to the tadpoles.
With $\chi(\Sigma)$ the Euler number of the four-cycle $\Sigma$ wrapped by the D$7$-brane, we find 
\eq{
\label{brane_charges}
  \arraycolsep2pt
  \begin{array}{lcl}
    \hspace*{4.25pt}Q_{{\rm D3}}&= & - N\,  e^{\mathsf A} e^{\mathsf B} m^{\mathsf C}\op \kappa_{\mathsf{ABC}} - 
    \tfrac{N}{24}\op \chi(\Sigma) \,, \\[7pt]
    \bigl[ Q_{{\rm D}5}\bigr]_a&=&+2 \op N \op \bigl(m^\alpha e^b+ e^\alpha m^b \bigr)\, 
    \kappa_{\alpha ab }\,, \\[8pt]
    \bigl[ Q_{{\rm D}7}\bigr]^\alpha &=&+2  \op N \,m^\alpha \,.
    \end{array}
}

%%%%%%%%%%%%%%%%%%%%%%%%%%%%%%%%%%%%%%%%%%%%%%%
%%%%%%%%%%%%%%%%%%%%%%%%%%%%%%%%%%%%%%%%%%%%%%%

\subsubsection{Freed-Witten anomalies}

Let us now turn to the Freed-Witten anomaly cancellation conditions. 
This issue has been discussed in similar configurations of D-branes and fluxes in 
\cite{Camara:2005dc, Villadoro:2006ia, LoaizaBrito:2006se, 
Aldazabal:2008zza, Aldazabal:2011yz}.

In the presence of geometric flux,
the Freed-Witten conditions guarantee that the cycle wrapped by the D-brane is still closed in
the deformed geometry. For the previously introduced  D$7$-brane  
wrapping $\Sigma=\Sigma_++\Sigma_-$  this means  that
\eq{
  \label{FWgeometric}
  \mathcal D [\Sigma]=m^a \Big(\tilde f^{\lambda}{}_a \alpha_{\lambda} -f_{\lambda\op a} \beta^{\lambda}\Big)
  + m^{\alpha} \Big(\tilde f^{\hat\lambda}{}_{\alpha} \alpha_{\hat\lambda} 
  -f_{{\hat\lambda}\op{\alpha}} \beta^{\hat\lambda}\Big)
  =0
  \,.
}         
In components, this relation can be expressed in the following way
\eq{
  \begin{array}{l}
  0 = m^a \op \tilde f^{\lambda}{}_a \,, \\[3pt]
  0 = m^a \op f_{\lambda\op a}\,,
  \end{array}
  \hspace{60pt}
  \begin{array}{l}
  0 = m^{\alpha} \op \tilde f^{\hat\lambda}{}_{\alpha} \,, \\[3pt]
  0 = m^{\alpha} \op f_{{\hat\lambda}\op{\alpha}} \,.
  \end{array}
}

In order to generalize \eqref{FWgeometric} to include
non-geometric $Q$- and $R$-fluxes, let us note that a $U(1)$ brane can result in a gauging 
of axionic shift symmetries.
This leads  to the so-called generalized Green-Schwarz mechanism,
which plays an important role for canceling possible chiral gauge anomalies
in four dimensions. 
The couplings relevant for the Green-Schwarz mechanism originate from the Chern-Simons action 
of a D$7$-brane and read
\eq{
  \label{CS-terms}
  \mathcal S_{\rm CS}\sim \hspace{10pt}& \int_{\mathbb R^{3,1}\times\Sigma} C_6\wedge F_2 - 
    \int_{\mathbb R^{3,1}\times\Sigma'} C_6\wedge F_2
     \\[4pt]
  +&\int_{\mathbb R^{3,1}\times\Sigma} C_4\wedge  {\cal E}\wedge F_2  
  -\int_{\mathbb R^{3,1}\times\Sigma'} C_4\wedge  {\cal E}'\wedge F_2 
  + \ldots\,,
}
where $F_2$ denotes the four-dimensional abelian gauge field and ${\cal E}$, as
before, denotes the internal background gauge field supported on the D$7$-brane.
The ellipsis indicate that there are additional terms in the Chern-Simons action, which are 
however not of importance here.

Let us now focus on the first line in \eqref{CS-terms} and 
expand the R-R form $C_6$ in the basis of even and odd four-forms shown in \eqref{basis_coho}.
Ignoring terms of different degree in the compact space, we find
\eq{
   C_6=C_{2,\alpha}\op \tilde \omega^\alpha+ C_{2,a} \op\tilde \omega^a + \ldots
}
Performing a dimensional reduction of the first line in \eqref{CS-terms} to four dimensions,
we obtain the St\"uckelberg mass terms
\eq{
   \mathcal S_{\rm CS}^{(1)}\sim \int_{\mathbb R^{3,1}}  2\op m^a \, C_{2,a}\wedge F_2\, .
}
As usual, such a term implies a gauging of shift symmetries of the
zero-forms dual to $C_{2,a}$ in four dimensions.
In the present setting, we have the relation (see for instance \cite{Plauschinn:2010zz})
\eq{
  C_{2,a} \;\longleftrightarrow \;- c^a \,,
}
where the four-dimensional scalars $c^a$ have been defined in \eqref{expansion_01}.
The gauging of the shift symmetry means that under a $U(1)$ gauge transformation
$A\to A+d\lambda$ the scalars transform as
\eq{
  \label{fw_09}
         c^a\to c^a+ m^a \lambda\, ,
}
where $A$ is the open string gauge field on the D-brane with field strength $F_2$.
Since gauge invariance is a fundamental property of all interactions,
we have to require that the flux induced superpotential \eqref{thebigW} is gauge
invariant. This imposes extra constraints on the fluxes, which are
the generalized Freed-Witten anomalies.
More concretely, in order for the superpotential \eqref{w_04} to be invariant
under (infinitesimal) transformations \eqref{fw_09}, we have to impose
\eq{   
  \label{qfluxFWa}
  \begin{array}{l@{\hspace{60pt}}l@{\hspace{40pt}}l}
  0 = m^a \tilde f^\lambda{}_a \,, &
  0= \kappa_{\alpha bc}\, m^b\, q_\lambda{}^\alpha \,, 
  \\[4pt]
  0= m^a f_{\lambda \op a} \,, &
  0= \kappa_{\alpha bc}\, m^b\,\tilde q^{\lambda\op\alpha} \,.
  \end{array}
}
Note that the first column corresponds to half of the conditions shown in \eqref{FWgeometric}.

We next turn to the second line in \eqref{CS-terms} and perform a similar analysis.
We expand the R-R four-form $C_4$ as
\eq{
   C_4=C_{2}^{\alpha}\op \omega_\alpha+ C_{2}^a \op  \omega_a + \ldots \,,
}
and dimensionally reduce the Chern-Simons terms. The resulting
St\"uckelberg mass terms take the form
\eq{
         \mathcal S_{\rm CS}^{(2)}\sim \int_{\mathbb R^{3,1}}
         2\,\Big( \kappa_{\alpha\beta\gamma} \op m^\beta e^\gamma+
         \kappa_{\alpha b c}\op m^b e^c\Big)  \, C_{2}^{\alpha}\wedge F_2 \, .
}
The four-dimensional two-forms $C_2^{\alpha}$ are dual to the four-dimensional 
scalars $\rho_{\alpha}$ appearing in the K\"ahler moduli $T_{\alpha}$, 
whose shift symmetry is again gauged. 
In particular, under open string $U(1)$ gauge transformations $A \to A + d\lambda$,
we have
\eq{
   \rho_{\alpha} \to \rho_{\alpha} + \Big( \kappa_{\alpha\beta\gamma}\op m^\beta e^\gamma+
   \kappa_{\alpha b c}\op m^b e^c\Big)\op  \lambda\, .
}  
The gauge invariance of the superpotential \eqref{w_04} together with the relations 
\eqref{qfluxFWa} then leads to the following 
Freed-Witten conditions
\eq{
\label{FWqfluxb}
  \begin{array}{l@{\hspace{40pt}}l}
  0= \kappa_{\alpha \beta\gamma}\, m^{\beta}\, e^{\gamma}\, q_\lambda{}^\alpha \,, 
  &
  0= \kappa_{\alpha \beta\gamma}\, m^{\beta}\,e^{\gamma}\,\tilde q^{\lambda\op\alpha} \,. 
  \end{array}
}

Let us conclude this section with the following two remarks on the generalized Freed-Witten
constraints derived above:
\begin{itemize}

\item The relations shown in \eqref{qfluxFWa} and \eqref{FWqfluxb}
imply that a pure ${\rm D}7$-brane placed on an orientifold-even
four-cycle with vanishing gauge flux satisfies the generalized 
Freed-Witten constraints. Such a brane would carry $SO/SP$
gauge group.

\item Once we deform it to a $U(1)$ brane, either geometrically or by gauge flux, 
we find extra non-trivial constraints. Note that this
generically also leads to a chiral matter spectrum on the ${\rm D}7$-branes.
Thus, we expect that the K\"ahler moduli governing the size
of the four-cycle wrapped by a (chiral)  ${\rm D}7$-brane cannot be stabilized by
turning on (non)-geometric fluxes.  We investigate this effect for certain
examples later. This situation is  reminiscent 
of  the freezing of K\"ahler moduli via brane instanton effects~\cite{Blumenhagen:2007sm}.

\end{itemize}

%%%%%%%%%%%%%%%%%%%%%%%%%%%%%%%%%%%%%%%%%%%%%%%
%%%%%%%%%%%%%%%%%%%%%%%%%%%%%%%%%%%%%%%%%%%%%%%
%%%%%%%%%%%%%%%%%%%%%%%%%%%%%%%%%%%%%%%%%%%%%%%
%%%%%%%%%%%%%%%%%%%%%%%%%%%%%%%%%%%%%%%%%%%%%%%

\subsection{S-dual  completion of non-geometric fluxes}
\label{sec_sdual}

With the additional geometric and non-geometric fluxes $F$, $Q$ and $R$,
the superpotential \eqref{s_pot_02} does not transform covariantly 
under S-duality. In particular, a non-vanishing $Q$-flux spoils the invariance of the
scalar potential \eqref{f_pot} under $SL(2,\mathbb Z)$; it has therefore 
been suggested to introduce a so-called $P$-flux
in order to restore the covariance of~$W$~\cite{Aldazabal:2006up}.
Further aspects of $P$-fluxes have been considered in 
\cite{Guarino:2008ik, Aldazabal:2008zza, Aldazabal:2010ef, Aldazabal:2011yz,Gao:2015nra}. 
Here we are mostly interested in extending the superpotential to include $G$-moduli.

To address this point, let us begin by recalling that the ten-dimensional type IIB action (in Einstein frame) is invariant 
under the following transformation\footnote{In the conventions employed here,  the combination $C_4 -\tfrac{1}{2}\op C_2\wedge B_2$ is invariant under $SL(2,\mathbb Z)$, and
the unusual factors of $i$ in the transformation of $S$ are due to $S = e^{-\phi} - i\op C_0$.}
\eq{
  \label{sl2z_01}
  S \to \frac{a\op S - i\op b}{i\op c\op S + d} \,, 
  \hspace{60pt}
  \binom{C_2}{B_2} \to 
  \left(\begin{matrix} a & b \\ c & d \end{matrix} \right)  \binom{C_2}{B_2} \,,
}
where the matrix with components $a,b,c,d$ is an element of $SL(2,\mathbb Z)$. The 
K\"ahler potential \eqref{k_pot} transforms under $SL(2,\mathbb Z)$, and thus the 
superpotential $W$ has to transform as well for the scalar F-term potential \eqref{f_pot}
to be invariant. More concretely, we have
\eq{
  \label{sl2z_02}
  K \to  K + \log\left( \bigl| i\op c\op S + d \bigr|^2 \right)
  \hspace{40pt}\Longrightarrow\hspace{40pt}
  W \to \frac{1}{i\op c\op S + d}\, W\,.
}  
Turning now to the moduli shown in table~\ref{table_moduli}, 
from \eqref{sl2z_01} we can determine the following transformations
\eq{
  G^a \to \frac{1}{i\op c\op S + d}\,G^a \,, \hspace{40pt}
  T_{\alpha} \to T_{\alpha} + \frac{i}{2} \, \frac{c}{i\op c\op S+ d} \op \kappa_{\alpha b c} G^b G^c \,.
}
Let us note that despite these somewhat involved transformations rules, the term $-2\log \mathcal V$ in the K\"ahler
potential \eqref{k_pot} stays invariant. This is to be expected since ${\cal V}=\frac{1}{6} \kappa_{\alpha\beta\gamma} t^\alpha t^\beta t^\gamma$  depends only on the two-cycle volumina $\{t^{\alpha}\}$ (in Einstein frame), which
do not transform under $SL(2,\mathbb Z)$.

Given the above results, we can now construct an $SL(2,\mathbb Z)$-covariant extension of the superpotential 
\eqref{s_pot_02}. In particular, in analogy to the $Q$-flux we introduce a so-called $P$-flux \cite{Aldazabal:2006up}
as a map
\eq{
  \label{p_flux}
  \renewcommand{\arraystretch}{1.2}
  \arraycolsep2pt
  \begin{array}{l@{\hspace{7pt}}c@{\hspace{12pt}}lcl}
  P\,\bullet & :& \mbox{$p$-form} &\to& \mbox{$(p-1)$-form} \,,
  \end{array}
}
which transforms in combination with $Q$ as a $SL(2,\mathbb Z)$ doublet
\eq{
  \label{sl2z_03}
  \binom{Q}{P} \to 
  \left(\begin{matrix} a & b \\ c & d \end{matrix} \right)  \binom{Q}{P} \,.
}
Note that the action of the $P$-flux \eqref{p_flux} is analogous to that of the $Q$-flux. In particular, 
extending \eqref{deffluxes} we have
\eq{
  \label{def_p-flux}
\arraycolsep2pt
\begin{array}{lcl@{\hspace{1.5pt}}l@{\hspace{40pt}}lcr@{\hspace{1.5pt}}l}
-P\bullet\alpha_\Lambda &=& p_{\Lambda}{}^{ A} \omega_{ A} \,,
&
&
-P\bullet\beta^\Lambda &=& \tilde p^{\Lambda \op A} \omega_{ A}
&, 
\\[8pt]
-P\bullet\omega_{ A}&=& 0\,, &
&
-P\bullet\tilde\omega^{ A} &=& -\tilde p^{\Lambda\op  A} \alpha_\Lambda &+  
\,p_{\Lambda}{}^{ A} \beta^\Lambda\,.
\end{array}
}
For the superpotential, we then require the transformation behavior shown in \eqref{sl2z_02}, which
leads us to the following expression
\eq{
  \label{w_04}
   W^{(3)} = \int_{\mathcal M} \Bigl[\hspace{18pt}  &\mathfrak F - i\op S \op H \\[-1pt]
   +\op & i\op G^a \, (F \circ \omega_a)  \\
   +\op & i\op T_{\alpha} \, \bigl( [ Q  -i\op S\op P]\bullet \tilde \omega^{\alpha}\bigr)
   + \frac{1}{2}\op \kappa_{\alpha bc}\op  G^b G^c \, \bigl(P\bullet \tilde\omega^{\alpha}\bigr)
   \hspace{12pt} \Bigr]_{3} \wedge \Omega_3 \,.  
}
Evaluating this expression leads to
\eq{
 W^{(3)}= W^{(2)} + 
 \Bigl(S\, T_{\alpha} + \frac{1}{2} \op \kappa_{\alpha
   bc}\op  G^b G^c \, \Bigr)\,( \,p_{\lambda}{}^{ \alpha} X^{\lambda}
 - \tilde p^{\lambda\op  \alpha} F_\lambda)\,.
}
Note that this superpotential contains new terms $S \op T_{\alpha}$, as well as  terms
quadratic in the  $G^a$ fields.
Furthermore, the term involving the geometric flux $F$ transforms 
covariantly under $SL(2,\mathbb Z)$, and therefore no additional flux parameters have 
to be introduced. This observation is particularly interesting, because it contradicts the 
common expectation that for every known flux one has to introduce a dual flux, when 
constructing a duality-invariant theory. One explanation might be that
the geometric flux involves solely the metric which does not transform
under S-duality.

%%%%%%%%%%%%%%%%%%%%%%%%%%%%%%%%%%%%%%%%%%%%%%%
%%%%%%%%%%%%%%%%%%%%%%%%%%%%%%%%%%%%%%%%%%%%%%%
%%%%%%%%%%%%%%%%%%%%%%%%%%%%%%%%%%%%%%%%%%%%%%%
%%%%%%%%%%%%%%%%%%%%%%%%%%%%%%%%%%%%%%%%%%%%%%%
%%%%%%%%%%%%%%%%%%%%%%%%%%%%%%%%%%%%%%%%%%%%%%%
%%%%%%%%%%%%%%%%%%%%%%%%%%%%%%%%%%%%%%%%%%%%%%%
%%%%%%%%%%%%%%%%%%%%%%%%%%%%%%%%%%%%%%%%%%%%%%%
%%%%%%%%%%%%%%%%%%%%%%%%%%%%%%%%%%%%%%%%%%%%%%%

\section{Non-supersymmetric flux  vacua}
\label{sec_simplemodel}

In this section, we analyze the flux induced scalar potential in
several examples with increasing
complexity. We will not be completely specific in the sense
that we do not present a concrete Calabi-Yau three-fold with a 
fixed orientifold projection. Instead, we only specify the data at the
level of supergravity by taking a set of moduli together with their
K\"ahler potential and superpotential.  We believe that this approach 
gives a representative picture of the structure of mathematically
well-defined models of background geometries. 
The intention is to learn about the structure of the flux landscape, in
particular about the space of stable non-supersymmetric minima of
the scalar potential. 
Since in later sections we apply our findings to string
phenomenology and string cosmology, in this section we 
focus on the following properties:
\begin{itemize}
\item{Vacua should be non-supersymmetric and tachyon-free,
        so that after uplifting they can lead to stable de Sitter
        vacua.}
\item{The moduli should be stabilized in the perturbative regime, i.e.
     at weak string coupling and large radius.}
\item{All saxionic moduli should be stabilized with axions providing
    candidates for the inflaton and possibly dark radiation}
\end{itemize}

The aim of this section is to gain some insight into a 
subset of generic and well-treatable models, which show a particular 
scaling behavior with the fluxes.
The ratios of the latter provide  relations among the various mass
scales of the background, such as the string scale, the Kaluza-Klein
scale, the moduli-mass scale, the inflaton-mass scale and the
soft-term scale.
After briefly describing some generalities, we consider particular examples. 
First, we discuss models without complex structure moduli, in which K\"ahler-moduli 
stabilization can occur due to non-geometric fluxes. Then, we study cases
with all types of moduli. 
Readers who are primarily interested in the phenomenology can skip parts of this section and come 
back later when a certain model is used later.

%%%%%%%%%%%%%%%%%%%%%%%%%%%%%%%%%%%%%%%%%%%%%%%
%%%%%%%%%%%%%%%%%%%%%%%%%%%%%%%%%%%%%%%%%%%%%%%
%%%%%%%%%%%%%%%%%%%%%%%%%%%%%%%%%%%%%%%%%%%%%%%
%%%%%%%%%%%%%%%%%%%%%%%%%%%%%%%%%%%%%%%%%%%%%%%

\subsection{Generalities}
\label{ss:generalities}

Our starting point is the ${\cal N}=1$ supergravity scalar potential \eqref{f_pot}, which we recall for
convenience
\eq{
  \label{VF}
  V = \frac{M_\text{Pl}^4}{4 \pi}\,\, e^{K} \Bigl( K^{I\ov J} D_IW D_{\ov J}\ov W - 3 \op\bigl|W\bigr|^2 \Bigr) \, ,
}
where  $K_{I\ov J} = \partial_I \partial_{\ov J}\op K$, and
$D_I W = \partial_I W + (\partial_I K)\op W$. The indices run over the moduli fields displayed in table \ref{table_moduli}, and for ease of notation we set
\eq{
\label{reimSG}
S := s + i\op c \, , \hspace{40pt} G^a := \psi^a + i\op \eta^a \, . 
}
The K\"ahler potential is determined from \eqref{k_pot}, and 
turning on fluxes induces superpotentials of the form shown in \eqref{thebigW}.
When $S$-dual $P$-flux is included, there are additional terms derived from \eqref{w_04}. 

The Planck mass in \eqref{VF} is 
$M_\text{Pl} = (8\pi G)^{-1/2} \approx 2.435 \cdot 10^{18} {\rm GeV}$ in our conventions. 
As usual the string mass is $M_{\rm s}=(\alpha')^{-{1\over 2}}$, and in terms of
$M_\text{Pl}$ the string and Kaluza-Klein scales can be expressed as
\eq{
\label{stringandKKscale}
       M_{\rm s}= {\sqrt{\pi} M_{\rm Pl}\over s^{1\over 4}\,{\cal V}^{1\over 2}}\,, 
       \hspace{40pt}
       M_{\rm KK}= {M_{\rm Pl}\over \sqrt{4\pi}\, {\cal V}^{2\over 3}} \, ,
}
where $s=e^{-\phi}$ (see e.g. \cite{Blumenhagen:2009gk}). 
Recall that ${\cal V}$ is the volume of the Calabi-Yau manifold in Einstein frame
measured in string units, namely ${\cal V}={\rm Vol}/\ell_s^6$ with $\ell_s=2\pi\sqrt{\alpha'}$.
 
Supersymmetric extrema of the potential \eqref{VF} are generically AdS. Such extrema are stable
even if the Hessian of the potential has negative eigenvalues. Indeed, as it is well known, for AdS vacua 
tachyonic fluctuations are stable provided they satisfy the Breitenlohner-Freedman bound
\cite{Breitenlohner:1982bm}
\eq{
\label{BFbound}
m^2 \, M^2_{\rm Pl}\geq \frac{3}4 \, V_0 \, ,
}
where $V_0$ is the value of the potential at the extremum and $m^2$ is the physical mass.
For supersymmetric extrema the bound is always verified. At a given extremum of a potential $V$
the squared physical masses for the canonically normalized fields can be computed as the eigenvalues of 
the matrix
\eq{
\label{physmass}
(M^2)^i_j = K^{ik} V_{kj}
\, ,}
with $V_{kj} = \frac12 \partial_k \partial_j V$ \cite{Conlon:2007gk}.
It is also useful to introduce the gravitino mass term
\eq{
\label{m32}
M^2_{\frac32} = e^{K_0}\op  |W_0|^2 \op {M_{\rm Pl}^2\over {4\pi}}\, ,
}
with $K_0$ and $W_0$ denoting the value of the K\"ahler and superpotential in the minimum.
The scale of supersymmetry breaking is determined by the non-vanishing F-terms 
$F^{I}=  e^{K\over 2} K^{I\ov J} D_{\ov J} \ov W$. 
In general there are numerical factors coming from $e^K$ that are fixed by the geometry which we do not specify here.

We use conventions in which the flux parameters entering in the superpotential are quantized.
These fluxes are constrained by the Bianchi identities \eqref{Bianchi}, and  they induce tadpoles
\eqref{tadpole_01} for the R-R $p$-form potentials that can be cancelled by couplings to appropriate sources. 
In particular, magnetized ${\rm D}7$-branes give a specific contribution to the R-R tadpoles as indicated in
\eqref{brane_charges}. 
In turn, including D-branes leads to additional Freed-Witten cancellation conditions on the fluxes,
c.f. \eqref{qfluxFWa} and \eqref{FWqfluxb}.

Non-geometric fluxes were originally considered in \cite{Shelton:2005cf}. Subsequently many authors
have looked into the question of moduli stabilization due to a scalar potential induced by non-geometric
fluxes \cite{
Aldazabal:2006up,
Villadoro:2006ia,
Shelton:2006fd,
Micu:2007rd,
Palti:2007pm,
Gray:2008zs,
Font:2008vd,
Guarino:2008ik,
deCarlos:2009fq,
Caviezel:2009tu,
deCarlos:2009qm,
Aldazabal:2011yz,
Dibitetto:2011gm,
Danielsson:2012by,
Damian:2013dq,
Damian:2013dwa,
Blaback:2013ht}.
Similar vacua have also been constructed and analyzed in detail in the T-dual type IIA language
\cite{Derendinger:2004jn, Villadoro:2005cu, DeWolfe:2005uu, Camara:2005dc, Grana:2006kf, 
Hertzberg:2007wc, Aldazabal:2007sn, Palti:2008mg, Caviezel:2008ik, Haque:2008jz, 
Caviezel:2008tf, Flauger:2008ad, Danielsson:2009ff, Danielsson:2010bc, Danielsson:2011au, Danielsson:2012et}. 
In this paper we direct our search to non-supersymmetric and tachyon-free vacua,
that can give rise to de Sitter after uplifting. Moreover, we will look for vacua 
in which all saxionic moduli are stabilized whereas axions 
furnish viable candidates for the inflaton.

%%%%%%%%%%%%%%%%%%%%%%%%%%%%%%%%%%%%%%%%%%%%%%%
%%%%%%%%%%%%%%%%%%%%%%%%%%%%%%%%%%%%%%%%%%%%%%%
%%%%%%%%%%%%%%%%%%%%%%%%%%%%%%%%%%%%%%%%%%%%%%%
%%%%%%%%%%%%%%%%%%%%%%%%%%%%%%%%%%%%%%%%%%%%%%%

\subsection{Models without complex structure moduli}

In this section we describe examples in which the fields are the axio-dilaton
and up to two K\"ahler moduli, while complex structure moduli are absent.

%%%%%%%%%%%%%%%%%%%%%%%%%%%%%%%%%%%%%%%%%%%%%%%
%%%%%%%%%%%%%%%%%%%%%%%%%%%%%%%%%%%%%%%%%%%%%%%

\subsubsection{Model A}
\label{sss_A}

Let us consider the simple case of a CY manifold with $h_-^{2,1}=0$
and  $h_+^{1,1}=1$. One can consider this model as the isotropic
six-torus with frozen complex structure modulus.
In this situation the K\"ahler potential is given by
\eq{
               K=-3 \log ( T+\ov T )-\log (S+\ov S) \,.
}
We also consider NS-NS flux $h_0=h$,
the non-geometric flux $q _{0}{}^{1}= q$, and the R-R three-form flux
$\tilde{\mathfrak f}^0=\tilde{\mathfrak f}$. These fluxes satisfy the Bianchi identities \eqref{Bianchi},
and are subject to the quantization condition $\tilde{\mathfrak f},h,q\in \mathbb Z$.
From \eqref{thebigW} we determine the corresponding superpotential as
\eq{
          W=  i\op \tilde {\mathfrak f}+i \op h\op S   + i\op q\op T\, ,
}
where we have set $X^0=1$ and $F_0=i$. 
The resulting scalar potential takes the very simple form
\eq{
\label{scalarpotentialexa}
     V = \frac{M_\text{Pl}^4}{4\pi \cdot 2^4} \,\, \left[
{(hs - \tilde{\mathfrak f} )^2 \over s \tau^3} - {6 h q s+ 2 q \tilde{\mathfrak f}\over
  s \tau^2} - {5 q^2\over 3 s \tau} +{1\over s \tau^3}\left( h c +q \rho\right)^2
	\right] ,
}
which only depends on the following linear combination of axions 
\eq{
  \label{model_A_02}
  \theta=h \op c+q\op \rho \,. 
}
Hence, the orthogonal linear combination of axions is not stabilized by
the potential \eqref{scalarpotentialexa}.

\begin{figure}[ht]
  \centering
  \includegraphics[width=0.6\textwidth]{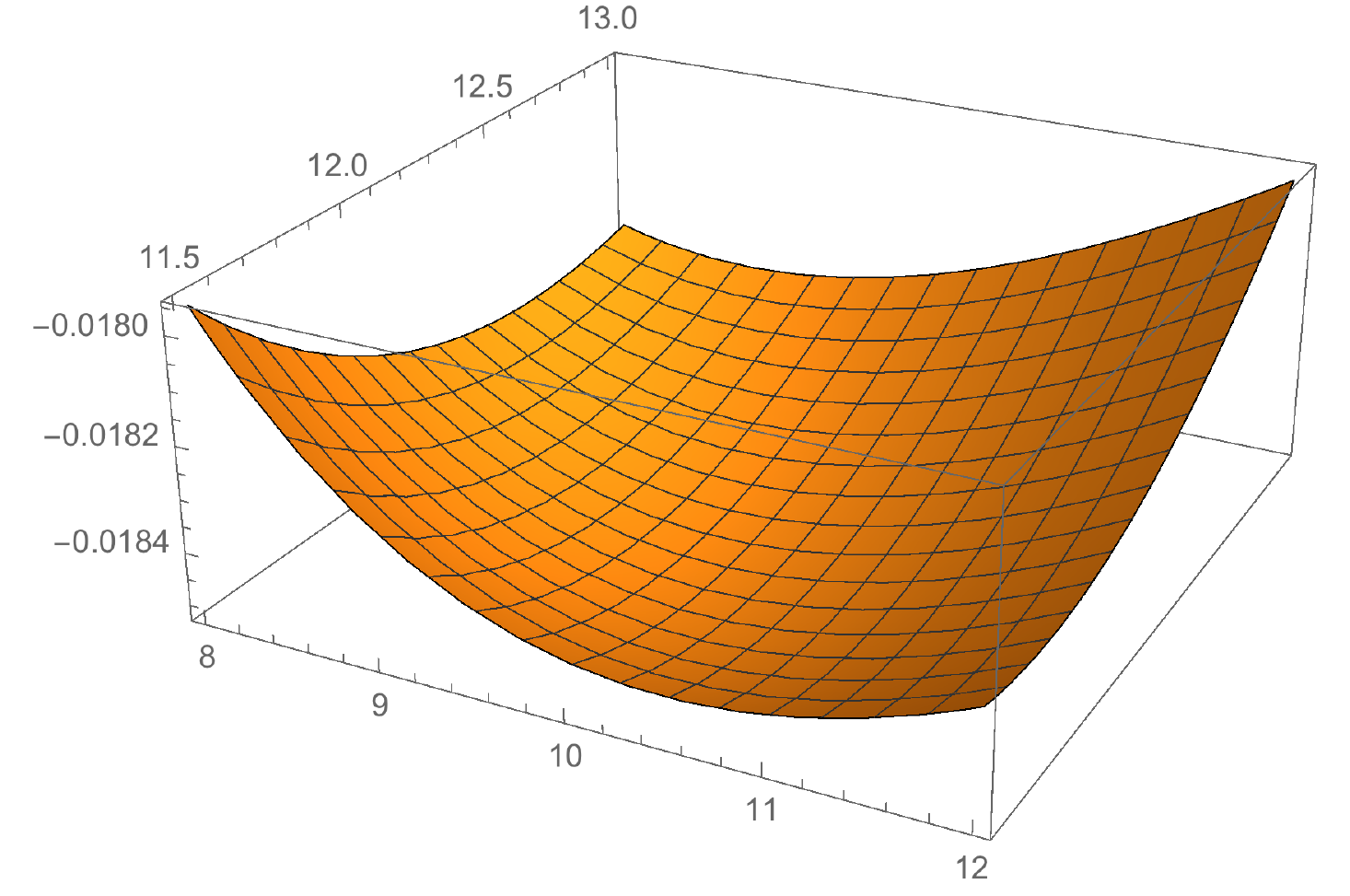}
  \begin{picture}(0,0)
  \put(-260,100){\footnotesize $V$}
  \put(-200,145){\footnotesize $\tau$}
  \put(-160,14){\footnotesize $s$}
  \end{picture}
  \caption{\small The scalar potential $V$ in units of ${M_\text{Pl}^4 \over 4 \pi \cdot 2^4}$ for $h=q=1$, 
  $\tilde{\mathfrak f}=10$, showing the  expected stable minimum at $s_0=10$ and $\tau_0=12$. }
  \label{KKLTana}
\end{figure}
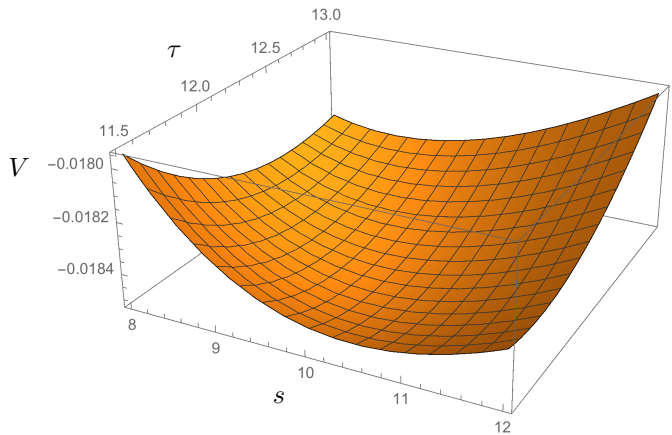

The extremal points of \eqref{scalarpotentialexa} are obtained by solving for  $\partial_s V=\partial_\tau V
=\partial_\theta V=0$, and we find the three solutions shown in table
\ref{table_extr}.
%%%%%%%%%%%%
%%%%%%%%%%%%
\begin{table}[ht]
  \centering
  \renewcommand{\arraystretch}{1.3}
  \begin{tabular}{|c|c|c|c|c|}
  \hline
   solution& $(s,\tau,\theta)$ & susy &  tachyons & $\Lambda$ \\
  \hline\hline
  & & & &\\[-0.5cm]
  1 & $(-{\tilde{\mathfrak f}\over 2 h}, -{3 \tilde{\mathfrak f}\over 2 q},0)$ & yes & yes & AdS \\[0.2cm]
  2 & $({\tilde{\mathfrak f}\over 8 h}, {3 \tilde{\mathfrak f}\over 8 q},0)$ & no & yes & AdS \\[0.2cm]
  3 & $(-{\tilde{\mathfrak f}\over h}, -{6 \tilde{\mathfrak f}\over 5 q},0)$ & no & no & AdS \\[0.2cm]
 \hline
     \end{tabular} 
     \caption{\small Extrema of the scalar potential \eqref{scalarpotentialexa} for model A.}
      \label{table_extr}
\end{table}
%%%%%%%%%%%%
%%%%%%%%%%%%
Note that the fluxes must be chosen so that the values
of $s$ and $\tau$ are inside the physical domain $s,\tau>0$.
\begin{itemize}

\item The first solution in table \ref{table_extr} is the supersymmetric one, since here
$D_T W= D_S W=0$. As $W$ does not depend on one axionic
direction, the no-go theorem of \cite{Conlon:2006tq} implies that the saxionic
partner is tachyonic, which can indeed be confirmed for this minimum. However, as
expected in the supersymmetric case, the Breitenlohner-Freedman bound \eqref{BFbound} is
satisfied and the minimum is stable.

\item Solution three of  table \ref{table_extr} is non-supersymmetric and has no tachyonic directions. Such
AdS vacua we call strictly stable, as
they  can be uplifted  to  stable de Sitter vacua.
Moreover, for $|\tilde{\mathfrak f}/h|\gg 1$ and $|\tilde{\mathfrak f}/q|\gg 1$ we obtain weak string
coupling
and large radius, so that it is justified to ignore higher-order
corrections to the scalar potential. Note that the scaling of the stabilized
moduli with the fluxes implies that all terms in the superpotential
are of the same order.
For $\theta=0$ , the potential has the shape shown in figure \ref{KKLTana}.

\end{itemize}

%%%%%%%%%%%%
%%%%%%%%%%%%
%%%%%%%%%%%%
%%%%%%%%%%%%

So far, one linear combination of axions remains unstabilized. Let us
analyze whether this modulus can be given a parametrically-light mass by
turning on more general fluxes. If that can be achieved, this axion
is a good candidate for realizing F-term axion monodromy inflation.
From \eqref{thebigW} we determine  the most general superpotential (without $P$-flux) 
as follows
\eq{
  \label{model_a_01}
     W=-{\mathfrak f} + i \op \tilde{\mathfrak f}+ i (h -i\op \tilde h)\, S   + i(q-i\op \tilde q)\, T \, ,
}
with ${\mathfrak f} = {\mathfrak f}_{0}$, $\tilde h=\tilde h^{0}$ and $\tilde q=\tilde
q^{0\op 1}$.
The only non-trivial Bianchi identity following from \eqref{Bianchi} reads
\eq{
     \tilde h\, q- h\, \tilde q =0\, ,
}
so that the superpotential \eqref{model_a_01} reduces to
\eq{ \label{superpotwithbianchi}
          W= -{\mathfrak f} + i\op  \tilde{\mathfrak f}+ \left(1-i\, {\tilde q\over q}\right) i (h S   +q T)\, .
}
Therefore, $W$ still depends only on the linear
combination of axions \eqref{model_A_02},
so that the orthogonal direction remains unfixed. 
In fact, the vacua of the superpotential \eqref{superpotwithbianchi} can be determined
analytically and share the same qualitative structure of the three minima shown in table~\ref{table_extr}.

Let us now turn to the contribution of the fluxes to the tadpoles. From the expressions shown 
in \eqref{tadpole_01} we find
\eq{
N_{\rm D3}^\text{flux}=\tilde{\mathfrak f}\, h\,,
\hspace{40pt}
N_{{\rm D5}}^\text{flux} =  0 \, , \hspace{40pt} 
N_{\rm D7}^\text{flux}=\tilde{\mathfrak f}\, q \, .
} 
For the extrema 1 and 3 in table \ref{table_extr}, we have $N_{\rm D3}^\text{flux}<0$ and $N_{\rm D7}^\text{flux}<0$,
whereas it is the opposite in extremum 2.
Note that since in this model the $Q$-flux $q _{0}{}^{1}$ has been turned on,
the FW-condition forbids to wrap a magnetized ${\rm D}7$-brane on the 
single homological four-cycle $\Sigma_1=[\omega_1]$. Therefore,
for non-trivial flux, the ${\rm D}7$-brane tadpole can only be cancelled
by a number of un-magnetized ${\rm D}7$-branes wrapping $\Sigma_1$ 
and giving SO/SP gauge symmetry. A chiral gauge/matter
sector is not possible.

We finally compute the mass eigenvalues and eigenstates for the
canonically normalized fields and compare them to the string
and Kaluza-Klein scales. 
Evaluating the physical mass matrix \eqref{physmass} for the non-supersymmetric tachyon free minimum 
(solution 3 of table~\ref{table_extr}) gives 
\eq{  
     M^2= {M_\text{Pl}^2 \over 4\pi \cdot 2^4} \,\,
		{5 \op q^2\over 54 \op\tilde{\mathfrak f}^2}\left(\begin{matrix}   {60 h q} &
         {12 h^2} & 0 &  0\\
        {25\, q^2 } &  {25 h q } & 0 & 0 \\
     0 & 0 &  {12 h q} & {12 h^2} \\
   0 & 0 & {25 q^2  } &  {25 h q }
\end{matrix}\right) .
}
The mass eigenvalues can be written as
\eq{ \label{massessimplemodel}
M_{{\rm mod},i}^2=\mu_i  \op {h q^3\over \tilde{\mathfrak f}^2}\, {M_{\rm Pl}^2 \over 4 \pi \cdot 2^4} \,,
}
with the numerical values 
\eq{ 
\label{massesA}
\mu_i=\left(
{25(17+\sqrt{97})\over 108},
{25(17-\sqrt{97})\over 108}\, ; \, 
{185\over 54}, 
0 
\right) \, \approx \, (6.2, 1.7\, ; \, 3.4, 0)\,.
}
The eigenvectors of the first (last) two masses are combinations of saxions (axions).
The massless state is the axionic combination $(q \op c -h\op \rho)$,
and the three massive states are parametrically of the same
mass. The gravitino shares the same flux dependence as the moduli \eqref{massessimplemodel}
with the numerical prefactor $\mu_{3 \over 2} = {5\over 6} \approx 0.833$. 
This model should be considered as our simplest prototype example and
we will come back to it throughout this paper.

%%%%%%%%%%%%%%%%%%%%%%%%%%%%%%%%%%%%%%%%%%%%%%%
%%%%%%%%%%%%%%%%%%%%%%%%%%%%%%%%%%%%%%%%%%%%%%%

\subsubsection{Model B: inclusion of  $H^{1,1}_-$ moduli}
\label{sec_h11minusex}

Let us consider a generalization of the previous model and add
a $G$-modulus as well as a geometric flux. The underlying CY manifold therefore has to have Hodge numbers
$h_-^{2,1}=0$ and  $h_+^{1,1}=h_-^{1,1}=1$.
Using \eqref{defkaehler}, the K\"ahler potential is
\eq{
\label{model_B_02}
               K=-3 \log \Big( (T+\ov T)+{\kappa\over 4\op(S+\ov S)}\,
               (G+\ov G)^2 \Big)-\log (S+\ov S) \,,
}
where for later convenience we have set $\kappa := 2\op \kappa_{\alpha
a b}$ for $\alpha=a=b=1$. 
We turn on fluxes such that the superpotential \eqref{thebigW} becomes
\eq{
\label{superpotwithgmod}
          W=  i \op \tilde {\mathfrak f}+i \op h\op S   + i\op q\op T - i\op f\op G\, ,
}
 with $\tilde{\mathfrak{f}}=\tilde{\mathfrak{f}}^{0}$, $h=h_{0}$, $q=q_{0}{}^{1}$, and $f=f_{{0}\op1}$.
For this set of fluxes, the contribution to the tadpoles \eqref{tadpole_01} is given by
\eq{ 
\label{model_B_01}
N_{\rm D3}^\text{flux} = \tilde{\mathfrak f}\op h\, , \hspace{40pt}
N_{{\rm D5}}^\text{flux} =  -\tilde{\mathfrak f}\op f \, , \hspace{40pt} 
N_{\rm D7}^\text{flux} = \tilde{\mathfrak f} \op q \,.
}
The signs of these tadpoles depend on the signs of the fluxes, which are in turn fixed by the condition that
the saxions have positive vevs. In the most interesting vacuum discussed below we must demand 
$q, h < 0 < f, \tilde{\mathfrak f}$ for which all tadpole contributions in \eqref{model_B_01} are negative.

The scalar potential can be computed from \eqref{model_B_02} and \eqref{superpotwithgmod},
for which we find three AdS extrema.
One of them is supersymmetric and tachyonic, another one  non-supersymmetric
and tachyonic and the other one   non-supersymmetric and non-tachyonic.
In the following, we focus on the non-supersymmetric non-tachyonic extremum
which is characterized by
\eq{
\label{valuesmodgmod}
    &\tau = -{(6  + x) \over 5 (1+x) } \, {\tilde {\mathfrak f} \over q} \,,  \hspace{40pt}
    s=	- {1\over x +1 }\, {\tilde{\mathfrak f} \over h} \,, \hspace{40pt}
    \psi={2x \over x + 1} \, {\tilde{\mathfrak f} \over f} \,, \\[5pt]
    &0=q\op \rho - f\op\eta + h \op c \,,
}
where we remind the reader that our conventions for the modulus $G$ are shown in \eqref{reimSG}.
In \eqref{valuesmodgmod}, we have simplified the formulas by introducing the parameter
\eq{
x = {f^2\over {\kappa\op h \op q }} \,.
}
As expected, the superpotential \eqref{superpotwithgmod} fixes only one linear combination of
axions. 
Furthermore, notice that all the saxion vevs in \eqref{valuesmodgmod} 
scale with $\tilde {\mathfrak f}$, which has to be large to be in the perturbative regime.
In the minimum specified by \eqref{valuesmodgmod}, 
the superpotential becomes $x$-independent and we are left with
\eq{
W_0=- {6\op i\over 5} \, \tilde{\mathfrak f}\,.
}
We also note that for the other extrema in this model, we find a similar scaling with 
the flux, namely  $W_0 \sim  \tilde{\mathfrak f}$.
The scalar potential at the above-mentioned minimum is given by
\eq{
V_0 = - \,{ 2^2 \cdot M_\text{Pl}^4 \over 4 \pi } \,\,
\frac{25 }{864}
\,\, {h\op q^3 \over \tilde{\mathfrak f}^2}\,\, (1+x) \,,
}
where the dependence on $x$ originates from the $e^K$ factor.
The masses are given by the following expression 
\eq{
\label{model_B_03}
M_{{\rm mod},i}^2 = \mu_i   \,{ 2^2 \cdot M_\text{Pl}^2 \over 4 \pi }
\,\, {h\op q^3 \over \tilde{\mathfrak f}^2}\,\, (1+x) \,,
}
with the numerical coefficients 
\eq{
\mu_i \approx \bigl(\,0.097,\,0.026,\, 0\,;\,0.054,\,0,\,0 \bigr) \,.
}
The first three entries correspond to (linear combinations of)
saxionic moduli, while the last three entries are axionic
combinations. A novel feature is the appearance of a massless saxion
in the direction of $(f \op\tau + q \op\psi)$.
The gravitino mass has the same flux dependence \eqref{model_B_03}, 
with the numerical factor given by $\mu_{3 \over 2} = \frac{ 5 }{384}\approx 0.013$.

%%%%%%%%%%%%%%%%%%%%%%%%%%%%%%%%%%%%%%%%%%%%%%%
%%%%%%%%%%%%%%%%%%%%%%%%%%%%%%%%%%%%%%%%%%%%%%%

\subsubsection{Models with two K\"ahler moduli, $h^{1,1}_+=2$}
\label{ss:2kahler}

Next, we  investigate the effect of having several K\"ahler moduli. To this end, we consider
models, whose K\"ahler moduli sectors can be thought to be  based on
the K$3$-fibration ${\mathbb P}_{1,1,2,2,2}[8]$, 
and the swiss-cheese manifold ${\mathbb P}_{1,1,1,6,9}[18]$. 
It turns out that the vacuum solutions of the $h^{1,1}_+=1$ example generalize, but the spectrum contains additional tachyons.

\paragraph*{K3-fibration}
\label{sss:k3fib}

In the K\"ahler sector of ${\mathbb P}_{1,1,2,2,2}[8]$ 
the intersection numbers are such that the K\"ahler potential splits into sums and is given by
\eq{
              K=-2 \log ( T_1+\ov T_1 ) - \log (T_2 + \ov T_2 ) -\log (S+\ov S) \,,
} 
where for simplicity we have set $h^{2,1}_-=0$. Fluxes are chosen such that the superpotential 
\eqref{thebigW} takes the form
\eq{
          W=i \op \tilde {\mathfrak f}+i \op h\op S   + i\op q_1\op T_1+ i\op  q_2\op T_2\, ,
}
with $\tilde{\mathfrak{f}}=\tilde{\mathfrak{f}}^0$, $h=h_0$ and -- for ease of notation -- with $q_i=q_0{}^i$.
The resulting scalar potential has four AdS vacua
summarized in table~\ref{table_milka}, three of which are generalizations of those in table~\ref{table_extr}.
The stabilized axion is  
$\theta =q_1 \rho_1 + q_2 \rho_2  + hc $,
and  the potential does not depend on the two orthogonal axion combinations which thus remain  
unstabilized. 

%%%%%%%%%%%
%%%%%%%%%%%
\begin{table}[t]
  \centering
  \renewcommand{\arraystretch}{1.3}
  \begin{tabular}{|c|c|c|c|c|}
  \hline
   solution & $(s,\tau_1,\tau_2,\theta)$ & susy &  tachyons & $\Lambda$ \\
  \hline \hline
  & & & &\\[-0.5cm]
  1 & $(-{\tilde{\mathfrak f}\over 2 h}, -{ \tilde{\mathfrak f}\over q_1}, -{ \tilde{\mathfrak f}\over 2 q_2},0)$ & yes & 2 & AdS \\[0.2cm]
  2 & $({\tilde{\mathfrak f}\over 8 h},{ \tilde{\mathfrak f}\over 4 q_1}, { \tilde{\mathfrak f}\over 8 q_2},0)$ & no & 2 & AdS \\[0.2cm]
  3 & $(-{\tilde{\mathfrak f}\over h}, -{4 \tilde{\mathfrak f}\over 5 q_1}, -{ 2\tilde{\mathfrak f}\over  5 q_2},0)$ & no & 1 & AdS \\[0.2cm]
  4 & $(-{2 \tilde{\mathfrak f}\over 5 h}, -{4 \tilde{\mathfrak f}\over 5 q_1},-{ \tilde{\mathfrak f}\over q_2},0)$ & no & 1 & AdS \\[0.2cm]
 \hline
\end{tabular} 
\caption{\small Extrema of the scalar potential in the K3-fibration model.}
\label{table_milka}
\end{table}
%%%%%%%%%%%
%%%%%%%%%%%

The physical masses of the fields scale with the fluxes in the following way 
\eq{
\label{massk3}
M_{{\rm mod},i}^2 =\mu_i { h\op q_1^2 \op q_2\over {\tilde{\mathfrak f}^2}}{M_{pl}^2\over {4\pi} \cdot 2^4} \, ,
}
where the numerical factors $\mu_i$ depend on the specific solution.
The cosmological constant is negative and has
the same relation to the fluxes as the physical masses. The supersymmetric case contains, as expected, 
two tachyons
above the Breitenlohner-Freedman bound; for the non-supersymmetric vacua, tachyons are below the bound.  
In vacua 1, 2 and 3 there is a tachyon given by the combination of
saxions $\tau_{\rm tac} = q_2\op \tau_1 -q_1\op  \tau_2$. In section \ref{ss:dtermup} we will see that 
this tachyon can be lifted by adding a D-term to the F-term potential.

Turning to the tadpole conditions, according to \eqref{tadpole_01} in this model the flux contributions are given by
\eq{
\label{tadk3}
N_{\rm D3}^{\rm flux} = \tilde{\mathfrak f}\op h\,, \hspace{40pt}
\left[N_{\rm D7}^{\rm flux}\right]^1 = \tilde{\mathfrak f}\op q_1 \,, \hspace{40pt}
\left[N_{\rm D7}^{\rm flux}\right]^2 = \tilde{\mathfrak f} \op q_2\, .
}
For the vacua 1, 2 and 3 to have positive vevs for the saxions, we take for concreteness 
$\tilde{\mathfrak f} < 0$ and the remaining fluxes positive. The contributions \eqref{tadk3} to the 
flux tadpoles are then all negative.

\paragraph*{Swiss cheese}
\label{sss:swiss}

As a second example we discuss the K\"ahler sector of the swiss-cheese Calabi-Yau ${\mathbb P}_{1,1,1,6,9}[18]$ with K\"ahler potential
\eq{
K=-\log(S+\ov S)-2\log\Bigl((T_1+\ov T_1)^{3/2}-(T_2+\ov T_2)^{3/2}\Bigr) \,.
}
As in the K3-fibration example, for the superpotential we choose 
\eq{
W=i\op \tilde{\mathfrak{f}}+ i\op h\op S+ i\op q_1\op T_1+ i\op q_2\op T_2 \,,
}
and the complex structure sector of ${\mathbb P}_{1,1,1,6,9}[18]$ is again set to zero for simplification.

In case that the  non-geometric fluxes satisfy $q_1\op q_2 < 0$, we find four extrema of the potential with data summarized in 
table \ref{table_kokos}. The linear combination of axions $\theta=q_1 \op\rho_1+q_2\op\rho_2 + h\op c$ is stabilized, 
but two orthogonal combinations remain unstabilized.
%%%%%%%%%%%
%%%%%%%%%%%
\begin{table}[t]
  \centering
  \renewcommand{\arraystretch}{1.3}
  \begin{tabular}{|c|c|c|c|c|}
  \hline
   solution & $(s,\tau_1,\tau_2,\theta)$ & susy &  tachyons & $\Lambda$ \\
  \hline \hline
  & & & &\\[-0.5cm]
  1 & $(-{\tilde{\mathfrak f}\over 2\, h}, -{3 \,\tilde{\mathfrak f}\, q_1^2 \over 2\, (q_1^3+q_2^3)}, -{3 \,\tilde{\mathfrak f} \,q_2^2\over 2 \,(q_1^3+q_2^3)},0)$ & yes & 2 & AdS \\[0.2cm]
  2 & $({\tilde{\mathfrak f}\over 8\, h}, {3 \,\tilde{\mathfrak f}\, q_1^2 \over 8 \,(q_1^3+q_2^3)},{3\, \tilde{\mathfrak f}\, q_2^2\over 8\, (q_1^3+q_2^3)},0)$ & no & 2 & AdS \\[0.2cm]
  3 & $(-{\tilde{\mathfrak f}\over h}, -{6\, \tilde{\mathfrak f}\, q_1^2 \over 5\, (q_1^3+q_2^3)}, -{6\, \tilde{\mathfrak f}\, q_2^2\over 5 \,(q_1^3+q_2^3)},0)$ & no & 1 & AdS \\[0.2cm]
 \hline
\end{tabular} 
\caption{\small Extrema of the `swiss-cheese' scalar potential.}
\label{table_kokos}
 \end{table} 
%%%%%%%%%%%
%%%%%%%%%%%

The physical mass eigenvalues exhibit the following scaling with the flux parameters
\eq{
\label{massswiss}
M_{{\rm mod},i}^2=\mu_i {h\op (q_1^3+q_2^3)\over \tilde{\mathfrak{f}^2}} {M_{pl}^2\over{4\pi}}
\, , }
with vacuum-dependent numerical coefficients $\mu_i$. 
The vacua are analogous to those in the examples discussed so far. In particular,  as in the two previous models
with $h^{1,1}_+=2$,  there is an additional tachyonic state given by a combination of the saxions of the two K\"ahler moduli, either $T$ and $G$, or $T_1$ and $T_2$. 
In section \ref{ss:dtermup} we show that this tachyon can be uplifted  adding a D-term to the potential.

To have positive saxion vevs, for definiteness we take $h > 0$, $(q_1^3 + q_2^3)>0$ and
$\tilde{\mathfrak f} < 0$ in vacua 1 and 3, but $\tilde{\mathfrak f} > 0$ in vacuum 2. 
The flux tadpoles
are again given by \eqref{tadk3}. Then, in vacua 1 and 3 the D3-brane
tadpole is  $N_{\rm D3}^{\rm flux} < 0$.
Satisfying $q_1 q_2 <0 $ via  $q_1 > 0$ and $q_2 < 0$,
the two D7-brane tadpoles have signs
$\left[N_{\rm D7}^{\rm flux}\right]^1 < 0$, and  $\left[N_{\rm
    D7}^{\rm flux}\right]^2 > 0$.

%%%%%%%%%%%%%%%%%%%%%%%%%%%%%%%%%%%%%%%%%%%%%%%
%%%%%%%%%%%%%%%%%%%%%%%%%%%%%%%%%%%%%%%%%%%%%%%
%%%%%%%%%%%%%%%%%%%%%%%%%%%%%%%%%%%%%%%%%%%%%%%
%%%%%%%%%%%%%%%%%%%%%%%%%%%%%%%%%%%%%%%%%%%%%%%

\subsection{Flux scaling minima with complex structure moduli}

Let us now consider models which have  complex structure moduli,
and analyze whether they admit strictly stable,
non-supersymmetric minima of the flux-scaling type encountered
before.

%%%%%%%%%%%%%%%%%%%%%%%%%%%%%%%%%%%%%%%%%%%%%%%
%%%%%%%%%%%%%%%%%%%%%%%%%%%%%%%%%%%%%%%%%%%%%%%

\subsubsection{Model C}
\label{sss:C}

To begin, we analyze 
a model with $h^{2,1}_-=1$ and $h^{1,1}_+=1$, for which
the K\"ahler potential in the large complex structure limit can be written as
\eq{
\label{KSTU}
K=-3 \log (T+\ov T)-\log (S+\ov S)-3\log\big(U + \ov U\big)\, .
}
 One can view  this model as the isotropic
six-torus.
In the superpotential we have now more fluxes available, which of
course have to satisfy the Bianchi identities.
For the flux superpotential \eqref{s_pot_02} we choose 
\eq{
\label{superpotC}
W=- {\mathfrak f}_0 - 3\op \tilde{\mathfrak f}^1  U^2 - h\op U\op S - q\op U\op  T \,,
}
where we note that $U^2$ denotes the square of the modulus $U$,
and where $h := h_1$ and $q := q_1$.
Note that this superpotential only depends on the linear combination of axions 
$(h\op c + q \op\rho)$, and thus leaves its orthogonal combination unstabilized.
The latter is a possible inflaton candidate, and we  analyze below whether by turning on additional fluxes
a (parametrically small) mass can be generated.

Analyzing the  scalar potential following from \eqref{KSTU} and \eqref{superpotC}, we find 
two interesting extrema.
The first one is the supersymmetric AdS minimum with values
of the moduli
\eq{
\label{adsminimum}
\arraycolsep30pt
\begin{array}{@{}lll@{}}
 \displaystyle \tau = - 18 \op v \, {\tilde{\mathfrak f}^1 \over q} \,, &
 \displaystyle  s =  - 6 \op v \, {\tilde{\mathfrak f}^1 \over h} \,, &
 \displaystyle  v^2 =  {1\over 9}\, { {\mathfrak f}_0\over  \tilde{\mathfrak f}^1}\,, \\[16pt]
0=h\op c + q\op \rho \,, & u = 0 \,.
\end{array}
}   
Since one axion is unstabilized, this extremum contains tachyons which are
above the Breitenlohner-Freedman bound.
The second extremum is a non-super\-sym\-met\-ric tachyon-free AdS
minimum with frozen moduli
\eq{
\label{stablemodelC}
\arraycolsep22pt
\begin{array}{@{}lll@{}}
\displaystyle \tau = - 15\op v \, {\tilde{\mathfrak f}^1 \over q} \,, &
\displaystyle  s =  - 12 \op v \, {\tilde{\mathfrak f}^1 \over h} \,, &
\displaystyle  v^2 =  {1\over 3\cdot 10^{1\over 2}} \,{ {\mathfrak f}_0\over  \tilde{\mathfrak f}^1}\,, \\[16pt]
0=h\op c + q\op \rho\,, &
u = 0 \,.
\end{array}
}   
For $h, q < 0 < {\mathfrak f}_0, \tilde {\mathfrak f}^1$, all moduli are in the physical regime. 
The scaling of the moduli with the fluxes can already be
detected from the superpotential, which is also
the case for the other extrema we found. For this scaling, all terms
in $W$ are of the same order ${\mathfrak f}_0$. 
The contribution of the fluxes to the tadpoles are
\eq{
N_{\rm D3}^\text{flux}=\tilde{\mathfrak f}^1\op h\,,
\hspace{40pt}
N_{{\rm D5}}^\text{flux} =  0 \, , \hspace{40pt} 
N_{\rm D7}^\text{flux}=\tilde{\mathfrak f}^1\op q \,.
} 
Note that the flux ${\mathfrak f}_0$
does not contribute to  any of the tadpoles. Therefore, by scaling 
${\mathfrak f}_0\gg \tilde{\mathfrak f}^1, h, q \sim O(1)$, we can
ensure that all moduli are fixed in the perturbative regime.

Let us now analyze the non-tachyonic model \eqref{stablemodelC} in more detail.
The moduli masses in the canonically normalized basis are 
\eq{   M_{{\rm mod},i}^2=\mu_i\,  {h q^3 \over ({\mathfrak f}_0)^{3\over 2}
    (\tilde{\mathfrak f}^1)^{1\over 2}} \, {M_{\rm Pl}^2 \over 4 \pi \cdot 2^7} \,,
}
with numerical values
\eq{ \mu \approx  \bigl(\,2.1,\, 0.37,\,0.25\,;\,1.3,\,0.013 ,\,0 \bigr) \,.
}
The first three eigenstates are saxions and the last three are axions.
The massless mode is the axionic combination $(q \op c - h\op\rho)$.
Note that the lightest massive mode is axionic, and although not
parametrically light, its mass is numerically light.
In fact, it is by a factor of $1/5$ smaller than the second-lightest
massive state, which is  purely saxionic. For the gravitino mass the 
flux dependence is the same as for the moduli masses, with the numerical
prefactor given by $\mu_{3\over 2}  \approx 0.152$. 

%%%%%%%%%%%%%%%%%%%%%%%%%%%%%%%%%%%%%%%%%%%%%%%
%%%%%%%%%%%%%%%%%%%%%%%%%%%%%%%%%%%%%%%%%%%%%%%

\subsubsection{Model D}
\label{sss:D}

We next consider a variation of Model C, with the same K\"ahler
potential 
\eq{
K=-3 \log (T+\ov T)-\log (S+\ov S)-3\log\big(U +\ov U\big)\, ,
}
but with a different superpotential
\eq{ \label{modelDsuperpot}
W= i\hat{\mathfrak f}_1\op U+ i \op \tilde{\mathfrak f}^0  U^3+ 3\op i \op \tilde h^1 \op U^2\op S+3\op i \op \tilde q^1
\op U^2\op  T \,.
}
Note that here we introduced $\hat{\mathfrak f}_1 = -{\mathfrak f}_1$ for notational convenience.

For this model, we find extrema of the scalar potential in which the flux dependence of the moduli is 
governed  by the following overall scaling of the superpotential
\eq{ 
\label{scalingD}
W_0 \sim \frac{(\hat{\mathfrak f}_1)^{3 \over 2}}{(\tilde{\mathfrak f}^0)^{1 \over 2}} \, .
}
For instance,  there exists  a supersymmetric AdS minimum with values of the moduli
\eq{
\arraycolsep22pt
\begin{array}{@{}lll@{}}
\displaystyle \tau = -{2 \over 3}  \, {\tilde{\mathfrak f}^0 \over \tilde q^1}\,v \,, &
\displaystyle s =  -{2 \over 9}  \, {\tilde{\mathfrak f}^0 \over \tilde h^1}\,v \,, &
\displaystyle v^2 =  3 \, { \hat{\mathfrak f}_1\over  \tilde{\mathfrak f}^0}\,, 
\\[16pt]
0=\tilde h^1 \op c+ \tilde q^1 \op \rho \,, &
u = 0 \,.
\end{array}
}   
For this solution there are tachyons in the spectrum, that however fulfill the Breitenlohner-Freedman 
bound.
The value of the potential at the minimum is given by
\eq{
\label{valueminsusy}
   V_0=-{27 \cdot 3^{1\over 2}} \, {\tilde h^1\,
      (\tilde q^1)^3\over  (\hat{\mathfrak f}_1)^{1\over 2} (\tilde{\mathfrak f}^0)^{3\over
     2}} \,\, {M_\text{Pl}^4 \over 4\pi \cdot 2^7}\,.
}

We also find a non-supersymmetric strictly stable AdS minimum with frozen moduli
\eq{
\label{stablemodelD}
\arraycolsep20pt
\begin{array}{@{}lll@{}}
\displaystyle \tau = -{5^{1\over 2} \over 3\cdot 2^{1\over 2} }  \, {\tilde{\mathfrak f}^0 \over \tilde q^1}\,v \,, &
\displaystyle s =  -{2^{3 \over 2} \over  3\cdot 2^{1\over 2}}  \, {\tilde{\mathfrak f}^0 \over \tilde h^1}\,v\,, &
\displaystyle v^2 =  10^{1\over 2} \, { \hat{\mathfrak f}_1\over  \tilde{\mathfrak f}^0} \,, 
\\[16pt]
0=\tilde h^1\op c+ \tilde q^1 \op\rho\,, &
u = 0 \,.
\end{array}
}    
Note that for $\tilde q^1, \tilde h^1 < 0 < \hat{\mathfrak f}_1, \tilde{\mathfrak f}^0$, the moduli are in their physical regime.
The non-trivial contribution of the fluxes to the tadpoles given by
\eq{
\label{fluxtadD}
N_{\rm D3}^\text{flux}=\tilde{h}^1\, \hat{\mathfrak f}_1\,,\hspace{40pt}
N_{\rm D7}^\text{flux}=\tilde q^1 \hat{\mathfrak f}_1\,, 
} 
which in the physical regime are  negative.
We furthermore note that the flux $\hat {\mathfrak f}_0$ does not enter into the tadpoles.  
Thus, to guarantee weak coupling and large radius we can take  large $\hat {\mathfrak f}_0$, together with 
$\hat {\mathfrak f}_1,  \tilde h^1, \tilde q^1 \sim O(1)$.
The value of the potential of the non-tachyonic model \eqref{stablemodelD} in the minimum is found to be
\eq{
\label{valueminD}
    V_0=-{216 \cdot 2^{3 \over 4} \over  5^{5 \over 4}} {\tilde h^1\,
      (\tilde q^1)^3\over  (\hat {\mathfrak f}_1)^{1\over 2} (\tilde{\mathfrak f}^0)^{3\over
     2}}\, {M_\text{Pl}^4 \over 4\pi \cdot 2^7} \, ,
}
which is smaller than the potential \eqref{valueminsusy} in the supersymmetric extremum.
The moduli masses in the canonically normalized basis are
\eq{   
  M_{{\rm mod},i}^2=\mu_i  \, {\tilde h^1 (\tilde q^1)^3 \over (\hat{\mathfrak f}_1)^{1\over 2}
    (\tilde{\mathfrak f}^0)^{3\over 2}} \, {M_\text{Pl}^2 \over 4\pi \cdot 2^7} \,,
}
with 
\eq{ 
\mu_i= \bigl(\,291,\, 52, \, 35\, ; \,210,\, 1.8,\, 0\, \bigr) \,.
}
The first three eigenstates are saxions and the last three are axions.
Note that again the lightest massive mode is an axion, whose mass is
smaller  than the lightest saxion by a factor $1/5$.
The gravitino mass is again of the same order as the fluxes with
numerical prefactor $\mu_{3\over 2} \approx 114$. 

It is interesting to notice that the isotropic torus model D is related to model C by the transformation
$U  \to 1/U$ under which
\eq{
\label{udualCD}
W \to -\frac{i}{U^3} \left[ 
-\hat{\mathfrak f}_1 U ^2 - \, \tilde{\mathfrak f}^0  - 3\, \tilde h^1 U\, S - 3\, \tilde q^1 U\,  T \,
\right] = -\frac{i}{U^3} W^\prime
}
Hence, $e^K |W|^2 = e^K |W^\prime|^2$ and the resulting scalar potential is basically the same as
in model C because $W^\prime$ above has the same form as the superpotential in \eqref{superpotC}.
Indeed, notice that the vevs in both models, for instance \eqref{stablemodelC} and \eqref{stablemodelD}, match under $U \to 1/U$
and appropriate redefinition of the fluxes involved. This kind of transformation was exploited in \cite{Font:2008vd,deCarlos:2009fq} to classify the allowed superpotentials induced by non-geometric fluxes. Moreover, duality symmetries in moduli space allow to
fix the moduli vevs, thereby simplifying the search for vacua
\cite{Dibitetto:2011gm,Blaback:2013qza}.  
In non-toroidal models the transformation $U  \to 1/U$ is not expected to be a duality,
but still it can be used as a solution-generating technique.

%%%%%%%%%%%%%%%%%%%%%%%%%%%%%%%%%%%%%%%%%%%%%%%
%%%%%%%%%%%%%%%%%%%%%%%%%%%%%%%%%%%%%%%%%%%%%%%

\subsubsection{Freezing axionic $H^{1,1}_-$ moduli}
\label{freezing}

Let us consider the  case of a CY manifold with $h_-^{2,1}=1$
and  $h_+^{1,1}=h_-^{1,1}=1$, for which the K\"ahler potential reads
\eq{
  K=-3 \log \Big( (T+\ov T)+ {\kappa  \over 4\op(S + \ov S)}\, (G+\ov G)^2
  \Big)-\log (S+\ov S) -3\log\big(U + \ov U\big).
}
Although the resulting K\"ahler metric is off diagonal, 
we can still find extrema by extending the superpotentials of models C and D
to include a term depending on the $G$ modulus. 
Here, we just present  the generalization 
of model D. Turning on an additional geometric flux $\tilde f^1$ leads to 
\eq{ 
W= i\op \hat{\mathfrak f}_1 \op U - i \op \tilde{\mathfrak f}^0\op  U^3+3\op i \op \tilde h^1 \op U^2\op S
+3\op i \op \tilde q^1 \op U^2\op T  - 3\op i \op\tilde f^1\op U^2 \op G \,,
}
where we again introduced $\hat{\mathfrak f}_1 = - {\mathfrak f}_1$. 
Similarly to  model D, we obtain extrema with the flux scaling of $W_0$ shown in \eqref{scalingD}. 
In particular, we find a strictly stable non-supersymmetric AdS minimum characterized by 
\eq{ 
\begin{array}{l@{\hspace{50pt}}l}
\displaystyle  \tau = - \frac{1}{3\cdot 10^{1\over 2}} \, \frac{x+5}{x+1} \, {\tilde {\mathfrak f}^0 \over \tilde q^1}\, v \,, &
\displaystyle  s= - \frac{4}{3\cdot 10^{1\over 2}} \, {1 \over 1 + x} \, {\tilde {\mathfrak f}^0 \over \tilde h^1}\, v \,,
\\[16pt] 
\displaystyle  v^2 = 10^{1\over 2}\: \frac{\hat{\mathfrak f}_1 }{ \tilde{\mathfrak f}^0 }\, , &
\displaystyle  \psi = - \frac{8}{3\cdot 10^{1\over 2}} \, {x \over 1 + x} \, {\tilde {\mathfrak f}^0 \over \tilde f^1}\, v \, ,
\\[20pt]
0=\tilde h^1 \op c + \tilde q^1 \op \rho -  \tilde f^1\op \eta \,, &
u = 0 \,,
\end{array}
}
where
\eq{
x = {(\tilde f^1)^2\over {\kappa \op\tilde h^1\op \tilde q^1 }} \,.
}
To ensure that the moduli are in the physical regime, we require
$\tilde q^1, \tilde h^1 < 0 < \hat{\mathfrak f}_1, \tilde{\mathfrak f}^0$. 
The flux induced tadpoles, which are again given by \eqref{fluxtadD}, are negative and do not depend on
$\hat {\mathfrak f}_0$. We can then achieve weak coupling as well as large radius taking $\hat {\mathfrak f}_0 \gg 1$, and
$\hat {\mathfrak f}_1,  \tilde h^1, \tilde q^1 \sim O(1)$. 
The value of the scalar potential at the minimum is
\eq{
 V_0 = -\frac{27}{20 \cdot 10^{1\over 4} } \: \frac{\tilde h^1\, (\tilde q^1)^3\,\, (1+x)}{ ( \hat{\mathfrak f}_1 )^{1 \over 2} \, ( \tilde {\mathfrak f}^0 )^{3 \over 2}} \: \frac{M_\text{Pl}^4}{4\pi \cdot 2} \,,
}
and the mass eigenvalues of the fields are 
\eq{
M_{{\rm mod},i}^2 = \mu_i \, \frac{\tilde h^1\,  (\tilde q^1)^3 \,\, (1+x)}{ ( \hat{\mathfrak f}_1 )^{1 \over 2} \, ( \tilde {\mathfrak f}^0 )^{3 \over 2}} \: \frac{M_\text{Pl}^2}{4\pi \cdot 2} \,,
}
with the numerical prefactors given by 
\eq{
\mu_i = \bigl(\,4.6,\, 0.82,\, 0.55,\,0\,;\, 3.3,\,0.028,\, 0 ,\,0  \bigr) \,, 
\hspace{30pt}
\mu_{3 \over 2}= 0.333 \,.
}
The lightest eigenstate is still axionic, but the mass gap to the second-lightest state has decreased as compared to model D. 
As in model B,  a saxionic combination of $\tau$ and $\psi$ has become
massless. It would be interesting to know whether this is a generic feature of
non-tachyonic, non-supersymmetric minima for models with odd-moduli.

%%%%%%%%%%%%%%%%%%%%%%%%%%%%%%%%%%%%%%%%%%%%%%%
%%%%%%%%%%%%%%%%%%%%%%%%%%%%%%%%%%%%%%%%%%%%%%%

\subsubsection{Stabilization with non-geometric $P$-fluxes}
\label{sss:P}

We now want to present models including the S-dual $P$-fluxes discussed in section~\ref{sec_sdual}. 
The allowed $P$-fluxes are in general constrained by Bianchi identities. The models we analyze fulfill the constraints
derived for instance in \cite{Aldazabal:2006up}.
We will consider examples with and without odd K\"ahler moduli $G_a$.

\paragraph*{$\mathbf{h_-^{1,1}= 0}$}
We come back to model C with $h_-^{1,2} = 1$ and $h_+^{1,1} = 1$, for which 
the K\"ahler potential is given by 
\eqref{KSTU}. The new ingredient is an additional $P$-flux.
As an illustrative example we consider the superpotential
\eq{
\label{wpflux}
W = \hat {\mathfrak f} - 3\op \tilde{\mathfrak f} \op U^2 - h \op S \op U + p \op S \op T  \,,
}
with $\hat{\mathfrak f} := - {\mathfrak f}_0$, $\tilde{\mathfrak f} := \tilde{\mathfrak f}^1$,  $h:= h_1$ and $p:= p_0$. We find the same  structure of minima as in the other examples with the same Hodge numbers. 
Note that the superpotential is chosen in such a way that every modulus is stabilized. 
For $h$ negative and other fluxes positive, we find a tachyon-free
{\it supersymmetric} AdS vacuum at
\eq{ \label{Pfluxminimum}
\arraycolsep20pt
\begin{array}{lll}
\displaystyle \tau =  \frac{3}{2} \, \frac{h}{p}\, v \,, &
\displaystyle s = - \frac{12}{5} \, \frac{\tilde{\mathfrak f}}{h} \, v \, , &
\displaystyle  v^2 = \frac{5}{9}  \,\frac{ \hat {\mathfrak f}}{\tilde{\mathfrak f}} \,,
\\[16pt]
\rho  = 0 \,, &  c  = 0 \,, &  u = 0 \,.
\end{array}
}
The scalar potential and the superpotential at the minimum read
\eq{ \label{Pfluxminimum2}
V_0 =  \, \frac{288}{5^{5\over 2}}\,\, \frac{p^3 \, (\tilde {\mathfrak f})^{5\over 2}}{h^2 \, (\hat{\mathfrak f})^{3 \over 2}}\: 
{M_\text{Pl}^4 \over 4\pi \cdot 2^7} \, , 
\hspace{40pt}
W_0 =  - {4 \over 3}\,  \hat{\mathfrak f} \,,
}
and the masses are given by 
\eq{\label{Pfluxminimum3}
M_{{\rm mod},i}^2 = - \mu_i \,\frac{p^3 \, (\tilde {\mathfrak f})^{5\over 2}}{h^2 \, (\hat{\mathfrak f})^{3 \over 2}}\: 
{M_\text{Pl}^2 \over 4\pi \cdot 2^7} \,,
}
where the numerical prefactors take the values 
\eq{
\mu_i = \bigl(\,52, \,31,\, 15\,; \,73,\, 17,\, 5.7\,\bigr) \,, \hspace{40pt}
\mu_{3 \over 2} = 1.72\,. 
}
To be in the physical regime we demand $h, p < 0$ such that the masses are positive and the cosmological constant is negative.

The non-supersymmetric minimum is analogous to the above, except that in \eqref{Pfluxminimum} 
a minus appears in front of 
$\hat {\mathfrak f}$ in  $v^2$. Thus $ \hat{\mathfrak f}$ should now be taken negative. In this case $W_0 =  10\op\hat{\mathfrak f} /3$, while
the scalar potential and the flux dependence of the masses are the same as in the supersymmetric vacuum, 
except for the extra minus sign. The numerical prefactors of the masses are now 
$\mu_{3 \over 2} = 10.7$ for the gravitino 
and $\mu_i = (52,\, 31,\, 15\,;\, 45,\, -3.4,\, 19)$ for the moduli,
signaling the appearance of one tachyon above the Breitenlohner-Freedman bound.

\paragraph*{$\mathbf{h_-^{1,1}= 1}$}
It is interesting to extend the previous model by adding one odd K\"ahler modulus $G = \psi + i \eta$. In this case the K\"ahler potential is given by \eqref{model_B_02}. According to the general superpotential \eqref{w_04} the same $p$ flux in \eqref{wpflux} then leads to an additional term quadratic in $G$, namely
\eq{
\label{wpflux2}
W = \hat {\mathfrak f} - 3\op \tilde{\mathfrak f} \op U^2 - h \op S \op U + p \left( \op S \op T + {\kappa \over 4 } \, G^2 \right) \,,
}
where as in the K\"ahler potential $\kappa = 2 \kappa_{\alpha a b}$ with $\alpha = a = b = 1$.
It is important to notice that the $G^2$ and $S T$ terms are generated by the same $P$-flux. 
Since the K\"ahler potential and the superpotential differ from the previous ones only by terms depending on $G$ there are
supersymmetric and non-supersymmetric minima with the axionic odd moduli stabilized at $\psi = \eta = 0$.
The remaining moduli still take the values \eqref{Pfluxminimum} in the supersymmetric minimum, while
in the non-supersymmetric counterpart the difference is again a minus in front of $\hat {\mathfrak f}_0$. 
The potential and the superpotential at the minima are still given by \eqref{Pfluxminimum2}. 

The $G$ modulus decouples from the $S, T, U$ moduli in the canonically normalized mass matrix, thereby leading to the same 
masses in the $S, T, U$ sector as before. On the other hand, $\eta$ and $\psi$ turn out to be eigenstates of the canonically normalized mass matrix. The corresponding mass eigenvalues are of the form \eqref{Pfluxminimum3} with numerical prefactors
$(\mu_\psi ,\mu_\eta) =(0, -3.4)$ and $(0,17)$, for the supersymmetric and non-supersymmetric extrema, respectively.
Therefore both cases are now plagued with a tachyon and a massless saxion.
There exist additional extrema with unstabilized $\psi \neq 0$ showing the same qualitative behavior.

%%%%%%%%%%%%%%%%%%%%%%%%%%%%%%%%%%%%%%%%%%%%%%%
%%%%%%%%%%%%%%%%%%%%%%%%%%%%%%%%%%%%%%%%%%%%%%%

\subsubsection{A simple example with $h^{2,1}_-=2$} 
\label{sss:2U}

As the final example, let us discuss the case of more than one
complex structure modulus. 
We choose $h^{2,1}_-=2$, $h^{1,1}_{+} = 1$ and  $h^{1,1}_{-}= 0$
and work with a particularly simple prepotential such that K\"ahler potential reads
\eq{
      K=-2 \log ( U_1+\ov U_1 ) - \log ( U_2 + \ov U_2 ) - \log (S+\ov S)-3 \log(T+\ov T) \,.
}  
This corresponds to the mirror dual of the K\"ahler sector of ${\mathbb P}_{1,1,2,2,2}[8]$, 
which was discussed in section~\ref{ss:2kahler}.
For the superpotential we take
\eq{
  W= -\mathfrak{f}_0 -\left( h\op S + q\op T +\tilde{\mathfrak{f}}_2\op U_1 + 2 \op\tilde{\mathfrak{f}}_1\op U_2  \right) U_1
  \,,
}
where we have set $h_1 = h$ and $q_1 = q$.  In the extrema we obtain $u_1=0$, thus the superpotential depends 
effectively on the axionic combination 
$\theta = h \op c + q\op \rho + 2\op \tilde{\mathfrak{f}}_1\op u_2$, and the two orthogonal axions are unstabilized. 
In addition, all extrema contain at least one additional tachyon. As expected from the no-go theorem of \cite{Conlon:2006tq}, the supersymmetric minimum has in fact two tachyonic states, although above the Breitenlohner-Freedman bound. 
A representative non-supersymmetric AdS extremum is given by
\eq{
\label{min2U}
\arraycolsep15pt
\begin{array}{@{}l@{\hspace{15pt}}lll@{}}
\displaystyle \tau =2 \,  \frac{ \tilde{\mathfrak{f}}_2}{q}\, v_1\,, &
\displaystyle s = {2 \over 3} \,  \frac{ \tilde{\mathfrak{f}}_2}{h}\, v_1\,, &
\displaystyle v_1^2 = \frac{\mathfrak{f}_0}{\tilde{\mathfrak{f}}_2}\,,  &
\displaystyle v_2^2 = {1 \over 3} \, {\tilde{\mathfrak{f}}_2 \over  \tilde{\mathfrak{f}}_1} \, v_1 \,,
\\[16pt]
0=h\op c + q \op\rho + 2 \op\tilde{\mathfrak{f}}_1 \op u_2 \,, &
u_1 = 0 \,,
\end{array}
}
with 
\eq{
V_0 =- 3\,\,\frac{\tilde{\mathfrak{f}}_1 h\op  q^3}{\mathfrak{f}_0^{\frac{3}{2}}\, \tilde{\mathfrak{f}}_2^{\frac{3}{2}}}\,\, {M_\text{Pl}^4 \over 4\pi \cdot 2^7} \,,
\hspace{40pt}
W_0 = -{16 \over 3} \,\, {\mathfrak f}_0 \, .
}
The eigenstates of the normalized mass matrix can be computed to be 
\eq{
M_{{\rm mod},i}^2 = \mu_i \, \frac{\tilde{\mathfrak{f}}_1 \op h\op q^3}{\mathfrak{f}_0^{\frac{3}{2}}\, 
\tilde{\mathfrak{f}}_2^{\frac{3}{2}}}\:
{M_\text{Pl}^2 \over 4\pi \cdot 2^7}  \,,
}
where the numerical prefactor is $\mu_{3 \over 2} \approx 16$ for the gravitino mass and $ \mu_i = (18,\,18,\,-2,\, -2\, ; \, 10,\,10,\, 0,\,0)$ for the moduli masses. 
The massless eigenstates refer to the axionic combinations $(2 \op\tilde{\mathfrak{f}}_1 \op c - h\op u_2)$ and $( h \rho - qc)$, respectively, and the tachyonic directions are $( h \op \tau - q\op s)$ and $(2\op s\op \tilde{\mathfrak{f}}_1 - h\op  v_2)$. One can verify that the tachyons lie below the Breitenlohner-Freedman bound.

Let us mention one new feature arising in this model.
Computing the values of the auxiliary fields $F^i=e^{K\over 2}
K^{i\ov j} F_{\ov j}$ in the minimum \eqref{min2U}, we obtain
\eq{\label{FtermsforU}
\arraycolsep2pt
\begin{array}{@{}lcl@{\hspace{40pt}}lcl@{}}
F^S &=& \displaystyle e^{K \over 2} \,\frac{16\,({\mathfrak f}_0)^{3 \over 2}\, (\tilde{\mathfrak f}^2)^{1 \over 2}}{3h}\, , 
&
F^T &=& \displaystyle e^{K \over 2} \, \frac{16({\mathfrak f}_0)^{3 \over 2}\, (\tilde{\mathfrak f}^2)^{1 \over 2}}{q}\, , 
\\[10pt]
F^{U_1} &=& 0  , 
&
F^{U_2} &=&\displaystyle e^{K \over 2} \, \frac{8({\mathfrak f}_0)^{3 \over 2}\, (\tilde{\mathfrak f}^2)^{1 \over 2}}{3\tilde{\mathfrak f}^1}
\,.
\end{array}
}
This exemplifies that accidentally it can happen that an F-term vanishes
in a certain minimum. As we discuss in section 5,
such a result is essential  for realizing  sequestered supersymmetry
breaking on the Standard Model  branes.

%%%%%%%%%%%%%%%%%%%%%%%%%%%%%%%%%%%%%%%%%%%%%%%
%%%%%%%%%%%%%%%%%%%%%%%%%%%%%%%%%%%%%%%%%%%%%%%
%%%%%%%%%%%%%%%%%%%%%%%%%%%%%%%%%%%%%%%%%%%%%%%
%%%%%%%%%%%%%%%%%%%%%%%%%%%%%%%%%%%%%%%%%%%%%%%

\subsection{General properties of the flux scaling minima}
\label{ss:systematics}

In this section we summarize the salient aspects of the models constructed previously. We first explain the 
systematics behind our search of vacua and then address more specific features.

The defining property of the models is a common flux scaling of $W$ which in turn implies a common flux scaling of $V$ and the
moduli masses. This is potentially powerful to achieve parametric control over the hierarchies among the relevant scales
$M_{\rm s}$, $M_{\rm KK}$ and the moduli masses.
On the other hand, parametrically controlled hierarchies among the different moduli masses are then excluded. 
Later in section 6, we will try to circumvent this problem  by introducing
additional fluxes  in $W$ that break the scaling.

The strategy is to choose a superpotential dictated by a particular scaling. In practice this means that only a subset of the
allowed fluxes is turned on to ensure that the moduli vevs scale in a simple way. A typical example is model~C in
section \ref{sss:C}. In this case, at the extrema $W_0 \sim {\mathfrak f}_0$, whereas $sv \sim {\mathfrak f}_0/h$,
$\tau v \sim {\mathfrak f}_0/q$ and $v^2 \sim {\mathfrak f}_0/\tilde{\mathfrak f}^1$, 
as we can easily read from \eqref{superpotC}. 

Generically, for $n$ complex moduli it suffices to switch on $n+1$ flux parameters. For instance, to stabilize
$T^\alpha$ we include one flux of type $q_\lambda^\alpha$ or one of type $\tilde q_\lambda^\alpha$. Similarly,
for $S$ we take one $h_\lambda$ or one $\tilde h_\lambda$. For the
complex structure moduli we need one R-R flux of type 
${\mathfrak f}_\lambda$ and one $\tilde{\mathfrak f}_\lambda$. 
Of course, we have to be careful that the chosen NS-NS, R-R  and non-geometric fluxes
satisfy the Bianchi identities. 
We observe that in the studied  examples
 an off diagonal K\"ahler metric did not spoil the scaling of the potential and the masses. 
We want to stress that the scaling strategy can be used to efficiently engineer models with the desired pattern of masses. 
This will become apparent when we discuss the moduli spectroscopy in
section \ref{ss:modulispectrosopy} and  axion monodromy inflation
in section 6.

All models include NS-NS and R-R three-form fluxes that lead to stabilization of the real parts of the axio-dilaton and the
complex structure moduli. Further addition of non-geometric fluxes, as
well as of geometric ones when $h^{1,1}_- \neq 0$,
enables the stabilization of the K\"ahler moduli saxions. In general the fluxes always allow the existence of supersymmetric 
AdS vacua which often have tachyons above the Breitenlohner-Freedman
bound. 
One of the main  and maybe unexpected results  of this paper is that we have also found 
non-supersymmetric, non-tachyonic AdS vacua. We have seen that, once we
introduce more K\"ahler or more complex structure moduli, new
tachyonic modes appear. A natural question is whether one can identify an
uplift mechanism for these tachyons, hence  enlarging the space of 
good models. In section 4, we will see that a D-term uplift exists
for tachyons appearing for multiple K\"ahler moduli. For the tachyons
appearing for multiple complex structure moduli, such an uplift is
still an open question.

Equipped with a class of tachyon-free, non-supersymmetric AdS minima,
as in the LVS scenario, one can proceed to perform string
phenomenology studies in order to explore what
particle physics predictions can be made.
This  will be analyzed in more detail in chapter \ref{sec_pheno}. 
Recall that in these models (without $P$-flux) typically only one axionic combination is fixed. In chapter \ref{sec_cosmo} 
we examine how the axion sector can give rise to large field inflation.

A common feature of the models is the existence of R-R tadpoles due to the fluxes. Interestingly, in most examples
of AdS vacua we find that $N_{\rm D3}^{\rm flux}$ and $N_{\rm D7}^{\rm flux}$ are negative, as it happens in 
related T-dual type IIA models \cite{Camara:2005dc, Aldazabal:2007sn}. 
Thus, the flux tadpoles can be compensated by introducing
${\rm D}3$- and ${\rm D}7$-branes instead of ${\rm O}3$- and ${\rm O}7$-planes. Magnetized ${\rm D}7$-branes 
that induce ${\rm D}3$-charge are in principle allowed but they are constrained by cancellation of Freed-Witten anomalies. 

The R-R fluxes play a special role in the models since they determine the vevs of the moduli.
This is indeed the case for the complex structure moduli. On the other hand, 
the scale of the dilaton and the K\"ahler moduli is also set by the R-R fluxes upon taking the NS-NS, non-geometric
and geometric fluxes to be $O(1)$. Moreover, as we have seen, in
models with complex structure moduli
there are R-R fluxes that do not enter the flux tadpoles at all. It turns out that such fluxes can be chosen large enough to 
attain moduli vevs leading to large volume and small string coupling. In section \ref{sec_pheno} we will discuss
to what extent these fluxes are diluted in order to guarantee a reliable supergravity approximation.

%%%%%%%%%%%%%%%%%%%%%%%%%%%%%%%%%%%%%%%%%%%%%%%
%%%%%%%%%%%%%%%%%%%%%%%%%%%%%%%%%%%%%%%%%%%%%%%
%%%%%%%%%%%%%%%%%%%%%%%%%%%%%%%%%%%%%%%%%%%%%%%
%%%%%%%%%%%%%%%%%%%%%%%%%%%%%%%%%%%%%%%%%%%%%%%
%%%%%%%%%%%%%%%%%%%%%%%%%%%%%%%%%%%%%%%%%%%%%%%
%%%%%%%%%%%%%%%%%%%%%%%%%%%%%%%%%%%%%%%%%%%%%%%
%%%%%%%%%%%%%%%%%%%%%%%%%%%%%%%%%%%%%%%%%%%%%%%
%%%%%%%%%%%%%%%%%%%%%%%%%%%%%%%%%%%%%%%%%%%%%%%

\section{Uplift mechanisms}
\label{sec_uplift}
The models studied in the last section face two problems: all of them have a negative cosmological constant, 
and some of them have tachyonic mass eigenstates. To make an uplift to de Sitter possible, we therefore discuss a mechanism to uplift tachyonic K\"ahler moduli to a positive mass. Afterwards, we 
identify possible terms that uplift the cosmological constant and discuss their behavior.

%%%%%%%%%%%%%%%%%%%%%%%%%%%%%%%%%%%%%%%%%%%%%%%
%%%%%%%%%%%%%%%%%%%%%%%%%%%%%%%%%%%%%%%%%%%%%%%
%%%%%%%%%%%%%%%%%%%%%%%%%%%%%%%%%%%%%%%%%%%%%%%
%%%%%%%%%%%%%%%%%%%%%%%%%%%%%%%%%%%%%%%%%%%%%%%

\subsection{D-term uplifting of the tachyon}
\label{ss:dtermup}

To uplift tachyons one could  think that  taking perturbative and non-perturbative
corrections to $K$ and $W$ into account might help.
However, since we have taken care of freezing the moduli in the
perturbative regime, these corrections are generically suppressed
against the tree-level values. Of course, this also holds for the
tachyonic mass. The second and more natural option is to have an
additional  positive-definite contribution such as  a D-term potential. Thus,
in the following we study how a D-term of a stack of ${\rm D}7$-branes 
contributes to moduli stabilization and the mass terms. An analogous mechanism 
to uplift tachyons via D-terms from D-branes was proposed in \cite{Villadoro:2006ia}.

\paragraph*{K3-fibration}

To show how the D-term uplift works,  we perform our analysis in a concrete model.
In particular, we consider  the K3-fibration with $h^{1,1}_+=2$ and $h^{2,1}_-=0$ 
studied in section \ref{sss:k3fib}. The K\"ahler 
potential is given by
\eq{
   K=-2 \log  (T_1+\ov T_1) - \log  (T_2+\ov T_2)-\log (S+\ov S) \,,
}
whereas the superpotential is taken to be
\eq{  W=i\op \tilde{\mathfrak f} + i\op h\op S  +i\op q_1\op T_1 +i\op q_2\op T_2\,.
} 
Recall that in this model the supersymmetric AdS minimum is at
\eq{
\label{susyvaluesmin}
   \tau_1=-{\tilde {\mathfrak f}\over q_1}\,,\quad \tau_2=-{\tilde{\mathfrak f}\over 2
     q_2}\,,\quad
    s=-{\tilde {\mathfrak f}\over 2 h}\,, \quad hc +q_1 \rho_1 + q_2\rho_2 =0\, ,
}
and that there also exists a non-supersymmetric AdS minimum at
\eq{
\label{nonsusyvaluesmin}
   \tau_1=-{4 \op\tilde{\mathfrak f}\over 5 q_1}\,,\quad \tau_2=-{ 2\op \tilde {\mathfrak f}\over 5
     q_2}\,,\quad
    s=-{\tilde{\mathfrak f}\over  h}\,, \quad hc +q_1 \rho_1 + q_2\rho_2 =0 \,,
}
which  has mass eigenvalues 
\eq{
\label{massnonsusymin} 
         {M_{{\rm mod},i}^2}=\mu_i  {h \op q_1^2\op q_2\over {\tilde {\mathfrak f}^2}}{ M_{\rm Pl}^2 \over {4\pi \cdot 2^4}}\,,
}
with $\mu_i=(-15,\,11,\,42\, ; 23,\,0,\,0)$.
The tachyonic mode corresponds to a  linear combination of K\"ahler saxions given by
$\tau_{\rm tac }=q_2 \tau_1 -q_1 \tau_2$. To obtain positive vevs for the saxions we take 
$\tilde{\mathfrak f} <0$, $h >0$, $q_1>0$ and $q_2>0$.

We now introduce a stack of $N$ ${\rm D}7$-branes equipped with
a $U(1)$ gauge flux with 
\eq{
\label{c1div}
[c_1(L)]= [{\mathcal E}]=l_1 D_1 +l_2 D_2 \, ,
} 
where $D_{1,2}$ are two (effective) divisors in ${\mathbb P}_{1,1,2,2,2}[8]$ and $l_{1,2}\in \mathbb Z$.
The D$7$-branes are  wrapping a four-cycle defined by 
\eq{
\label{4cycle}
\Sigma=m_1 D_1 + m_2 D_2 \,,
}
with $m_{1,2}\in \mathbb Z$, which leads to a D-term potential of the form
\eq{  
\label{dterm}
V_D={M_{\rm Pl}^4 \over 2 {\rm Re}(f)} \xi^2\, .
}
Here $\xi$ is the Fayet-Iliopoulos (FI) term of the $U(1)\subset U(N)$ carried by the branes,
which is given by 
\eq{
 \xi={1\over {\cal V}}\int_{\Sigma} J\wedge c_1(L)\,,
}
and in \eqref{dterm}  we have assumed that
all charged fields have vanishing vevs. 
The holomorphic gauge kinetic function for the ${\rm D}7$-branes is  $f=T+\chi\op S$, where
$\chi ={1\over 4\pi^2} \int F\wedge F$ denotes the instanton number of the gauge flux
on the ${\rm D}7$-branes. In the example at hand, the volume is ${\mathcal V} = (t^1)^2 t^2$.

The wrapping numbers $(m_1,m_2)$ and the gauge fluxes $(l_1, l_2)$ are constrained by
the generalized Freed-Witten anomaly cancellation conditions \eqref{FWqfluxb}, which in the present  case lead
to 
\eq{
\label{fwla}
m_1 l_1 q_2 + \bigl(l_1\op m_2 + l_2\op m_1\bigr) q_1=0 \, .}
Using this condition, we find that the FI-parameter can be expressed as
\eq{
\xi={m_1 \op l_1\over q_1 \sqrt{\tau_2}\, {\cal V}} \bigl(q_1\op \tau_1 -2\op q_2 \op\tau_2\bigr)\, .
}
Note that for a supersymmetric minimum, a vanishing F-term
implies a vanishing D-term. And indeed, the values
\eqref{susyvaluesmin} give a vanishing FI-term.
Moreover, $\xi$ also vanishes for the non-supersymmetric minimum in \eqref{nonsusyvaluesmin}.
Therefore, adding the D-term will not change the
position of either extremum, but due to its positive-definiteness it is
expected to add positive contributions to the squares of the saxion masses.

We  now study in more detail the effect of adding a D-term to the former F-term  scalar potential.
Concretely, we  add 
\eq{ V_D={k \over \tau_1^2 \tau_2} 
{\bigl(q_1 \op \tau_1 -2\op q_2\op \tau_2\bigr)^2\over \bigl(m_1 \op \tau_1 + m_2\op \tau_2\bigr) \tau_2 } \,,
}
which is obtained by substituting the various ingredients in \eqref{dterm}.
Here $k$ is a positive numerical prefactor and for the gauge kinetic function we only included the
string tree-level part $\text{Re}(f) = m_1\op \tau_1 + m_2\op \tau_2$. As expected, the position of both the supersymmetric \eqref{susyvaluesmin}
and the non-supersymmetric  \eqref{nonsusyvaluesmin} extrema
do not change. Moreover, {} from the resulting mass matrix it follows that 
 only the mass eigenvalue corresponding to the tachyonic
 saxion $\tau_{\rm tac}$ receives corrections and can become positive.
 In the supersymmetric case a tachyonic state will remain, although above the Breitenlohner-Freedman
bound. In the non-supersymmetric
extremum there is only one negative mass eigenvalue that receives corrections, which  
is given by (in units of $M_{\rm Pl}^4/(4\pi)$)\,\footnote{Note that in the
following, we have omitted the factor $M_{\rm Pl}^4/(4\pi)$ for ease of notation.
It can be re-installed easily by dimensional analysis.} 
\eq{ 
m_{\rm tac}^2 = -{15\op h\op q_1^2\op q_2\over {16 \, \tilde {\mathfrak f}^2}} 
-  {375\op q_1^3 \op q_2^3\op k \over {4 \, \tilde {\mathfrak f}^3\op (m_1 q_1 + 2 m_2 q_1)}} \, .
}
We observe that the mass can become positive because $\tilde{\mathfrak f} < 0$.
For instance, choosing $h=2$, and $q_1=q_2=m_1=m_2=1$, implies that
$m_{\rm tac}^2$ will turn positive provided $k > -3 \op \tilde {\mathfrak f}/50$. We could take
for instance  $\tilde {\mathfrak f}=-10$ and $k=1$.
Thus, the tachyonic mode can be uplifted while the masses of the
other moduli do not change. Moreover, as the
D-term vanishes in the minimum, the cosmological constant  $V_0$ does not change either.

\paragraph*{Swiss cheese}
Uplifting of tachyons by D-terms also works in the `swiss-cheese' model of section \ref{sss:swiss}, as we now
briefly describe. As in the previous example we introduce $N$ ${\rm D}7$-branes with a $U(1)$ gauge
field \eqref{c1div}  wrapping a four-cycle \eqref{4cycle}. The main difference now is that the
non-zero intersection numbers are $\kappa_{111}$ and $\kappa_{222}$ which are taken to be equal.
Moreover, we expand the K\"ahler form as
$[J]=t^1 D_1 + t^2 D_2$, with $t^2 < 0$. 
Then, up to normalization, $t^1 =\sqrt{\tau_1}$,
$t^2 = -\sqrt{\tau_2}$, and ${\mathcal V}=(\tau_1^{3/2} - \tau_2^{3/2})$. The FI parameter
is found to be
\eq{
\xi={1\over \mathcal V } \bigl (m_1 \op l_1 \op \sqrt{\tau_1} -m_2 \op l_2 \op \sqrt{\tau_2}\bigr)\, .
}
On the other hand,  the Freed-Witten anomaly cancellation condition \eqref{FWqfluxb} now implies 
\eq{
\label{fwlac}
m_1 \op l_1\op  q_1+ l_2\op  m_2 \op q_2=0 \, .}
Substituting in \eqref{dterm} then gives
\eq{ V_D={k\op \bigl (q_2 \op\sqrt{\tau_1} + q_1\op \sqrt{\tau_2} \bigr)^2\over (m_1 \op \tau_1 + m_2\op \tau_2) 
\,\bigl(\tau_1^{3/2} - \tau_2^{3/2}\bigr)^2} \,,
}
where $k$ is again some positive number, and for ${\rm Re} f$ we took only the tree-level contribution
of the gauge kinetic function. The important
point is that $V_D$ vanishes not only for the supersymmetric AdS extremum as expected, but also for the non-supersymmetric ones
in table \ref{table_kokos}, which all happen to have $\sqrt{\tau_1/\tau_2} = |q_1|/|q_2|$. Indeed, cancellation
occurs because necessarily $q_1 q_2 < 0$ for the vevs in table \ref{table_kokos} to correspond to
true extrema of the F-term potential.

Adding $V_D$ to the F-term potential we find that only the mass
eigenvalue corresponding to the tachyonic
direction $\tau_{\rm tac }=q_2\op \tau_1 -q_1\op \tau_2$ changes. In
the non-supersymmetric extremum
with only the tachyon $\tau_{\rm tac}$ (third in table
\ref{table_kokos}) the new mass eigenvalue is given by
(in units of $M_{\rm Pl}^4/(4\pi)$) 
\eq{
m_{\rm tac}^2 =
-{5 h (q_1^3 + q_2^3)\over {36 \tilde{\mathfrak f}^2}} + {125\, k\, (q_1^3 + q_2^3)^3 \over{
 324 \tilde{\mathfrak f}^3 q_1 q_2 (m_1 q_1^2 + m_2 q_2^2)}}\, .
 }
Notice that $m_{\rm tac}^2$ can be uplifted precisely because $q_1\op q_2 <0$, while $h>0$, $\tilde{\mathfrak f}<0$, and
$(q_1^3 + q_2^3)>0$ to keep the saxion vevs positive.

To summarize, we have   identified a tachyon uplift mechanism, where a D-term on a ${\rm D}7$-brane, the
Freed-Witten anomaly conditions, and the nature of the non-supersymmetric 
minimum nicely conspire to  give a positive shift only for the tachyon mass.
Let us emphasize that for this uplift mechanism to work, it is
essential that $\xi$ vanishes not only for the supersymmetric minimum,
but also for the non-supersymmetric one.
Note that in concrete string model building, one will also have to take into account
tadpole cancellation conditions.

%%%%%%%%%%%%%%%%%%%%%%%%%%%%%%%%%%%%%%%%%%%%%%%
%%%%%%%%%%%%%%%%%%%%%%%%%%%%%%%%%%%%%%%%%%%%%%%
%%%%%%%%%%%%%%%%%%%%%%%%%%%%%%%%%%%%%%%%%%%%%%%
%%%%%%%%%%%%%%%%%%%%%%%%%%%%%%%%%%%%%%%%%%%%%%%

\subsection{Uplift of cosmological constant}

Eventually, also the cosmological constant needs to be uplifted so
that the vacuum becomes de Sitter. The common mechanism is to
add an extra sector to the theory, which changes  
the values of the moduli in the minimum in a controlled way, but adding
a substantial contribution to the vacuum energy.
In the KKLT \cite{Kachru:2003aw} and the LVS
\cite{Balasubramanian:2005zx} 
scenario, this can for instance be achieved
by adding anti-${\rm D}3$-branes which provide  a positive-definite contribution
to the potential
\eq{
              V_{\rm up}={\varepsilon\over {\cal V}^\alpha} \,,
}
where $\alpha=2$ for a $\ov{{\rm D}3}$-brane in the bulk and $\alpha=4/3$ for 
a brane located in a warped throat. In the LVS scenario,
the F-term contribution to the potential in the AdS minimum scales
as $V_F\sim {\cal V}_0^{-3}$ so that for the  $\ov{{\rm D}3}$-brane
to compete one needs $\varepsilon\sim {\cal V}_0^{\alpha-3}$ which is small
and provides the parameter controlling the slight shift of the minimum after
including the uplift potential.

Let us discuss whether such an uplift mechanism also works for the 
tree-level minima we are working with here. For concreteness, let us first
consider  this question for model A that was discussed in section \ref{sss_A}. To simplify notation, we define $\hat {\mathfrak f} := - \tilde {\mathfrak f}$ such that all fluxes in this section must be chosen positive to be in the physical regime. 
Looking at the terms in the scalar potential
\eqref{scalarpotentialexa}, we realize that in the minimum all terms
scale as $h\op q^3/\hat{\mathfrak f}^2$. Taking into account that 
for perturbative control we need $\hat{\mathfrak f} \gg h,q$, we
are led to an uplift term of the form
\eq{
            V_{\rm up}={\varepsilon\over 16\, {\tau}^\beta}
           \hspace{40pt} {\rm
              with}\quad  0<\beta<2 \,,
}       
in order to have $\varepsilon\sim \hat{\mathfrak f}^{\beta-2}$
small. Therefore, the two types of $\ov{{\rm D}3}$-brane uplifts mentioned above do
not work in our case.

\begin{figure}[t]
  \centering
  \vspace*{10pt}
  \includegraphics[width=0.6\textwidth]{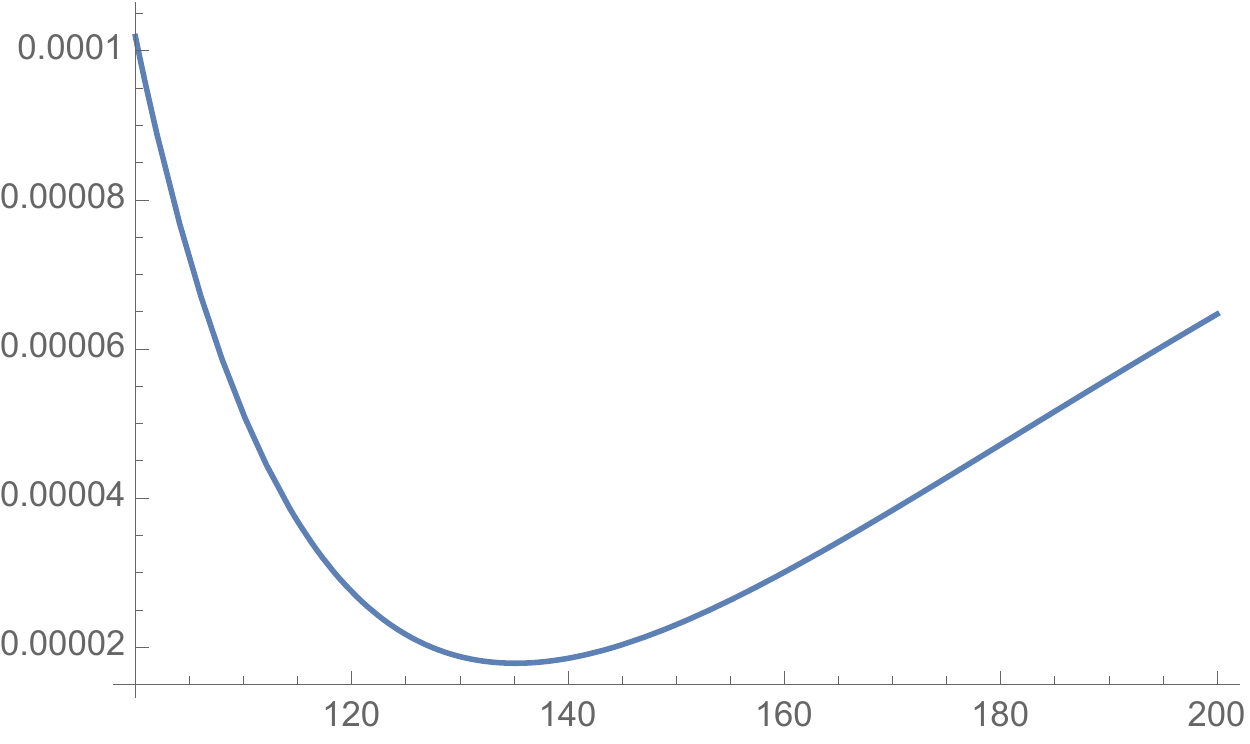}
  \begin{picture}(0,0)
  \put(-1,10){$\tau$}
  \put(-228,150){$V$}
  \end{picture}
  \caption{\small The exact potential $V(\tau,s_0,c_0,\rho_0)$ for $h=2,q=1$ and  $\hat{\mathfrak
      f}=100$ and the uplift term $V_{\rm up}=0.0013/(16\, \tau^{1\over 4})$.
    Here the numerical minimum lies at $\tau_0=135.13$, $s_0=40.60$
    and  $2c_0+\rho_0=0$. The uplifted minimum is de Sitter.}
  \label{fig_deSitter}
\end{figure}
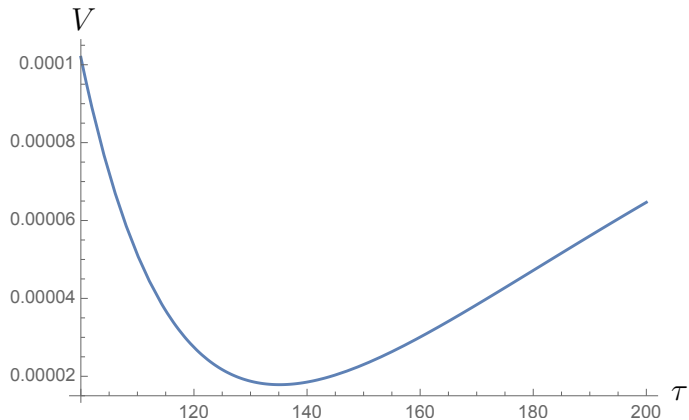

We can nonetheless study the above uplift for general $\beta$. Working at linear
order in the small parameter $\varepsilon$, one can show that the values
in the stable, non-supersymmetric minimum are shifted as
\eq{
\label{minvalcorr}
            \tau_0&= {6 \op \hat{\mathfrak f}\over 5\op q} +\varepsilon \,{3^{2-\beta} \beta 
              \over 5^{2-\beta}\op 2^{\beta+1}}
            { \hat{\mathfrak f}^3\over h\op q^4} 
  \left( {q\over \hat{\mathfrak f}}\right)^\beta + O(\varepsilon^2)\,,\\[0.1cm]
  s_0&={\hat{\mathfrak f}\over h} -\varepsilon\, {3^{2-\beta} \beta 
              \over 5^{2-\beta} \op 2^{\beta+1}}
            {\hat{\mathfrak f}^3\over h^2\op q^3} 
  \left( {q\over \hat{\mathfrak f}}\right)^\beta + O(\varepsilon^2) \,.
}
The value of the scalar potential at the minimum gets shifted as
\eq{
           V_0=-{25 \op h \op q^3  \over 216\, \hat{\mathfrak f}^2}
            +{\varepsilon\over 16} 
            \left({5 q\over 6\hat{\mathfrak f}}\right)^\beta + O(\varepsilon^2) \,.
}
Therefore, we could  uplift to $V_0=0$ for 
\eq{
\label{estiepsi}
\varepsilon \simeq {2^{\beta+1}  
              5^{2-\beta} \over  3^{3-\beta}}
            {h q^3 \over \hat{\mathfrak f}^2} 
  \left( {\hat{\mathfrak f}\over q }\right)^\beta \,,
}
which is small in the
perturbative regime $\hat{\mathfrak f}\gg h,q$. 
Inserting this value back into \eqref{minvalcorr}, we find
\eq{
         \tau_0={\hat{\mathfrak f}\over  q} \left({6\over 5}+{\beta\over 3}\right)\,,\qquad
     s_0\simeq {\hat{\mathfrak f}\over h} \left(1-{\beta\over 3}\right) ,
}
so that the correction term
is of the same order as the initial value. Therefore, it is not clear
whether the $O(\varepsilon^2)$ corrections are actually subleading.
Performing a numerical analysis we find that, indeed, choosing
$\varepsilon$ sufficiently  large  to uplift to a de-Sitter vacuum \eqref{estiepsi}, the minimum
gets destabilized for $\beta \gtrsim 1/4$. The same numerical behavior
is found for model B.
For $\beta=1/4$ and a specific choice of fluxes,  figure
\ref{fig_deSitter} shows a  plot of the potential around the
uplifted de Sitter minimum. We note that no linear approximation in $\varepsilon$
was done here.

%%%%%%%%%%%
%%%%%%%%%%%
%%%%%%%%%%%
%%%%%%%%%%%

Let us also check how the mass eigenvalues of model A change due to the  uplift.
Recall from \eqref{massesA} that the masses of the moduli in the stable AdS vacuum 
scale as $M_{{\rm mod},i}^2=\mu_i\op {h q^3\over \hat{\mathfrak f}^2} \, {M_\text{Pl}^2 \over 4\pi 2^4}\,$ with 
the numerical factors $\mu_i = (6.2,\, 1.7\, ; \, 3.4,\, 0 )$. In the uplifted
Minkowski vacuum the scaling remains the same and the numerical
factors decrease slightly to $\mu^{\rm up}_i =(5.3,\, 0.8\,;\, 2.9,\,0)$.

Of course, the question now is which string theoretical  effect can generate an
effective uplift potential of the required type. As we have discussed, the introduction of
just $\ov{{\rm D}3}$-branes does not give the appropriate power 
 $\beta$. One could envision more complicated uplift sectors, where
also matter-field contributions to D-terms and F-terms play a
role\footnote{For a D-term with such  low $\beta$ the modular weight of the
  matter field K\"ahler metric needs to be positive. For abelian
  heterotic orbifolds this  can occur for the twisted
  sector fields\cite{Ibanez:1992hc} .}. However, for
our purposes here, we just state that the uplift procedure is in
principle possible, but needs more care than in the KKLT and LVS scenarios.

%%%%%%%%%%%%%%%%%%%%%%%%%%%%%%%%%%%%%%%%%%%%%%%
%%%%%%%%%%%%%%%%%%%%%%%%%%%%%%%%%%%%%%%%%%%%%%%
%%%%%%%%%%%%%%%%%%%%%%%%%%%%%%%%%%%%%%%%%%%%%%%
%%%%%%%%%%%%%%%%%%%%%%%%%%%%%%%%%%%%%%%%%%%%%%%
%%%%%%%%%%%%%%%%%%%%%%%%%%%%%%%%%%%%%%%%%%%%%%%
%%%%%%%%%%%%%%%%%%%%%%%%%%%%%%%%%%%%%%%%%%%%%%%
%%%%%%%%%%%%%%%%%%%%%%%%%%%%%%%%%%%%%%%%%%%%%%%
%%%%%%%%%%%%%%%%%%%%%%%%%%%%%%%%%%%%%%%%%%%%%%%

\section{Physical aspects of the scaling vacua}
\label{sec_pheno}

In this section, we study several phenomenological aspects and problems of 
the flux vacua constructed above.
After discussing issues concerning the dilute-flux limit, we investigate the mass hierarchies of our models in several examples. 
We then consider particle-physics questions, in particular, we compute soft terms for a MSSM-like  D-brane setup.

%%%%%%%%%%%%%%%%%%%%%%%%%%%%%%%%%%%%%%%%%%%%%%%
%%%%%%%%%%%%%%%%%%%%%%%%%%%%%%%%%%%%%%%%%%%%%%%
%%%%%%%%%%%%%%%%%%%%%%%%%%%%%%%%%%%%%%%%%%%%%%%
%%%%%%%%%%%%%%%%%%%%%%%%%%%%%%%%%%%%%%%%%%%%%%%

\subsection{A note on the dilute flux limit}

Describing the string flux compactifications investigated above in an
orientifolded ${\cal N}=2$ gauged-supergravity 
framework  can only be an approximation, where 
fluxes are considered as (small) perturbations around the flux-less
Calabi-Yau geometry. The superpotential and the induced scalar
potential describe in an effective four-dimensional framework, how
the system reacts upon turning on fluxes that give extra contributions
to the ten-dimensional equations of motion. 
Usually, one hopes that the appearance of 
minima of $V_{\rm eff}$ signals new solutions of the 
full ten-dimensional string equations of motion.

In the case that only  NS-NS and R-R three-form fluxes are turned on,
it has been shown that the backreaction gives a warped Calabi-Yau 
geometry \cite{Giddings:2001yu}. Since, due to the no-scale structure, the K\"ahler moduli 
are unstabilized, one can take the large-volume limit, in which the 
fluxes become diluted and the effective gauged supergravity
description becomes a controlled approximation.

For the models discussed here, also  the K\"ahler moduli  are 
stabilized by turning on non-geometric fluxes, so that the
backreacted geometry is not explicitly known.  
Even though the  K\"ahler moduli are stabilized in terms of fluxes in 
the perturbative regime,  a priori it is not  clear whether a 
dilute-flux limit really exists. In fact, the expectation from
generalized
geometry and double field theory  is that non-geometric flux changes
the space from a smooth manifold to a T-fold  where the transition
functions between two charts are given by a T-duality transformation
(for reviews see \cite{Aldazabal:2013sca,Berman:2013eva,Hohm:2013bwa}).
Since the latter identifies small and big  radii, it would be
surprising if a dilute-flux limit did exist because the geometry could then be
better and better approximated by a flat torus or Ricci-flat
Calabi-Yau space, respectively.

Let us investigate this point for the class of models  presented in
this paper. For that purpose, we focus on model A as being realized
on the isotropic six-torus with fixed complex structure modulus $U=1$.
We then consider the flux kinetic terms in the ten-dimensional Einstein-frame action, including also the
non-geometric $Q$-flux \cite{Andriot:2011uh,Geissbuhler:2013uka,Blumenhagen:2013hva}
\eq{
\label{oxiaction2}
        S={1\over 2 \kappa^2_{10}}\int  d^{10} x\,  \sqrt{-g} \Big(  {\cal L}^{HH}+ {\cal L}^{QQ}_1 + {\cal L}^{QQ}_2
           +{\cal L}^{HQ} +{\cal L}^{\rm RR} \Big) \,,
}
with the various contributions given by 
\eq{
\label{tyepIIBoxidize}
\arraycolsep2pt
\begin{array}{@{}lcl@{\hspace{40pt}}lcl@{}}
\displaystyle {\cal L}^{HH}&=&
\displaystyle -{e^{-\phi}\over 12} {H}_{ijk}\,  {H}_{i'j'k'}\, g^{ii'} g^{jj'} g^{kk'} \,, 
&
\displaystyle {\cal L}^{HQ} &=& 
\displaystyle {1\over 2} {H}_{mni} \, {Q}_{i'}{}^{mn}\, g^{ii'} \,, 
\\[12pt]
\displaystyle {\cal L}^{QQ}_1&=&
\displaystyle - {e^{\phi}\over 4} {Q}_k{}^{ij}\, {Q}_{k'}{}^{i'j'}\,g_{ii'} g_{jj'} g^{kk'} \,, 
&
\displaystyle {\cal L}^{QQ}_2&=&
\displaystyle -{e^{\phi}\over 2} {Q}_m{}^{ni}\,  {Q}_{n}{}^{mi'}\, g_{ii'} \,, 
\\[12pt]
\displaystyle  {\cal L}^{\rm RR}&=&
\displaystyle   -{e^{\phi}\over 12}\, {\mathfrak F}_{ijk}\,  {\mathfrak F}_{i'j'k'}\, g^{ii'} g^{jj'} g^{kk'} \,.
\end{array}
}
In \cite{Blumenhagen:2013hva} it was shown explicitly that this action coincides with the one
derived in double field theory. Moreover,
upon dimensional reduction it gives the scalar potential generated by
the superpotential and tree-level K\"ahler potential reviewed in
section 2.
With the fluxes being integers, in model A the dilaton and
metric behave as
\eq{
       e^{-\phi}\sim s\sim {\hat {\mathfrak f}\over h}\,,\qquad 
       g\sim \sqrt{\tau}\sim {\hat{\mathfrak f}^{1\over 2}\over q^{1\over 2}}
\,,\qquad 
       g^{-1}\sim {q^{1\over 2} \over \hat{\mathfrak f}^{1\over 2}}\,,
}
where $\hat{\mathfrak f}=-\tilde{\mathfrak f}$.
Hence, all the kinetic terms in \eqref{tyepIIBoxidize} scale in the
same way as
\eq{ 
    {\cal L}^{HH}\sim {\cal L}^{QQ}_1\sim {\cal L}^{QQ}_2\sim {\cal
      L}^{HQ} \sim {\cal L}^{\rm RR}\sim {h q^{3\over 2}\over \hat{\mathfrak f}^{1\over 2}}\,.
}
Therefore, in the large radius limit, $\hat{\mathfrak f}\gg 1$, all terms
are suppressed and one could think that there exists a dilute flux
limit.
However, in order to control the backreaction of the fluxes on the
geometry,  the  essential quantity is not the action but the energy-momentum tensor
$T_{ij}=\frac{1}{\sqrt{-g}}{\delta S\over \delta g^{ij}}$,
appearing on the right-hand-side of the Einstein equation.
Now, it turns out that all contributions to $T_{ij}$ scale in the same way, namely
\eq{
         T^{HH}_{ij}\sim T^{QQ}_{1\, ij}\sim T^{QQ}_{2\, ij}\sim T^{HQ}_{ij}\sim T^{\rm
           RR}_{ij}\sim h\op q\, .
}
Therefore, the backreaction of the fluxes on the metric is of order
one and is not diluted in the limit $\hat{\mathfrak f}\gg 1$.
On the other hand, the backreaction is also not substantially large,
i.e. we do not have $T_{ij}\sim \hat{\mathfrak f}^p$ for some positive
power $p$. Therefore, it can be claimed that we are on the boundary of
controlling/non-controlling the backreaction.

The upshot is that the existence of a full string theory uplift of the
solutions found using the effective supergravity action, is on a less firm ground
than for no-scale models with only NS-NS and R-R three-form fluxes $H$
and ${\mathfrak F}$. Nevertheless, absence of an argument for dilute fluxes
logically does not rule out that the effective string (double
field) theory still provides
a sort of consistent truncation of the full dynamics of the
theory. This is still an unsettled open question in double field
theory.

%%%%%%%%%%%%%%%%%%%%%%%%%%%%%%%%%%%%%%%%%%%%%%%
%%%%%%%%%%%%%%%%%%%%%%%%%%%%%%%%%%%%%%%%%%%%%%%
%%%%%%%%%%%%%%%%%%%%%%%%%%%%%%%%%%%%%%%%%%%%%%%
%%%%%%%%%%%%%%%%%%%%%%%%%%%%%%%%%%%%%%%%%%%%%%%

\subsection{Moduli spectroscopy}\label{ss:modulispectrosopy}

We now look at the moduli spectroscopy of the flux-scaling vacua, i.e.
we investigate whether one has control over the desired hierarchy of
mass scales
\eq{ \label{masshierarchya}
M_{\rm Pl}> M_{\rm s}>M_{\rm KK}> M_{\rm mod}\,.
}
The first two hierarchies are evident. In order to trust the four-dimensional supergravity approximation we are
using, the masses of the moduli should also be smaller than the
Kaluza-Klein scale. One can define an   additional mass-scale which is
related to the energy-density in the uplift potential
\eq{
M_{\rm  up}=(V_{\rm up})^{1\over 4}= |V_0|^{1\over 4}\,.
}
In the flux-scaling models we had large  fluxes $f_L$ guaranteeing that the
moduli are in their perturbative regime and other fluxes $f_S$ that we
usually choose to be of order one. Moreover, there are further  order one
coefficients entering the K\"ahler potential, once we specify a
concrete  Calabi-Yau manifold. 

Let us now formalize what we mean by
parametrical control:
A scale $M_1$ is called {\it parametrically larger} than a scale $M_2$,
denoted as $M_1\parag M_2$, if it occurs that $M_2/M_1\to 0$ for $f_L\to \infty$.
The two scales are called {\it parametrically equal}, $M_1\parasim M_2$, if $M_2/M_1\to O(1)$ for $f_L\to \infty$.
This distinguishes  the case where one has   parametric control over
the relative size of two mass scales from the case when their relative
size is just a numerical coincidence. It can happen that even though
$M_1\parasim M_2$  one of the order one fluxes $f_S$ can guarantee
parametric control. If that is the case we mention it explicitly.
It is also possible that in our examples it just happens that the
numerical prefactors are such that $M_1>M_2$. In this case, we say
that $M_1$ is numerically larger than $M_2$ and denote it as $M_1\numg M_2$.

We observe that in all the models we have studied, we have demanded
that moduli are stabilized in the perturbative regimes as $\tau,s,v \parag 1$, which lead us to the following 
relations for the mass scales
\eq{
\label{generalparas}
                 M_{\rm up}^2\parasim M_{\rm mod}\, M_{\rm
                   Pl}\,,\qquad\qquad
                M_{\rm up}\parag M_{\rm s}\,.
}
The first relation can be viewed as a generic prediction for this
class of models, where the second relation rather indicates that
the energy density in the uplift potentially exceeds the scale where
we can confidently use the effective supergravity description.
In Appendix \ref{app_a} we provide a model, showing that 
one can have  $M_{\rm up}\paras M_{\rm s}$, once one gives up the
requirement that all $\tau,s,v \parag 1$.
We will come back to this point in section 6, when we discuss
inflation. In that context the two relations receive a different interpretation.

The relative sizes of $M_{\rm s},M_{\rm KK}$ and $M_{\rm mod}$ turn
out to be  model dependent. Let us discuss three representative examples.

\paragraph*{Model A}
We first discuss model A of section~\ref{sss_A}. Using \eqref{stringandKKscale} we can calculate the Kaluza-Klein and the string scale. For the tachyon-free vacuum 3 of model A we obtain
\eq{
         {M_{\rm s}\over M_{\rm KK}}=2\pi \left({2 \tau\over s}\right)^{1\over
           4}=2\pi \left({12\over 5}\right)^{1\over 4} \left({h\over
             q}\right)^{1\over4} \,.
}
Therefore, to have the string scale parametrically higher than the
KK-scale, we need to require $h>q$. This means $\tau>s$ so that $\alpha'$-corrections 
to the tree-level K\"ahler potential are indeed
subleading.
The ratio of the KK-scale to the moduli mass scale comes out as 
\eq{
     { M_{\rm KK}\over M_{\rm mod}}= \, {10\over 6 \sqrt{\mu_i\, h q }} \,,
}
Since both ratios do not depend on the very large flux $\hat{\mathfrak  f}$, we
would write that $M_{\rm s}\parasim M_{\rm KK} \parasim M_{\rm mod}$.
However, by choosing for the order one fluxes $h>q$ we can at least
guarantee $M_{\rm s}\parag M_{\rm KK}$.
However, the 
KK-scale is not separated from the flux induced moduli masses.

\paragraph*{Model D}

The same problem appears for model D of section~\ref{sss:D}. The scales for the tachyon-free vacuum are
\eq{
  M^2_{\rm s}=\mu_{\rm s} \,{(\tilde h^1)^{1\over 2} (\tilde q^1)^{3\over 2}\over 
   \hat{\mathfrak f}_1
    \tilde{\mathfrak f}^0} \,{M^2_{\rm Pl} \over 4\pi \cdot  2^7}\,,\hspace{40pt}
   M^2_{\rm KK}= \mu_{\rm KK} \,{(\tilde q^1)^{2}\over 
   \hat{\mathfrak f}_1\,
    \tilde{\mathfrak f}^0} \,{M^2_{\rm Pl} \over 4\pi  \cdot 2^7}\,,
} 
with $\mu_{\rm s}=2274$ and $\mu_{\rm KK}=36$.
For the ratio of the Kaluza-Klein scale and the moduli masses we find 
\eq{ \label{massratioD}
       { M^2_{\rm mod}  \over  M^2_{\rm KK}} = \mu \,\,  {\,
         (\tilde h^1)^{} (\tilde q^1)^{} (\hat{\mathfrak f}_1)^{1\over 2}
\over  (\tilde{\mathfrak f}^0)^{1\over 2}}\,,
}
with the prefactor $\mu$ of order one. Recall that we had to scale
$\hat{\mathfrak f}_1 \gg \tilde h^1, \tilde q^1, \tilde{\mathfrak f}^0
\approx O(1)$ in order to be in the weak-coupling and large-radius
regime. 
Then the KK-scale becomes parametrically lighter than the heavy
moduli,  $M_{\rm mod} \parag M_{\rm KK}$.

\paragraph*{Model C}

\begin{figure}[t]
  \centering
  \includegraphics[width=0.6\textwidth]{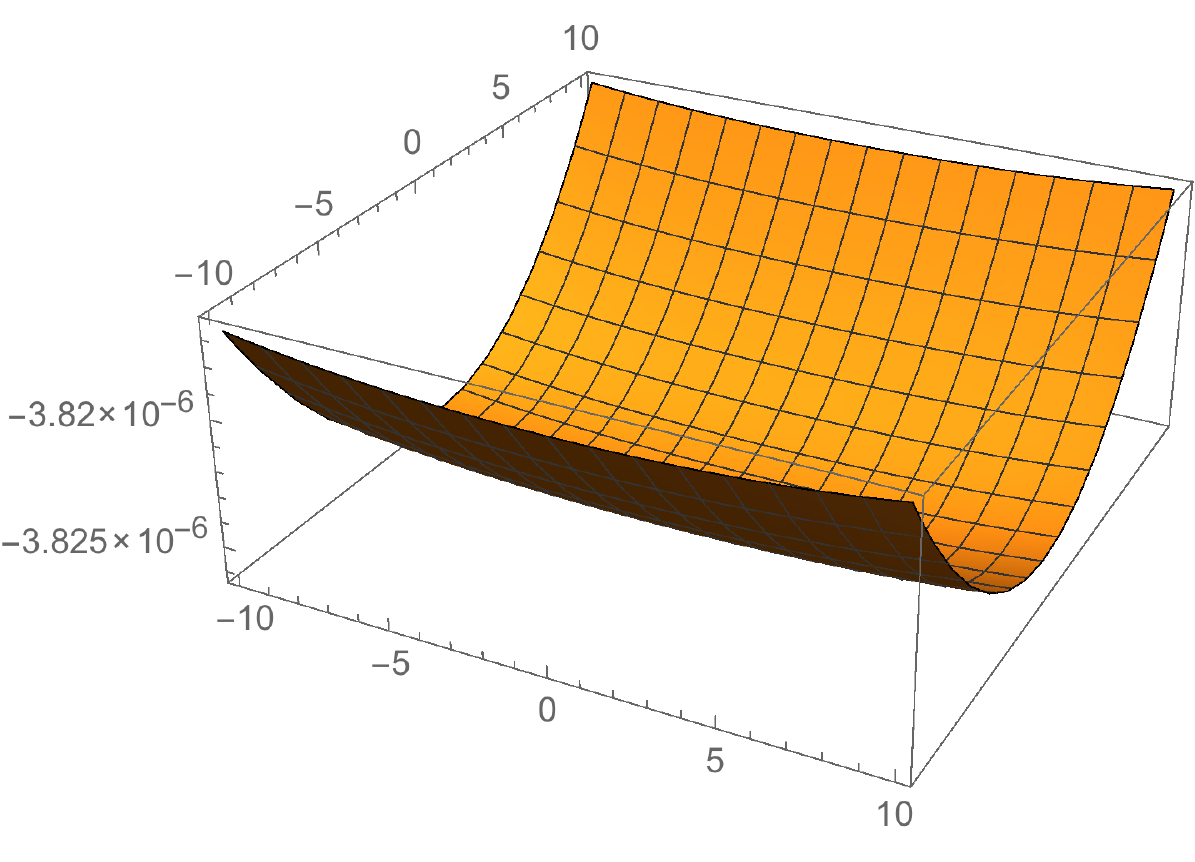}
  \begin{picture}(0,0)
    \put(-25,50){$y$}
    \put(-150,20){$x$}
  \put(-214,130){$V$}
  \end{picture}
  \caption{\small The potential $V(x,y)$ around the minimum, where $x$
    is pointing in  the direction of the lightest axionic modulus and 
   $y$  in the direction of the lightest saxionic modulus. }
  \label{fig_2}
\end{figure}
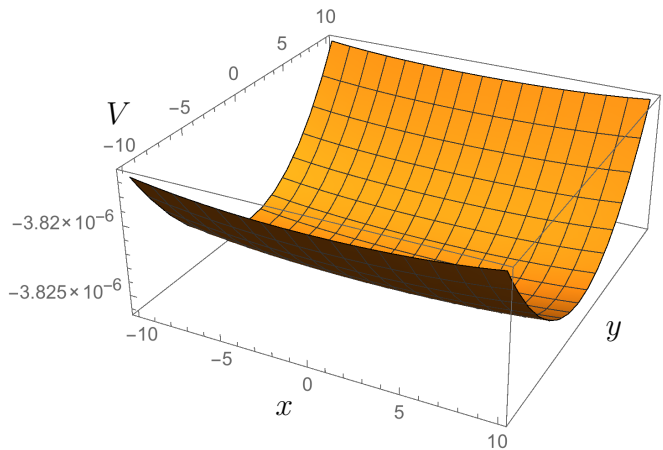

Finally, let us present a model where  parametric control over the ratios in principle is not in conflict with the perturbative regime. Using \eqref{stringandKKscale} in model C, the string and Kaluza-Klein scale are computed as
\eq{
  M^2_{\rm s}=\mu_{\rm s}\, {h^{1\over 2} q^{3\over 2}\over 
   {\mathfrak f}_0\op
    \tilde{\mathfrak f}^1}\, {M^2_{\rm Pl} \over 4\pi \cdot 2^7}\,,\hspace{40pt}
   M^2_{\rm KK}= \mu_{\rm KK} \,{q^{2}\over 
   {\mathfrak f}_0\,
    \tilde{\mathfrak f}^1}\, {M^2_{\rm Pl} \over 4\pi \cdot 2^7}\,,
} 
with $\mu_s=84$ and $\mu_{\rm KK}=1.4$. For the ratio of the Kaluza-Klein and the string  scale we obtain
\eq{
\label{hannover96}
       { M^2_{\rm KK} \over M^2_{\rm s} }= 0.016  \left({q\over
         h}\right)^{1\over 2}\,,
}
whereas   the ratio of the moduli masses and the Kaluza-Klein scale  is
\eq{
 { M^2_{\rm mod} \over M^2_{\rm KK}}\sim  { 
         h^{}\op q^{}\op (\tilde{\mathfrak f}^1)^{1\over
           2}\over
     {\mathfrak f}_0^{1\over 2}}\,,
}
so that for large enough ${\mathfrak f}_0$ one can ensure that the moduli
are lighter than the Kaluza-Klein scale, i.e.  $M_{\rm KK} \parag
M_{\rm mod}$.
To summarize, for this model we find the following controlled
hierarchy  of mass scales
\eq{ \label{masshierarchyb}
M_{\rm Pl}\,\parag\, M_{\rm up} \,\parag\, M_{\rm s}\,\parag\, M_{\rm KK}\,\parag\, M_{\rm mod}\,.
}
Since all scales differ only by a relative factor of $O(10)$, they are very
sensitive to numerical prefactors.   
For concreteness let us  make the choice
\eq{
\label{valuesmodelC}
 {\mathfrak f}_0=3200\,,\qquad \tilde{\mathfrak f}^1=1\,,\qquad h=-2\,,\qquad q=-1\,,
}
and analyze the moduli around the minimum with values
\eq{
     \tau = 275\,,\qquad  s = 110\,,\qquad  v = 18\,,\qquad  u=c=\rho=0\,.
}
Using $M_{\rm Pl}=2.44\cdot 10^{18}\,$GeV, the string and KK-scale  come out as
\eq{
    M_{\rm s}\sim 1.17 \cdot 10^{16}\,{\rm GeV}\,,\hspace{40pt} M_{\rm KK}\sim
  1.25\cdot 10^{15}\, {\rm GeV}\,.
}
The masses of the saxion moduli are 
\eq{
    M_i^{\rm sax}\sim  \bigl(2.9,\, 1.2,\,
    1.0\, \bigr)\cdot 10^{14} \,{\rm GeV}\,, 
}
and the masses of the two heavy axions are 
\eq{
       M_i^{\rm ax}\sim \bigl( 2.5,\, 0.23 \, \bigr) \cdot 10^{14}\,{\rm GeV}\,.
}
Note that the second axion is the lightest (massive) axion and
therefore could be a candidate for an inflaton. 
In figure \ref{fig_2} we show the potential around the minimum, in
the directions of the lightest and the second-lightest modulus.

\subsection{Tunneling between  flux branches}

Another potential problem could be the occurrence of substantial
tunneling between the various branches in the flux landscape. Such
tunnelings are induced by domain walls changing the value of the 
fluxes and were discussed to be a potential problem also to maintain
axion-monodromy inflation for a long enough period \cite{Brown:2015iha}.

Let us  estimate this tunneling rate for Model A using
the  formulas from Coleman-De Luccia for the thin-wall approximation \cite{Coleman:1977py,Coleman:1980aw}. The tunneling
amplitude can be computed from
\eq{
         P\sim e^{-S_{\rm CD}}\, , \qquad  
    S_{\rm CD}\sim {B\over (1+A^2)^2}\,,
}
with coefficients given by
\eq{
      B={27\pi^2 \sigma^4\over 2 (\Delta V)^3 }\,,\qquad
    A={\sqrt{3} \sigma \over 2\sqrt{ \Delta V} \, M_{\rm Pl} }\,.
}
Here $\sigma$ denotes the tension of the domain wall and
$\Delta V$ the potential difference 
on the left and the right hand side of the domain wall.     

For Model A the fluxes we have turned on are the NS-NS and R-R
three-form fluxes and the non-geometric Q-flux. 
Thus, following \cite{Marchesano:2014mla},  the corresponding domain walls  are the NS5-brane, the D5-brane
and the non-geometric $5^2_2$-brane \cite{LozanoTellechea:2000mc,deBoer:2010ud} wrapped on three-cycles of the
Calabi-Yau manifold. Due to the tadpole cancellation conditions,
we cannot change a single flux but only all three fluxes at the same
time. Therefore, we expect the corresponding domain wall to be
a bound state of those three kinds of branes. Let us compute the
action for each individual brane, where as we will see it is sufficient to
consider the NS5-brane and the D5-brane.

Via dimensional reduction of the corresponding world-volume actions,
we find
\eq{
                    \sigma_{\rm D5}\sim M_s^3 \, s^{1\over 4}\,
                    \tau^{3\over 4}\,,\qquad 
                      \sigma_{\rm NS5}\sim M_s^3 \, s^{5\over 4}\,
                    \tau^{3\over 4}\,.
}
Using $\Delta V\sim V\sim M_{\rm mod}^2\, M_{\rm pl}^2$ and the expressions
for $M_{\rm s}$  and $M_{\rm KK}$ for model A, we obtain for a
D5-domain wall
\eq{
       B_{\rm D5}\sim {M_{\rm s}^8\over M_{\rm Pl}^2\, M_{\rm
           mod}^6}\sim {1\over \hat{\mathfrak f}^2 h q^3}\,,\qquad
     A^2_{\rm D5}\sim {M_{\rm s}^4\over M_{\rm Pl}^2\, M_{\rm
           mod}^2}\sim {1\over \hat{\mathfrak f}^2} \,,
}
and for a NS5 domain wall
\eq{
       B_{\rm NS5}\sim {M_{\rm KK}^{12}\, M_{\rm Pl}^2\over M_{\rm
           s}^8 \, M_{\rm
           mod}^6}\sim {\hat{\mathfrak f}^2 \over h^5 q^3}\,,\qquad
     A^2_{\rm NS5}\sim {M_{\rm KK}^6\over M_{\rm s}^4\, M_{\rm
           mod}^2}\sim {1\over h^2}\,,
}
where $\hat{\mathfrak f}=-\tilde{\mathfrak f}$.
This is the result just for  unit one domain walls, i.e. those
changing $\Delta \hat{\frak f}=\Delta h=1$. As mentioned we
need to satisfy the tadpole cancellation conditions $h \op\hat{\frak
  f}={\rm const.}$ and $q\op \hat{\frak f}={\rm const}$. 
The solution to these constraints is $\hat{\frak f}\to \kappa
\hat{\frak f}$ while $h\to \kappa^{-1} h$ and  $q\to \kappa^{-1} q$.
For the vacuum energy to decrease, we need $\kappa>1$.
As a consequence, the relevant domain wall must change the three fluxes
as
 $\Delta \hat{\frak f}\sim \hat{\frak f}$, $\Delta h\sim h$ and $\Delta q\sim q$
so that also the  individual tensions
scale as $\sigma^{(\kappa)}_{\rm D5}\sim\hat{\frak f}\sigma_{\rm D5}$ and
$\sigma^{(\kappa)}_{\rm NS5}\sim h\sigma_{\rm NS5}$. Including these
factors, one finds
\eq{
     B_{\rm D5}^{(\kappa)}\sim B_{\rm NS5}^{(\kappa)} \sim {\hat{\mathfrak f}^2 \over h q^3}\,,\qquad
     (A^{(\kappa)})^2_{\rm D5}\sim   (A^{(\kappa)})^2_{\rm NS5}  \sim 1\,,
}
i.e. both branes have parametrically the same action.
Therefore,  assuming that
the  bound state is essentially at threshold, the tunneling amplitude
scales as 
\eq{   P\sim \exp\left(- \hat{\mathfrak f}^2 \right) ,
}
so that one has  parametric control to suppress such tunneling transitions.

%%%%%%%%%%%%%%%%%%%%%%%%%%%%%%%%%%%%%%%%%%%%%%%
%%%%%%%%%%%%%%%%%%%%%%%%%%%%%%%%%%%%%%%%%%%%%%%
%%%%%%%%%%%%%%%%%%%%%%%%%%%%%%%%%%%%%%%%%%%%%%%
%%%%%%%%%%%%%%%%%%%%%%%%%%%%%%%%%%%%%%%%%%%%%%%

\subsection{Soft masses of MSSM on   ${\rm D}7$-branes}

In this section, we study some of the particle-physics aspects of the
scaling-type minima presented in the previous sections.
Since the vacua generically break supersymmetry, we are in particular
interested in the induced supersymmetry breaking scale in some
${\rm D}7$-brane sector supporting the Standard Model.
We have already seen that the gravitino mass scale is of  the same
order as the moduli masses. For models of large-field inflation,
this scale is of the order $10^{14}-10^{15}\,$GeV, thus leading
to a very high supersymmetry breaking scale. 
Motivated by the sequestered scenario in the LVS framework \cite{Blumenhagen:2009gk},
we can therefore ask whether also here it is possible
to obtain soft-masses that are  smaller than the 
gravitino mass.

\subsubsection{Bulk  scenario}

We now add an extra four-cycle to the geometry and compute
the soft gaugino and sfermion masses for magnetized ${\rm D}7$-branes wrapping
that cycle and supporting the MSSM. 
For concreteness we consider the model C described in section \ref{sss:C}.
Due to the FW anomalies, we cannot place a magnetized  ${\rm D}7$-brane on the
single four-cycle. Thus, in order to also allow a sector where the
Standard Model can be supported, we need to introduce additional four-cycles
into the geometry.
In order to avoid brane deformation moduli, we  deform the geometry such that we introduce 
a further del Pezzo surface so that the volume form is given by the
familiar swiss-cheese type 
\eq{
{\cal  V}=\tau^{3\over 2}- \tau_s^{3\over 2} \,.
}
Moreover, to avoid Freed-Witten anomalies,  the superpotential
should  not depend on the K\"ahler modulus $T_s$.

We now assume that the MSSM is realized on stacks of  magnetized
${\rm D}7$-branes wrapping  the added four-cycle and proceed as in the analysis 
of the sequestered LVS scenario \cite{Blumenhagen:2009gk}. The axion
$\rho_s$ in $T_s$ becomes massive via the St\"uckelberg mechanism and
the K\"ahler modulus $\tau_s$ appears in  an induced
Fayet-Iliopoulos term that shrinks the del Pezzo surface to
zero size $\tau_s\sim 0$. Since $W$ does not depend on $T_s$ and
$K^{\ov T_s,i}\partial_iK =-2 \tau_s\sim 0$, we find $F^{T_s}=0$.

Focusing on the supersymmetry breaking strictly stable minimum \eqref{stablemodelC}, 
using the formalism from \cite{Brignole:1997dp}, let us compute the soft supersymmetry breaking masses on
the magnetized ${\rm D}7$-branes wrapping the small cycle.
The gaugino masses are then given by
\eq{
       M_a= {1\over 2}\op ({\rm Re} f_a)^{-1}   F^i \partial_i f_a \,,
}
with 
\eq{
       F^{i}=  e^{K\over 2} K^{i\ov j} D_{\ov j} \ov W \,,
}
and where $f_a=T_s+\chi_a S$ is the holomorphic gauge kinetic function
for the ${\rm D}7$-brane.
Here $\chi_a={1\over 4\pi^2} \int F_a\wedge F_a$ is again the instanton number of the gauge flux
on the ${\rm D}7$-branes.
Using  $F^{T_s}=0$ and evaluating $F^S$, we find for the gaugino masses
\eq{
     M^2_{a}=\mu_{a} \, {h_1 (q_1)^3 \over ({\mathfrak f}_0)^{3\over 2}
    (\tilde{\mathfrak f}^1)^{1\over 2}}\: {M_{\rm Pl}^2 \over 4\pi \cdot 2^4}\sim M^2_{\frac32} \,,
}
with $\mu_{a}=12$.
The sfermion masses are given by 
\eq{
       M^2_\alpha=M^2_{\frac32} +V_0-F^{\ov i}
       F^j \partial_{\ov i} \partial_j \log  Z_\alpha \,,
}
where $Z_\alpha$ is the K\"ahler metric for the matter field.
It was argued in \cite{Conlon:2006tj} that for magnetized branes on a small
shrinkable cycle, at tree-level one has $Z_\alpha=k_\alpha/\tau$.
Assuming also an uplift mechanism to $V_0=0$, the sfermion masses
become
\eq{
       M^2_\alpha=M^2_{\frac32} -{(F^{T})^2\over 4\op \tau^2}=
\mu_{\alpha} \,{h_1 (q_1)^3 \over ({\mathfrak f}_0)^{3\over 2}
    (\tilde{\mathfrak f}^1)^{1\over 2}} {M_{\rm Pl}^2 \over 4\pi \cdot 2^4} \sim M^2_{\frac32} \,,
}
with $\mu_\alpha=28$.
Thus, we realize that all soft masses are of the same order as the
gravitino mass. The shrinkable cycle itself therefore does not lead to
sequestering. Opposed to the sequestered LVS scenario \cite{Blumenhagen:2009gk},
the main difference is that here $F^S\ne 0$ already at tree-level.

\subsubsection{Sequestered scenario}

As we have seen, in order to achieve a suppression for the soft-masses,
we need $F^S=0$ at tree-level and we need to turn on a further small
correction that can induce  $F^S\ne 0$. 
In our model search so far we have not\footnote{In section \ref{sss:2U} with $h^{1,2}_- = 2$, we found a model with $F^{U_1} = 0$ (see \eqref{FtermsforU}), which does not help here as only $F^T$ and $F^S$ appear in the gaugino or sfermion masses. Nevertheless, it shows that accidentally a zero auxiliary field is in principle possible.} obtained a non-supersymmetric 
model with $F^S=0$.
Therefore, we now consider the non-supersymmetric minimum 
of model B discussed in section~\ref{sec_h11minusex}, where we will be able to enforce
$F^T=0$ for a subset of fluxes. If we now place an unmagnetized  ${\rm D}7$-brane on
the four-cycle, we have a toy model for the situation we are
interested in.

\paragraph*{Gravity-mediated gaugino  masses}
Let us estimate the size of the gaugino masses, once we take 
into account the $(\alpha')^3$-correction to the K\"ahler potential as
\eq{
\label{Kpert}
               K=-2 \log \left[ \Big( (T+\ov T)+{\kappa\over 4\op(S+\ov S)}
               (G+\ov G)^2 \Big)^{3 \over 2}  +
        {\xi_p\over 2} \op s ^{3\over
          2} \right] -\log (S+\ov S) \,,
}
where $\xi_p=-{\chi({\cal M}) \zeta(3)\over 2\, (2\pi)^3}$. The
superpotential \eqref{superpotwithgmod} for this model is not changed. 
The values of the auxiliary fields 
$F^i=e^{K\over 2}   K^{i\ov j} F_{\ov j}$ at the minimum are
\eq{
\label{VFLwolfsburg}
         F^T&= e^{K\over 2} {8\op i \over 25} \op {\tilde{\mathfrak f}^2 \over q}\, \frac{8x + 3}{1+x}
						\,, \hspace{40pt}
         F^S= -e^{K\over 2} {8\op i \over 5} \op { {\mathfrak
                      f}^2 \over  h}\, {1\over (1+x)} \,,\\
         F^G&= e^{K\over 2} {16 \op i \over 5} \op { {\mathfrak
                      f}^2 \over  f}\, {x \over x+1} \,, 
}
with $x = {f^2 \over \kappa hq}$. Thus, we see that we can force $F^T=0$ by choosing $8x + 3 = 0$. This is only possible for negative $\kappa$ since otherwise we would leave the physical regime and $s_0, \tau_0 < 0$. 

To assess the order of magnitude of the $\alpha'$-correction first notice that \eqref{stringandKKscale} implies 
$M_{\rm KK}/M_{\rm s} \sim  s^{1 \over 4}/{\mathcal V}^{1 \over 6}$ for the ratio of the KK and the string scale.
Using  that the volume is ${\cal V}\propto \left(2 \tau+{\kappa\over 2 s} \psi^2\right)^{3\over 2}$, and substituting 
the values of the moduli in the minimum \eqref{valuesmodgmod}, we then find
\eq{
\label{werder}
{M_{\rm KK}\over M_{\rm s}}\sim  
 \left( {q \over h }\, \right)^{1\over 4}\, ,
}
where $x = -{3 \over 8}$ has been used. Inserting in \eqref{Kpert} we then conclude that the $\alpha'$-correction
is small compared to the tree-level term provided $\xi_p (q/h)^{3/2}\ll 1 $, which can be achieved 
taking $h \gg q$.

A numerical analysis shows that in this regime the former
minimum gets slightly shifted and that the main contribution
to the new value of $F^T$, denoted $F_{\xi}^T$, comes from plugging in  the values
of the old minimum in the corrected expression for $F^T$.
Thus at linear order in $\xi_p$, the 
 induced vacuum expectation value of the 
auxiliary field $F^T$ is parametrically given by
\eq{  {F_{\xi}^T\over F^T_0}\sim \xi_p \left({q\over h}\right)^{3\over 2}
  \sim \xi_p \left({M_{\rm KK}\over M_{\rm s}}\right)^{6} \,,
 }         
where $F_0^T$ is the size of the tree-level   F-term
\eqref{VFLwolfsburg} for $x\ne -{3\over 8}$ and where we used \eqref{werder}.
The gravity-mediated  gaugino masses can now be expressed
as
\eq{
\label{werder2}
       M_a\sim    \left({M_{\rm KK}\over M_{\rm s}}\right)^{6}\,    M_{3\over 2} \,,
}
which is suppressed relative to the gravitino mass scale by a high
power of the ratio of the KK-scale  to the string scale.

\paragraph*{Anomaly-mediated gaugino  masses}

With the tree-level gravity-mediated gaugino masses vanishing at
leading order, the one-loop generated anomaly-mediated gaugino masses
are expected to be generically larger than the next-to-leading order
tree-level masses. In the sequestered LVS scenario, it turned out that
even the leading-order anomaly-mediated contribution vanishes due to
an extended no-scale structure. Let us estimate this contribution
in our model.
The anomaly-mediated gaugino masses are given by \cite{Bagger:1999rd} 
\eq{
  M_a^{\rm anom}=-{g^2\over 16 \pi^2} \Big( (3\op T_G-T_R)
                 M_{\frac32}-(T_G-&T_R) (\partial_i K) F^i \\
        &-{2\op T_R\over
                   d_R} F^i  \partial_i \log \det Z_{\alpha\beta} \Big),
}
where $T_G=N$ is the Dynkin index of the adjoint representation of
$U(N)$ and
$T_R$ is the Dynkin index of some matter representation $R$ of dimension $d_R$.   
 In our simple case of unmagnetized ${\rm D}7$-branes, there is no charged
 matter so that the above formula simplifies. 
Indeed, there is no cancellation between the first and second term and
we obtain
\eq{
          M_a^{\rm anom}= {1\over 16 \pi^2 {\rm Re}(f_a)} {8\, \over
            3}\, N M_{\frac32}={1\over (4\pi)^{3\over 2}}{16\, N\over
            9}{M_{\rm KK}\, M_{\frac32}\over M_{\rm Pl}}\,.
}                
Therefore, we still get a suppression, which generically will be
weaker
than the next-to-leading order gravity-mediated one \eqref{werder2}.
For instance, for $M_{\rm s}\sim 10^{16}\,$GeV, $M_{\rm KK}\sim
10^{15}\,$GeV and $M_{\frac32}\sim 10^{14}\,$GeV, we find
$M_a\sim 10^8\,$GeV and $M_a^{\rm anom}\sim 10^{11}\,$GeV.
Therefore, one can get gaugino masses in the intermediate regime.

As argued in \cite{Blumenhagen:2009gk}, the computation of other soft terms is
sensitive to higher-order corrections to the matter-field
metric and to the uplift, so that we are not pursuing this question here further.
Of course, what we  have presented is just a toy model, as the brane
wrapping the four-cycle is non-chiral and presumably will carry
extra massless deformation modes (that also have to be stabilized).
The purpose of our analysis was to show how one can arrange for a situation where the
gaugino masses are induced by higher-order corrections, and can
therefore be parametrically smaller than the gravitino mass scale.
This is important for string model building, if one wants to have the
supersymmetry breaking scale 
for the MSSM smaller than the GUT or inflation
scale. With the supersymmetry breaking scale in the intermediate regime, one
can realize the scenario of \cite{Ibanez:2012zg}, where gauge coupling unification
is obtained by the F-theory motivated scenario proposed in \cite{Blumenhagen:2008aw}.

%%%%%%%%%%%%%%%%%%%%%%%%%%%%%%%%%%%%%%%%%%%%%%%
%%%%%%%%%%%%%%%%%%%%%%%%%%%%%%%%%%%%%%%%%%%%%%%
%%%%%%%%%%%%%%%%%%%%%%%%%%%%%%%%%%%%%%%%%%%%%%%
%%%%%%%%%%%%%%%%%%%%%%%%%%%%%%%%%%%%%%%%%%%%%%%
%%%%%%%%%%%%%%%%%%%%%%%%%%%%%%%%%%%%%%%%%%%%%%%
%%%%%%%%%%%%%%%%%%%%%%%%%%%%%%%%%%%%%%%%%%%%%%%
%%%%%%%%%%%%%%%%%%%%%%%%%%%%%%%%%%%%%%%%%%%%%%%
%%%%%%%%%%%%%%%%%%%%%%%%%%%%%%%%%%%%%%%%%%%%%%%

\section{Axion monodromy inflation}
\label{sec_cosmo}

We now turn to string-cosmological properties of our models.
Recall that a large tensor-to-scalar ratio points towards
an inflationary scenario with the slow rolling occurring for large field
values $\Theta/M_{\rm Pl}\sim 1-10$. Therefore, in any UV complete
theory of gravity one has to control higher order corrections.
Axions with their perturbative shift-symmetries are good candidates
and various scenarios have been proposed, ranging from natural
inflation, over N-flation to axion monodromy inflation.
The latter can be naturally realized in string theory, where the very
same scalar potential that stabilizes the moduli can also give rise
to the axion potential. Here the shift symmetry of the 
axion is spontaneously broken by the choice
of fluxes in the background, thus giving rise to an effective 
potential and an unwrapping of the compact field range for the axion.

In this section we investigate  whether the flux scaling models we
have discussed  can provide
working examples to realize  F-term axion monodromy inflation
in set-ups with consistent moduli stabilization. As a matter of fact,
this was our initial motivation to look into this part of the 
string/gauged supergravity landscape in more detail.
Please recall the challenges for such a construction that have been listed in
the introduction.

By including all closed string moduli in the tree-level flux 
superpotential, we have available many of the axions that have been
put forward as inflaton candidates in the literature. In general
the eventual inflaton $\Theta$ will be a linear combination of some of the
following axions:
\begin{itemize}
\item There is the universal axion $c$ from the axio-dilaton
  superfield. It was proposed \cite{Blumenhagen:2014gta} that if  this axion is
  part of the inflaton, it can provide an appealing reheating mechanism.
\item The K\"ahler moduli $T_\alpha$ contain  the R-R four-form axions $\rho_\alpha$.
     As opposed to KKLT and the LVS scenario these moduli are
    also stabilized at tree-level by turning on non-geometric $Q$-flux
    \cite{Shelton:2005cf, Hassler:2014mla}.
\item In \cite{Marchesano:2014mla,McAllister:2014mpa} the proposal was to consider the two-forms $B_2$ or $C_2$ 
 as the inflaton. This  can be realized by generating an F-term
 potential for  odd K\"ahler moduli $G^a$ by turning on geometric flux.
\item In the large complex structure limit, extra geometric shift
  symmetries arise so that the quasi-axions ${\rm Re}(U)=u$ can also be considered
   as inflaton candidates \cite{Hebecker:2014eua}.
 \end{itemize}

The purpose of this section is not to construct a fully-fledged
cosmological
model, but, continuing the analysis of \cite{Blumenhagen:2014nba,Hebecker:2014kva}, 
to study the question whether an axion can 
realize large field inflation.  

In the following  the inflaton is
considered to be an initially  massless axion, which receives  a
parametrically smaller mass by turning on additional fluxes. Since
fluxes are of order one, the hierarchy occurs by turning on large
fluxes $\lambda$   for the heavy moduli. Thus, we have a (flux) parameter 
available by which we can control both the mass hierarchy of the
inflaton and the heavy moduli as
well as its backreaction on the other moduli. In fact it has been
shown in \cite{Hebecker:2014kva} that for $\lambda\gg 1$ the backreaction is under
control and that one obtains  the naive  polynomial scalar potential.
We find that this is in principle possible but that, in all examples
 we have looked at, the  KK-scale becomes lighter than the moduli
masses. 

In \cite{Blumenhagen:2015qda} we analyze a toy model for this
scenario, where the backreaction can be taken into account
analytically. There, changing a parameter analogous to $\lambda$ interpolates between chaotic,
linear  and Starobinsky-like inflation.

%%%%%%%%%%%%%%%%%%%%%%%%%%%%%%%%%%%%%%%%%%%%%%%
%%%%%%%%%%%%%%%%%%%%%%%%%%%%%%%%%%%%%%%%%%%%%%%
%%%%%%%%%%%%%%%%%%%%%%%%%%%%%%%%%%%%%%%%%%%%%%%
%%%%%%%%%%%%%%%%%%%%%%%%%%%%%%%%%%%%%%%%%%%%%%%

\subsection{Brief review on large field inflation}

In this section we review the basics of large field inflationary models. (For more details see
for instance \cite{Baumann:2014nda}.)
In general one can distinguish convex and concave scenarios. The prototype examples of the first type are
models  with polynomial scalar potentials, like for instance chaotic
inflation governed by a quadratic potential. Such models would have been
the best candidates to explain the BICEP2  \cite{Ade:2014xna} result $r=0.2$. 
However, due to the PLANCK 2015 data \cite{Ade:2015tva, Ade:2015lrj}
this is explained by the foreground
dust contamination of the signal and substituted by the upper bound
$r<0.113$. Moreover, the reported values for the spectral index and its running are 
$n_s=0.9667\pm 0.0040$ and $\alpha_{s}=-0.002\pm0.013$, respectively.
As a consequence, potentials $V\sim\Theta^p$ with $p\ge 2$ are disfavored.
Instead, the recent results point towards concave models.
The Bayesian analysis reviewed in \cite{Martin:2015dha} also indicates
that plateau-like potentials are the best class of models fitting the current data. 
Nonetheless, in the following we will investigate how polynomial inflation is realized in
our fluxed vacua.

Let us recall the cosmological data needed for our discussion.
For a polynomial potential appearing in the single field
 Lagrangian
\eq{
   {\cal L}= {1\over 2} \partial_\mu \Theta\,  \partial^\mu \Theta+
\mu^{4-p}\,\Theta^p \,,
}
the slow-roll parameters 
\eq{
   \epsilon={1\over 2} \left({V'\over V}\right)^2\,, \hspace{40pt}
  \eta=\left({V''\over V}\right),
}
can be computed  as
\eq{
    \epsilon = {1\over 2}\, {p^2\over \Theta^2}\,,\qquad
   \eta= {p\op (p-1)\over \Theta^2}\, .
}
The number of e-foldings is expressed as 
\eq{
   N_e&=\int^{\Theta_*}_{\Theta_{\rm end}} {V\over V'}\,
   d\Theta= {1\over p} \int^{\Theta_*}_{\Theta_{\rm end}}
      \Theta\, d\Theta
   \simeq  {\Theta_*^2\over 2\op p} \, .
}
Thus, we can write $N_e\sim {p\over 4\epsilon}$.
Therefore, for the spectral indices and the tensor-to-scalar ratio
one obtains 
\eq{
     n_s&=1+2\eta-6\epsilon \sim 1-{(p+2)\over 2 N_e}\,,\\
    n_t&=-2\epsilon\sim -{p\over 2 N_e}\,,\\
    r&=16\epsilon\sim {4p\over N_e}\, .
}
For a quadratic potential, that is $p=2$, and 60 e-foldings, this leads  to  
\eq{
     n_s\sim 0.967\, ,\qquad  n_t\sim -0.017\, ,\qquad r=0.133\,,
}   
which is excluded by the recent measurements from Planck and from BICEP2
at 95$\%$ confidence level \cite{Ade:2015tva, Ade:2015lrj}.
The amplitude of the scalar power spectrum ${\cal P}=2.142\cdot 10^{-9}$ can be
written as follows
\eq{
   {\cal P}\sim {H^2_{\rm inf}\over 8\pi^2\epsilon M^2_{\rm Pl}} \,,
}
which leads to a Hubble constant during inflation of $H_{\inf}\sim 9.14\cdot
10^{13}\,$GeV. Using $V_{\rm inf}=3M_{\rm Pl}^2\, H^2_{\inf}$,  we can 
extract the mass scale of inflation as 
$M_{\rm inf}=V_{\rm inf}^{1/4}\sim 1.96\cdot 10^{16}\,$GeV, 
which is of the order of the GUT scale. Finally, the mass of the axion 
$M^2_{\Theta}=3\eta H^2_{\rm inf}$ comes out as $M_{\Theta}\sim 1.45\cdot
10^{13}\,$GeV. 

From a stringy point of view, for realizing single field inflation in
a controlled way, one needs the hierarchy of
string theoretic and inflationary scales\footnote{For a recent
  discussion of these hierarchies see \cite{Mazumdar:2014qea}.}
\eq{ \label{inflationhierarchy}
M_{\rm Pl}> M_{\rm s}>M_{\rm KK}> M_{\rm inf}\sim M_{\rm mod} >H_{\rm inf}>|M_\Theta|\, .
}
As we have seen, for large field inflation we have $H\sim
10^{14}\,$GeV so that between the Hubble-scale and the Planck-scale
there are only four orders of magnitude for  all the other scales.
Clearly, to achieve and control such a sensitive hierarchy is 
a major challenge for string theory.

%%%%%%%%%%%%%%%%%%%%%%%%%%%%%%%%%%%%%%%%%%%%%%%
%%%%%%%%%%%%%%%%%%%%%%%%%%%%%%%%%%%%%%%%%%%%%%%
%%%%%%%%%%%%%%%%%%%%%%%%%%%%%%%%%%%%%%%%%%%%%%%
%%%%%%%%%%%%%%%%%%%%%%%%%%%%%%%%%%%%%%%%%%%%%%%

\subsection{Realization of polynomial inflation}

Following the procedure  suggested in \cite{Blumenhagen:2014nba},
the idea is to  generate a non-trivial  scalar potential for the axion
$\Theta$ (in the following called inflaton) by turning
on the additional fluxes $f_{\rm ax}$, while the former
fluxes are scaled by a large number $\lambda$.  Thus  the total
superpotential reads
\eq{
W_{\rm inf}=\lambda\op W+f_{\rm ax}\, \Delta W\,.
}
In \cite{Blumenhagen:2014nba} this procedure led to a parametrically
lighter axion mass.\footnote{An alternative idea with $\lambda=1$ was promoted in
  \cite{Hebecker:2014kva}, where certain assumptions about tunings in the string
landscape had to be  made.}
We will see that for our case, where now all moduli appear in $W$, 
the situation is different and more subtle.

\subsubsection{Parametrically light axion  for Model C}

As a first attempt we consider the superpotential
\eq{
\label{inflatonsuperA}
W_{\rm inf}=\lambda\Big(- {\mathfrak f}_0 - 3\tilde{\mathfrak f}^1  U^2 - h_1 U\, S - q_1 U\,  T \, \Big)+ i
 (h_0 S+q_0 T)\,.
}
Let us first proceed under the assumption that the mass for $\Theta$
can be parametrically smaller than the masses of all the other moduli
and that the backreaction of $\Delta W$ on the values of the moduli in
the old minimum \eqref{stablemodelC}
is negligible.
Then we can analyze the problem by first integrating out all heavy moduli
 and computing an effective potential for $\Theta\sim c$.
In practice this means determining the scalar potential induced by $W_{\rm inf}$
and then inserting  the values \eqref{stablemodelC} for the moduli.
In this way we obtain
\eq{
V_{\rm eff}(c)= {1\over  2^7}\Big( A \op(c-B)^2 +V_0 \Big) \,,
}
with 
\eq{
     A=\frac{2^{15\over 4}}{5^{5\over 4} \cdot 3^{1\over 2} }\, {q_1 h_1 (h_1 q_0-
       q_1 h_0)^2 \over  {\mathfrak f}_0^{7\over 2} (\tilde{\mathfrak
         f}^1)^{1\over 2}}\,,
    \hspace{40pt}
    B={q_1 {\mathfrak f}_0\lambda\over 2(
       q_1 h_0- h_1 q_0)} \,,
}
and
\eq{
    V_0=-\frac{7\cdot  2^{7\over 4}}{5^{5\over 4} \cdot 3^{3\over 2} } \, {\lambda^2 h_1 q_1^3
      \over {\mathfrak f}_0^{3\over 2} ( \tilde{\mathfrak
         f}^1)^{1\over 2}} + O({\mathfrak f}_0^{-{5\over 2}})\,.
}
Therefore, the inflaton receives a large vacuum expectation value. In
particular,  inserting its value back into the superpotential
\eqref{inflatonsuperA}
one realizes that the two terms $\lambda W$ and $f_{\rm ax} \Delta W$
scale in the same way with the fluxes. Therefore, one expects that the
backreaction on the  old minimum is substantial. This is confirmed
by a numerical analysis of the scalar potential.
As a consequence, the effect of this form of $\Delta W$ is not under parametric
control and therefore it is not a good candidate for a deformation. 
The problem is the resulting linear term in $\Theta$ in the
effective scalar potential, whose prefactor relative to the quadratic
term is generically of the  order $\lambda W_0$. To control
the backreaction on the former minimum, one needs an effective
potential, where the prefactor becomes zero, as was generically
the case in \cite{Blumenhagen:2014nba}.

Let us consider a different deformation of the superpotential, now
generated by turning on non-geometric $P$ flux
\eq{
W_{\rm inf}=W+\Delta W=\lambda\Big(- {\mathfrak f}_0 - 3\tilde{\mathfrak f}^1  U^2 - h_1 U\, S - q_1 U\,  T \, \Big)- p_0\, S\, T\,.
}
In this case the effective scalar potential becomes
\eq{
 V_{\rm       eff}={1\over 2^7}\Big( A\, c^4 + B\, c^2 +C\Big) \,,
}
with
\eq{
     A=\frac{2^{7\over 4}}{5^{5\over 4} \cdot 3^{1\over 2}} {p_0^2 h_1^3
       q_1  \over {\mathfrak f}_0^{7\over 2}  (\tilde{\mathfrak
         f}^1)^{1\over 2}}\,,\qquad
    B=\frac{2^{3\over 4}}{5^{9\over 4} \cdot 3^{1\over 2}}
   {p_0 h_1
       q_1  (20 \lambda h_1 q_1 +73\sqrt{10} p_0 \tilde{\mathfrak
         f}^1 )\over {\mathfrak f}_0^{5\over 2}  (\tilde{\mathfrak
         f}^1)^{1\over 2}} \,,
}
and
\eq{
  C= \frac{2^{15\over 4}}{5^{5\over 4} \cdot 3^{3\over 2}}
{ q_1  (-h_1^2 q_1^2  \lambda^2 +21 p_0 h_1 q_1 \tilde{\mathfrak
         f}^1 \lambda +90 p_0^2 (\tilde{\mathfrak
         f}^1)^2 )
  \over h_1 {\mathfrak f}_0^{3\over 2}  (\tilde{\mathfrak
         f}^1)^{1\over 2}} \,.
}
For all fluxes and $\lambda$ being positive, the effective potential
has a global minimum at $c=0$. Therefore, in this case we expect
that the backreaction of the $P$-flux term on the other moduli
can be made small.
In this minimum the mass of the canonically normalized inflaton is computed as 
\eq{
              M^2_{\Theta}=\mu_{\Theta} \,
							{p_0 q_1 (\tilde{\mathfrak
         f}^1)^{1\over 2}  (20 h_1 q_1 \lambda +73 \sqrt{10} p_0 \tilde{\mathfrak
         f}^1)\over h_1 {\mathfrak f}_0^{3\over 2}}\,\, {M_\text{Pl}^2 \over 4\pi \cdot 2^7}  \,,
}
with $\mu_{\Theta}=1.6$.
Therefore, in the regime $20 h_1 q_1 \lambda \gg 73 \sqrt{10} p_0 \tilde{\mathfrak f}^1$
the mass of the inflaton can be made  parametrically smaller than the mass of the heavy
moduli
\eq{   
\label{axmodulire}
{M^2_{\Theta}\over M^2_{\rm mod}}\sim  {p_0 \tilde{\mathfrak
      f}^1\over h_1 q_1 \lambda}\,.
}

In order to realize chaotic inflation with a quadratic potential, we
have to ensure that for $c\sim 10\op M_{\rm Pl}$ the quartic term in
the canonically normalized $V_{\rm eff}$ can be neglected.
Indeed, as long as  
\eq{
c^2\ll {q_1  h_1 \lambda \over 10\, p_0 \tilde {\mathfrak f}^1} \,,
}
 the effective potential
is dominated by the quadratic term, where the same combination
of fluxes as in \eqref{axmodulire} appears.
Therefore, we conclude that for sufficiently large $\lambda$ we can
gain  parametric control over the scales in the  inflaton sector while
also  guaranteeing a quadratic potential.
Note that the ratio of the KK-scale  and the moduli masses behaves as
\eq{
       { M^2_{\rm mod} \over M^2_{\rm KK}}\sim  { \lambda^2
         h_1^{} q_1^{} (\tilde{\mathfrak f}^1)^{1\over
           2}\over
     {\mathfrak f}_0^{1\over 2}}\, .
}
For large $\lambda$, it becomes impossible to keep
the KK-scale larger than the heavy moduli mass, while
still having a string scale of the order of the GUT scale. We can
summarize these findings for realizing quadratic inflation by
\eq{
        M_{\rm mod}\,\parag\, M_{\Theta}\quad \Longrightarrow\quad M_{\rm
          mod}\parag\, M_{\rm KK}\,.
}

\subsubsection{Axion potential for Model D}

As a second example  let us also consider Model D, which 
was   designed such that a deformation of the superpotential
of the type already mentioned in \eqref{inflatonsuperA}
can generate a parametrically small mass for the  so far
massless axion. We then choose the deformation
\eq{
\label{inflatonsuperB}
W_{\rm inf}=\lambda\Big(  \hat {\mathfrak f}_1 U+ i \, \tilde{\mathfrak f}^0  U^3+ 3i \, \tilde h^1 U^2\, S+3i \, \tilde q^1 U^2\,  T \,\Big)+ i
 (h_0 S+q_0 T)\, .
}
After integrating out the massive moduli we obtain an effective potential
\eq{
V_{\rm eff}(c)= {1\over
       2^7} \Big(A c^2 +V_0+O({\mathfrak f}_1^{-{3\over 2}}) \Big)
}
with 
\eq{
     A=\frac{324\cdot 2^{1\over 4}}{5^{11/4}} 
{\tilde q^1 \tilde h^1 (\tilde h^1 q_0- \tilde q^1 h_0)^2 \over 
 \hat{\mathfrak f}_1^{7\over 2} (\tilde{\mathfrak     f}^0)^{1\over 2}}\,, 
}
where $V_0\sim \hat{\mathfrak f}_1^{-{1\over 2}}$ is the value given in \eqref{valueminsusy} multiplied by $\lambda^2$.
Since this effective potential is minimized at $c=0$ and the mass of
the inflaton is parametrically smaller than the mass of the heavy
moduli, we expect that one can trust this approximation.
Taking into account the kinetic term for $c$ the mass for the canonically normalized inflaton  becomes 
\eq{   M_\Theta^2=\mu_\Theta \,{\tilde q^1 (\tilde h^1 q_0- \tilde q^1
    h_0)^2 \, (\tilde{\mathfrak f}^0)^{1\over 2}
 \over \tilde h^1\, (\hat{\mathfrak f}_1)^{5\over 2} }  \, {M_{\rm Pl}^2 \over 4\pi \cdot 2^7} \,,
}
with $\mu_\Theta=10$.
Therefore, due to
\eq{   
{M^2_{\Theta}\over M^2_{\rm mod}}\sim  {
(\tilde h^1 q_0- \tilde q^1 h_0)^2 \, (\tilde{\mathfrak  f}^0)^2\over
\lambda^2 (\tilde
h^1\, \tilde
q^1)^2\, \hat{\mathfrak f}_1^{2} }\,,
}
the mass of the inflaton can be made  parametrically smaller than the mass of the heavy
moduli by choosing the flux $\lambda{\mathfrak f}_1$ large enough. 
However, we come into conflict with the separation of the KK and moduli scales. Recalling the ratio of the Kaluza-Klein scale and the moduli masses \eqref{massratioD},
\eq{
       { M^2_{\rm mod}  \over M^2_{\rm KK}}\sim  { \lambda^2\,
         (\tilde h^1)^{} (\tilde q^1)^{} (\hat{\mathfrak f}_1)^{1\over 2}
\over (\tilde{\mathfrak f}^0)^{1\over 2}}\,,
}
we can derive the relation
\eq{
          {M^2_{\Theta}\over M^2_{\rm mod}}\cdot     { M^8_{\rm mod}  \over M^8_{\rm KK}}\
         \sim \lambda^6 (\tilde h^1)^{2} (\tilde q^1)^{2} (\tilde h^1
         q_0- \tilde q^1 h_0)^2\ge 1\,.
}
We want both quantities on the right to be smaller than
one, but this is not compatible with the mass scales
we derived for this model. 

We conclude from this analysis that our attempts to gain parametric control over axion masses were
only half-successful. By turning on additional fluxes it was possible  to find slightly shifted  minima, where the
lightest axion was parametrically smaller than all the other
moduli. However, it was not possible at the same time to keep the
heavy moduli lighter than the KK-scale. All this is reflected in the
simple formula
\eq{
        M_{\rm mod}\,\parag\, M_{\Theta}\quad \Longrightarrow\quad M_{\rm
          mod}\parag\, M_{\rm KK}\,.
}
We close this section with two remarks:
\begin{itemize}
\item{These tree-level flux induced potentials 
clearly provide a
generic framework for realizing F-term axion monodromy inflation in 
type IIB string theory, while also controlling the masses of 
other relevant moduli.  They contain all the closed string axions that
have been proposed in the literature as inflaton candidates.
The models discussed in this section involved the three axions
$\{\rho,c,u\}$. Note in particular, that in contrast to the no-go
results from \cite{Blumenhagen:2014nba}, for the generic superpotential involving also the
K\"ahler moduli,  the universal axion $c$ could also be present
in the linear combination for the inflaton. 
Moreover, by turning on also the geometric fluxes, generically the  
orientifold odd axions ${\rm Im}(G)=C_2$
would appear in the inflaton. We did not explicitly discuss an example
of this class, as the two models presented in sections \ref{sec_h11minusex} and \ref{freezing}  
were plagued by massless saxions. }

\item{In \cite{Blumenhagen:2015qda} we analyze backreaction issues.
While the models discussed in the present section are already
quite involved, in \cite{Blumenhagen:2015qda} a simple toy model based
on Model A is defined, for which the backreaction can be solved
analytically. 
For very large values of a parameter similar to $\lambda$, indeed the potential becomes
effectively quadratic while for decreasing values the effect of the
flattening becomes more and more visible. First, one gets an effective
linear potential while for values of $O(1)$ the potential becomes 
Starobinsky-like. 
However, in this regime also the hierarchy between
the inflaton and the heavy moduli mass diminishes so that one
is actually dealing with a model of multi-field inflation\footnote{We
  thank Francisco Pedro for pointing this out to us.}.}
\end{itemize}

%%%%%%%%%%%%%%%%%%%%%%%%%%%%%%%%%%%%%%%%%%%%%%%
%%%%%%%%%%%%%%%%%%%%%%%%%%%%%%%%%%%%%%%%%%%%%%%
%%%%%%%%%%%%%%%%%%%%%%%%%%%%%%%%%%%%%%%%%%%%%%%
%%%%%%%%%%%%%%%%%%%%%%%%%%%%%%%%%%%%%%%%%%%%%%%
%%%%%%%%%%%%%%%%%%%%%%%%%%%%%%%%%%%%%%%%%%%%%%%
%%%%%%%%%%%%%%%%%%%%%%%%%%%%%%%%%%%%%%%%%%%%%%%
%%%%%%%%%%%%%%%%%%%%%%%%%%%%%%%%%%%%%%%%%%%%%%%
%%%%%%%%%%%%%%%%%%%%%%%%%%%%%%%%%%%%%%%%%%%%%%%

\section{Conclusions and  outlook}

In this paper we have proposed a certain large scale scenario of  tree-level
moduli stabilization.  We considered the class of non-supersymmetric, strictly
stable minima of the scalar potential generated by type IIB
orientifolds on  CY three-folds with non-trivial geometric and non-geometric fluxes
turned on.
This gives the scalar potential of orientifolded ${\cal N}=2$ gauged
supergravity
and also involves the orientifold odd moduli made up by the
$B_2$ and $C_2$ two-forms.

We have presented an algorithm to construct
such minima. Their characteristic feature is a certain scaling
with the fluxes that allows to parametrically control many properties of
the vacuum. For instance, it is easy to guarantee that all moduli are
stabilized  in the perturbative regime where higher order corrections are suppressed.
We have started our investigation with simple models with 
a few moduli. By going to more involved models, we encountered the 
appearance of tachyonic states. For multiple K\"ahler moduli we
identified a general mechanism, involving the addition of certain D7-branes, which
allows to uplift a class of tachyonic modes. For models with multiple
complex structure moduli such a mechanism is still an open question.
We also mention that our model search is not exhaustive.

All the vacua considered are of AdS type. Since all moduli are
stabilized at string tree-level, identifying a proper uplift mechanism for the cosmological
constant is a more involved task.  We provided
a possible uplifting term but could not  justify how it could arise.
It would be interesting to really find a stringy realization of this type of uplift.
In the literature it has been asserted that there exist de Sitter
vacua  for non-geometric flux models. It would be interesting to
find out whether extending the set of fluxes in our models can lead to dS minima, while
maintaining other desirable properties.

We also addressed some phenomenological issues. Since all moduli
are stabilized at tree-level, the whole physics is expected to happen
at the high scale. For ultra-large fluxes one could in principle
lower the moduli masses and the gravitino mass scale.
We have computed soft supersymmetry
breaking masses on MSSM-like D$7$-brane set-ups. Generically, the 
supersymmetry breaking scale is given by the gravitino mass which is
of the same scale as the moduli masses. One can arrange for 
sequestering of the gravity mediated terms, but then anomaly mediation
happens to be the dominant contribution.  This  allows for a further
suppression of the soft masses down to e.g. the intermediate regime.

A technical problem is that models
with non-geometric fluxes do not admit  a proper dilute flux limit. Thus,
one cannot argue that the minima found in the effective
four-dimensional theory can be uplifted to true solutions of the ten-dimensional
string equations of motion or double field theory. This is an
open issue whose eventual clarification relies on further progress in
the understanding of non-geometric backgrounds in e.g. generalized
geometry and double field theory. Another generic feature is that 
to achieve parametric control over the perturbative regime, the uplift mass-scale is larger than the string scale.

In the final section, we applied our results to the study of inflation, more
concretely to F-term monodromy inflation with the inflaton given by an axion.
In particular we asked the question whether by an appropriate 
scaling of the fluxes one can engineer viable models with polynomial
potentials. We find that this is in principle possible via turning on additional
fluxes, though at the expense of introducing large flux quanta and
of making the moduli masses larger than the KK-scale.
One way of interpreting these difficulties we realized in controlling all mass scales
at the same time, is that maybe string theory, as a UV complete theory, wants
to tell us that one has to give up at least some of the usual order of
scales. A proper description of large field inflation in string theory might require
to take some of the Kaluza-Klein and string states into account from the very
beginning.

%%%%%%%%%%%%%%%%%%%%%%%%%%%%%%%%%%%%%%%%%%%%%%%
%%%%%%%%%%%%%%%%%%%%%%%%%%%%%%%%%%%%%%%%%%%%%%%
%%%%%%%%%%%%%%%%%%%%%%%%%%%%%%%%%%%%%%%%%%%%%%%
%%%%%%%%%%%%%%%%%%%%%%%%%%%%%%%%%%%%%%%%%%%%%%%
%%%%%%%%%%%%%%%%%%%%%%%%%%%%%%%%%%%%%%%%%%%%%%%
%%%%%%%%%%%%%%%%%%%%%%%%%%%%%%%%%%%%%%%%%%%%%%%
%%%%%%%%%%%%%%%%%%%%%%%%%%%%%%%%%%%%%%%%%%%%%%%
%%%%%%%%%%%%%%%%%%%%%%%%%%%%%%%%%%%%%%%%%%%%%%%

\vskip3em
\noindent
\centerline{\emph{Acknowledgments}}
\medskip

\noindent
We thank D.~Baumann, G.~Dall'Agata, K.~Choi, X.~Gao, A.~Hebecker, O.~Loaiza-Brito, 
A.~Lukas, D.~L\"ust, L.~Martucci,
A.~Mazumdar, F.~Pedro, F.~Quevedo, P.~Shukla,
S.~Theisen, P.~Vaudrevange  and T.~Weigand for helpful discussions.   
A.F. thanks the Alexander von Humboldt  Foundation for support.
E.P. is supported by the MIUR grant FIRB RBFR10QS5J
and the Padua University Project CPDA144437.

%%%%%%%%%%%%%%%%%%%%%%%%%%%%%%%%%%%%%%%%%%%%%%%
%%%%%%%%%%%%%%%%%%%%%%%%%%%%%%%%%%%%%%%%%%%%%%%
%%%%%%%%%%%%%%%%%%%%%%%%%%%%%%%%%%%%%%%%%%%%%%%
%%%%%%%%%%%%%%%%%%%%%%%%%%%%%%%%%%%%%%%%%%%%%%%

\clearpage
\appendix

\section{A model with parametric control $M_{\rm s}>M_{\rm up}$}
\label{app_a}

We again consider a model with $h^{2,1}_-=1$ and $h^{1,1}_+=1$, that can also be related to the isotropic six-torus
with K\"aher potential
\eq{
\label{KSTUC}
K=-3 \log (T+\ov T)-\log (S+\ov S)-3\log\big(U + \ov U\big)\, .
}
For the flux superpotential we now choose 
\eq{
\label{superpotCapp} 
W= \hat {\mathfrak f}_0 -  i\op {\mathfrak f}_1  U +  h\op U^3\op S + q\op U^3\op  T \,,
}
where $h={\tilde h}^0$,  $q={\tilde q}^0$ and $ \hat {\mathfrak f}_0 =  - {\mathfrak f}_0$. 
Analyzing the  scalar potential, we find a non-supersymmetric,
non-tachyonic minimum with stabilized  moduli 
\eq{
\label{adsminimumapp}
 \displaystyle \tau &= {5^{1\over 2}\over 2^{3\over 4}\cdot 3^{11 \over 4} }
{ {{\mathfrak f}}^3_1\over q \, \hat{\mathfrak f}_0^2}\,, \qquad
 \displaystyle  s = {2^{5\over 4} \cdot 5^{1\over 2}\over 3^{15\over 4}}
 {{{\mathfrak f}}^3_1\over h \,\hat{\mathfrak f}_0^2}\,, \qquad
  h c+ q \rho=-\, {2 \over 27}\, {{{\mathfrak f}}^3_1\over \hat{\mathfrak f}_0^2} \,,\\
 \displaystyle  v &= {3 \cdot 6^{1\over 4}\cdot 5^{1\over 2}\over (5+\sqrt 6)} 
  {\hat{\mathfrak f}_0\over {{\mathfrak f}}_1}\,,\qquad 
  \displaystyle  u = - {3 \cdot \sqrt 6\over (5+\sqrt 6)} 
  {\hat{\mathfrak f}_0\over {{\mathfrak f}}_1}\,.
}  
To stay in the physical region we take all fluxes positive. It is interesting to notice that
in this example the flux tadpoles are positive.

The value of the cosmological constant in the AdS minimum is
\eq{
   V_0=-{6561 \cdot 3^{ 1 \over 4} \over 50\cdot 5^{1\over 2}\cdot 2^{3\over 4}} {h\, q^3\,
     \hat{\mathfrak f}_0^7\over {{\mathfrak f}}_1^9} {M_{\rm Pl}^4\over 4\pi}\,.
}
For the ratio of the string scale and the uplift scale it then follows
\eq{
     {M^4_{\rm s}\over M^4_{\rm up}}={54 \cdot 6^{3\over 4} \cdot 5^{1\over 2}\cdot
     \pi^3}\; {\hat{\mathfrak f}_0\over {{\mathfrak f}}^3_1}\,,\qquad\quad
   {M^4_{\rm KK}\over M^4_{\rm up}}={9 \cdot 3^{3\over 4} \cdot 5^{1\over 2}
     \over 2^{5\over 4}\cdot \pi} {q\, \hat{\mathfrak f}_0\over h\,{{\mathfrak f}}^3_1} \,,
}
so that for $\hat {\mathfrak f}_0/ {{\mathfrak f}}^3_1>1$ we have gained
parametric control. However, in this regime parametrically we get
$\tau,s\paras 1$. 

To give a concrete example, let us 
choose the numerical values $\hat {\mathfrak f}_0=1$ and ${{\mathfrak f}}_1 = 3$, leaving $h$ and $q$ free.
For the moduli we readily obtain
\eq{
   \tau=\frac{1.75}{q}\,,\quad s=\frac{2.33}{h}\,,\quad v=0.47\,,\quad u=-0.33\,,\quad
   c+\rho= - 2 \,,
}
and  for the mass scales
\eq{
    {M_{\rm Pl}\over M_{\rm s}}=\frac{1.78}{h^{\frac14} q^{\frac34}}\,,\qquad 
     {M_{\rm s}\over M_{\rm up}}=4.8\,,\qquad 
 {M_{\rm KK}\over M_{\rm up}}=\frac{0.69 \, q^{\frac14}}{h^{\frac14}}\,. 
}
This shows  that it is not possible to have all moduli in the
perturbative regime while also attaining $M_{\rm up}<M_{\rm KK}$.
For the purpose of realizing models of large field inflation, this means
that we get parametric control of $M_{\rm inf}\paras M_{\rm KK}$ only for $\tau \paras 0.3$.

After all, independent of parametrically controlling certain ratios of
scales or not, in models
of  large field inflation the mass scales  are pushed to the threshold of having control.
Moreover, with all scales being close together, extra numerical factors matter.

\clearpage
\bibliography{references}  
\bibliographystyle{utphys}

%%%%%%%%%%%%%%%%%%%%%%%%%%%%%%%%%%%%%%%%%%%%%%%
%%%%%%%%%%%%%%%%%%%%%%%%%%%%%%%%%%%%%%%%%%%%%%%
%%%%%%%%%%%%%%%%%%%%%%%%%%%%%%%%%%%%%%%%%%%%%%%
%%%%%%%%%%%%%%%%%%%%%%%%%%%%%%%%%%%%%%%%%%%%%%%

\end{document}